\documentclass[11pt]{article}
\usepackage[margin=1in]{geometry}
\usepackage{url, dsfont, dsfont}
\usepackage{amsmath, prodint, amssymb, amsthm, amsfonts}
\usepackage{graphicx, setspace, caption, float, moreverb, subcaption, threeparttable, multicol, multirow, lscape, bbm, tikz, colortbl, dcolumn, float, enumerate, enumitem, booktabs, makecell}
\usepackage{algorithm,algorithmic}
\usepackage{authblk}
\usepackage[round]{natbib}
\usepackage{xcolor}
\usepackage[colorlinks=true, linkcolor=red, citecolor=blue, urlcolor=blue]{hyperref}
\usepackage{xr}
\usepackage{url}

\newcommand{\mc}{\mathcal}
\newcommand{\mb}{\mathbf}
\newcommand{\bd}{\boldsymbol}

\newcommand{\Ex}{\mathbb{E}}
\newcommand{\Vx}{\mathbb{V}}

\newcommand{\Px}{\mathbb{P}}

\newcommand{\Ix}{\mathbb{I}}
\newcommand{\stdest}{\widehat\theta^\mathrm{std}}
\newcommand{\surrest}{\widehat\theta^\mathrm{surr}}
\newcommand{\adptest}{\widehat\theta^{\mathrm{saces}}}
\newcommand{\independent}{\perp\!\!\!\perp}

\newtheorem{assumption}{Assumption}[section]
\newtheorem{theorem}{Theorem}[section]
\newtheorem{corollary}{Corollary}[section]
\newtheorem{condition}{Condition}[section]
\newtheorem{proposition}{Proposition}[section]
\newtheorem{lemma}{Lemma}[section]
\newtheorem{remark}{Remark}[section]

\numberwithin{equation}{section}
\numberwithin{figure}{section}
\numberwithin{table}{section}

\title{\bf\Large Causal Effect Estimation Under Uncertain Statistical Surrogacy}

\author[1]{Yi Liu}
\author[2]{Alexander W. Levis}
\author[3]{Shu Yang}
\author[4,5,*]{Larry Han}

\affil[1]{Public Health Sciences Division, Fred Hutch Cancer Center,
Seattle, WA, U.S.A.}

\affil[2]{Department of Biostatistics, Epidemiology and Informatics,
University of Pennsylvania, Philadelphia, PA, U.S.A.}

\affil[3]{Department of Statistics, North Carolina State University,
Raleigh, NC, U.S.A.}

\affil[4]{Department of Biostatistics, Brown University,\\
Providence, RI, U.S.A.}

\affil[5]{Vaccine and Infectious Disease Division, Fred Hutch Cancer Center,
Seattle, WA, U.S.A.}

\affil[*]{Correspondence: \texttt{larry\_han@brown.edu}}

\date{}

\begin{document}

\maketitle

\begin{abstract}\normalsize
Surrogate endpoints can improve the efficiency of causal effect estimation, but such gains may depend on structural assumptions. One such assumption is statistical surrogacy, which requires the primary outcome to be conditionally independent of treatment given the surrogate and baseline covariates. We study causal effect estimation with multiple candidate surrogates of uncertain validity under statistical surrogacy, considering settings where the primary outcome is either fully observed or subject to missingness. We characterize the asymptotic efficiency differences between surrogate-assisted and benchmark estimators, examine the bias induced by violations of statistical surrogacy, and identify conditions under which incorporating surrogate information leads to efficiency gains or losses. To address uncertainty in surrogate validity, we propose Surrogate-assisted Adaptive Causal effect EStimation (SACES), a procedure that anchors estimation to a robust benchmark while adaptively selecting among a finite set of surrogate-assisted estimators using an estimated mean squared error criterion. SACES aims to leverage efficiency gains when statistical surrogacy holds while preserving consistency when the assumption is violated. Simulations demonstrate that SACES achieves a favorable bias–variance trade-off across varying degrees of surrogacy violation. We illustrate the proposed methods using data from the ACTG 175 clinical trial to demonstrate how surrogate information can be adaptively incorporated into causal effect estimation. 
\end{abstract}

\textit{Some keywords.} Adaptive estimator selection; Efficiency gains; Semiparametrically efficient estimation; Surrogate endpoint. 

\newpage
\doublespacing

\section{Introduction}\label{sec:intro}

Causal inference often focuses on long-term or clinically meaningful outcomes that may be costly to measure, delayed, or subject to missingness. In such settings, surrogate outcomes, which can be measured earlier or more readily than the primary outcome, may provide valuable predictive information for treatment effect estimation. However, treatment effects on surrogate outcomes do not necessarily reflect those on the primary outcome. For example, aducanumab received accelerated approval for Alzheimer's disease based on reductions in amyloid plaque, despite conflicting findings on clinical outcomes across two phase III trials \citep{parast2025surrogate}. 

Beyond serving as alternative endpoints, surrogates can also be used as auxiliary information to improve the efficiency of estimating treatment effects on the primary outcome, particularly when the primary outcome is only partially observed. One commonly invoked condition is statistical surrogacy, which assumes that the primary outcome is conditionally independent of treatment given the surrogate and baseline covariates \citep{prentice1989surrogate}. This condition requires the surrogate and baseline covariates to capture all treatment-related information relevant to the primary outcome, a strong restriction whose plausibility may vary across candidate surrogate sets. Although imposing this condition can yield substantial efficiency gains, estimators that rely on it may be biased when it is violated. Moreover, its plausibility can be difficult to assess in practice, particularly when multiple candidate surrogates are available. 

In this paper, we study estimation of the average treatment effect (ATE) under uncertain statistical surrogacy, considering both fully observed and missing-at-random primary outcomes. We characterize the efficiency gains from imposing statistical surrogacy and the asymptotic bias arising from its violation. We also investigate how additional post-treatment information affects efficiency when outcomes are missing. Motivated by these results, we develop a Surrogate-assisted Adaptive Causal effect EStimation (SACES) procedure that adaptively selects among surrogate-assisted estimators while retaining an unrestricted estimator as a robust reference.

The remainder of this paper is organized as follows. Sections \ref{subsec:related} and \ref{subsec:contri} review related work and summarize our contributions. Section \ref{sec:prelim} introduces the causal framework. Section \ref{sec:EIF} examines the efficiency and bias consequences of surrogate-assisted estimation. Section \ref{sec:adaptive} presents SACES. Sections \ref{sec:simu} and \ref{sec:data} report simulation studies and an application to the ACTG 175 trial, respectively. Section \ref{sec:conclude} concludes.

\subsection{Related work}\label{subsec:related}

The study of surrogate outcomes has a long history in causal inference and clinical trials. Following the seminal work of \citet{prentice1989surrogate}, several frameworks have been developed for evaluating surrogacy, including causal mediation \citep{robins1992identifiability}, principal stratification \citep{frangakis2002principal}, the proportion of treatment effect explained \citep{freedman1992statistical,parast2016,wang2020model,han2022identifying}, and meta-analytic validation \citep{buyse2000validation,burzykowski2005evaluation}. Other work has considered combining multiple markers to improve their surrogacy properties \citep{wang2023robust}. While this literature primarily focuses on evaluating or constructing surrogate endpoints, we study how to leverage surrogate information for treatment effect estimation when statistical surrogacy is uncertain.

Another line of work uses surrogates as auxiliary information for estimating long-term treatment effects. \citet{athey2020combining} studied the combination of experimental and observational data using surrogate outcomes, while subsequent work developed semiparametric efficiency theory and efficient estimators for related settings \citep{cheng2020robust,chen2023semiparametric,kallus2025role}. Extensions have also considered individualized treatment effects and prediction sets \citep{gao2024role}. In particular, \citet{kallus2025role} characterize efficiency gains from fully observed surrogates when the primary outcome is partially observed, without imposing statistical surrogacy. Related efficiency considerations arise in \citet{rotnitzky2020efficient}, who study how covariate adjustment affects estimation efficiency. In contrast, we investigate the efficiency gains and potential bias arising from imposing statistical surrogacy and examine how additional post-treatment information affects efficiency in the missing-outcome setting.

Finally, our proposed SACES procedure is related to the literature on adaptive estimator selection under bias--variance trade-offs. \citet{rothenhausler2024model} develops a general framework for selecting among potentially biased estimators using a consistent reference estimator. Building on this framework, we develop SACES to adaptively select among a finite collection of surrogate-assisted estimators, without requiring any particular candidate surrogate set to satisfy statistical surrogacy. 

\subsection{Our contributions}\label{subsec:contri}

Our main contribution is SACES, a data-adaptive procedure for estimating the ATE when the validity of statistical surrogacy is uncertain across multiple candidate surrogate sets. Rather than committing to a particular surrogate specification, SACES adaptively selects among a finite collection of surrogate-assisted estimators, using an unrestricted efficient estimator that does not impose statistical surrogacy as a robust reference. Importantly, adapting a general estimator selection framework to this setting is not straightforward, as it requires identifying an appropriate robust benchmark and constructing candidate estimators with well-characterized efficiency and bias properties under uncertain statistical surrogacy. Building on the estimator selection framework of \citet{rothenhausler2024model}, SACES employs a modified estimated mean squared error (MSE) criterion to balance potential efficiency gains against bias arising from violations of statistical surrogacy. Under regularity conditions, SACES achieves a risk no greater than that of the unrestricted reference estimator up to an asymptotically negligible remainder, providing a principled approach to leveraging surrogate information while limiting the consequences of invalid surrogate specifications.

To establish this theoretical foundation, we investigate the efficiency and bias consequences of imposing statistical surrogacy when the primary outcome is either fully observed or missing at random (MAR). Building on existing semiparametric efficiency theory, we characterize the efficiency gains from imposing statistical surrogacy and derive the asymptotic bias of the corresponding estimators when this condition fails. We further establish that, in the unrestricted MAR model, incorporating additional post-treatment information does not necessarily improve efficiency. Specifically, information predictive of the primary outcome can reduce the semiparametric efficiency bound, whereas information predictive only of outcome missingness can increase it. Our results provide a theoretical basis for identifying a suitable robust benchmark and constructing appropriate surrogate-assisted candidate estimators.

\section{Preliminaries}\label{sec:prelim}

\subsection{Notation, causal estimand, and identification assumptions}\label{subsec:setup}

Consider data from either a randomized trial or an observational study. Throughout, we work under the potential outcomes framework \citep{Neyman1923, Rubin1974}. For each study participant, let $\mb X$ denote baseline covariates and let $A\in\{0,1\}$ denote treatment assignment, with $A=0$ for control and $A=1$ for active treatment. Let $Y(a)$ denote the potential primary outcome under treatment level $a$, for $a=0,1$. We assume consistency, so that the observed primary outcome is
$Y = Y(A) = A Y(1) + (1-A)Y(0).$ We also assume no interference and well-defined treatment levels, collectively referred to as the Stable Unit Treatment Value Assumption (SUTVA).

We additionally consider post-treatment surrogate information. Let $Q(a)$ denote the potential surrogate value under treatment level $a$, and let
$Q = Q(A) = A Q(1) + (1-A)Q(0)$ denote the observed surrogate under consistency. The surrogate may be scalar or multivariate; for example, $Q \equiv (Q_1,\ldots,Q_T)$ for a finite number of candidate surrogates. In practice, $Q$ represents intermediate information that is typically easier and/or earlier to measure than the primary outcome $Y$. 

The observed data for each participant are $\mc O=(\mb X,A,Q,Y)\sim \Px$, and a random sample of size $n$ from $\mathbb{P}$ is denoted by $\{\mc O_i\}_{i=1}^n$. Our target causal estimand is the average treatment effect,
$\theta = \Ex\{Y(1)-Y(0)\}.$ To identify $\theta$ from the observed data distribution $\mathbb{P}$, we make the following assumptions. 

\begin{assumption}[Unconfoundedness]\label{asp:unconf}
    (i) $Q(a)\independent A\mid \mb X$; and (ii) $Y(a)\independent A\mid \mb X,Q(a)$ for $a=0,1$, where ``$\independent$'' denotes independence. 
\end{assumption}
Note that Assumption \ref{asp:unconf} is equivalent to $(Y(a),Q(a))\independent A\mid\mb X$. We prove this equivalence in Appendix \ref{app:proof}. 

\begin{assumption}[Positivity on treatment propensity scores]\label{asp:positivity}
    Let $\pi(a\mid\mb X) = \Px(A=a\mid\mb X)$. Then, there exists a positive constant $c$ such that $0<c\leq\pi(a\mid\mb X)\leq 1-c<1$ with probability 1 for $a=0,1$. 
\end{assumption}

\subsection{Statistical surrogacy condition}\label{subsec:statsurr}

We consider the following statistical surrogacy condition, motivated by classical criteria for surrogate endpoints \citep{prentice1989surrogate} and subsequent work on surrogate-based causal effect estimation \citep{athey2026surrogate}.

\begin{condition}[Statistical surrogacy]\label{cond:statsurr}
The primary outcome is conditionally independent of treatment given baseline covariates and the observed surrogate:
$Y \independent A \mid \mb X,Q.$
\end{condition}

\begin{figure}[ht]
    \centering
    \includegraphics[width=0.7\linewidth]{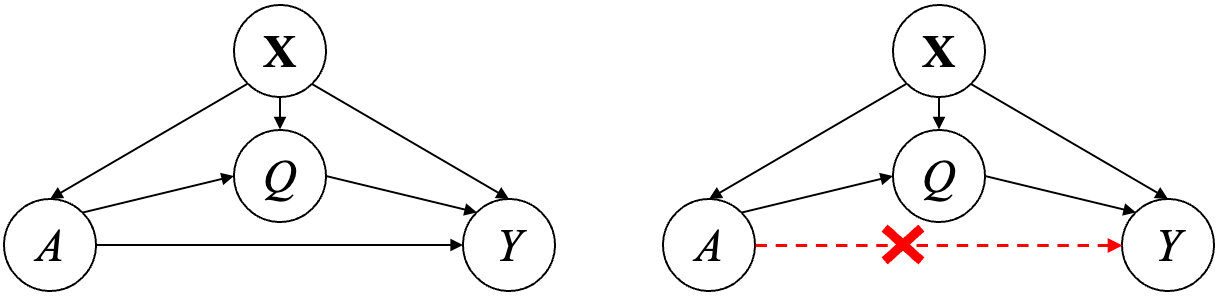}
    \caption{Directed acyclic graphs for the fully observed outcome setting. The left panel shows the unrestricted graph, in which treatment may remain associated with $Y$ after conditioning on $(\mb X,Q)$; the right panel shows the statistical-surrogacy graph, in which $Y\independent A\mid\mb X,Q$ and the $A\to Y$ edge is absent.}
    \label{fig:DAG-full}
\end{figure}

Compared with the classical operational criterion of \citet{prentice1989surrogate}, which conditions on the observed surrogate alone, Condition \ref{cond:statsurr} additionally conditions on baseline covariates. 
In particular, Condition~\ref{cond:statsurr} implies that
$\Ex(Y\mid A=1,\mb X,Q)=\Ex(Y\mid A=0,\mb X,Q) \equiv \Ex(Y \mid \mathbf{X}, Q).$ Thus, after conditioning on baseline covariates and the surrogate, treatment assignment provides no additional predictive information about the primary outcome. Consequently, the treatment-specific outcome regression
$\Ex(Y\mid A=a,\mb X,Q)$ can be represented by a common regression function of $(\mb X,Q)$ across treatment arms. Figure \ref{fig:DAG-full} illustrates this restriction by contrasting the unrestricted setting with the statistical surrogacy setting.


The appeal of Condition \ref{cond:statsurr} is twofold. First, it enables efficiency gains: if $Q$ explains a substantial portion of the variability in $Y$, conditioning on $Q$ reduces residual noise and improves precision of treatment effect estimates. Second, in many applications $Y$ may be costly, time-consuming, or infeasible to measure on all participants, while surrogates $Q$ are often easier and/or earlier to collect. Leveraging $Q$ under the surrogacy condition allows for estimators that are both more efficient and more robust to missing outcome data. We specifically explore the setting of missing outcome data in Section \ref{subsec:missoutEIF}. We next characterize the corresponding efficiency gains under both fully observed and missing outcome settings. 

\section{Efficiency and Bias in Surrogate-Assisted Estimation}\label{sec:EIF}

We examine the efficiency gains from imposing statistical surrogacy in ATE estimation and the potential bias arising from violations of this condition. We consider settings where the primary outcome is either fully observed or MAR. Building on existing semiparametric efficient influence function (EIF) theory \citep{athey2026surrogate, kallus2025role}, we characterize the asymptotic efficiency differences between estimators with and without imposing Condition \ref{cond:statsurr} and quantify the bias induced when the condition is violated. We further investigate how the choice of post-treatment information affects estimation efficiency in the unrestricted MAR model. Appendix \ref{app:proof} provides a review of the preliminaries of EIF theory, the proposed estimators, their multiple robustness properties and convergence rates in relation to those of the nuisance function estimators, and other supporting technical results. 

\subsection{Fully observed outcomes}\label{subsec:fulloutEIF}

Suppose the primary outcome is fully observed, with $\mc O=(\mb X,A,Q,Y)$. Let $\theta_a=\Ex\{Y(a)\}$ for $a=0,1$, so that $\theta=\theta_1-\theta_0$. We introduce the following nuisance functions: $\pi(a\mid\mb X)=\Px(A=a\mid\mb X)$, $\widetilde\pi(a\mid\mb X,Q)=\Px(A=a\mid\mb X,Q)$, $\widetilde\nu(\mb X,Q)=\Ex(Y\mid\mb X,Q)$, and $\widetilde\mu(a,\mb X)=\Ex\{\widetilde\nu(\mb X,Q)\mid A=a,\mb X\}$.

Let $\stdest$ denote the EIF-based estimator under the unrestricted model, which does not impose Condition \ref{cond:statsurr}. In the fully observed outcome setting, this estimator coincides with the standard augmented inverse probability weighted (AIPW) estimator \citep{robins1994estimation,kallus2025role}. In contrast, let $\surrest$ denote the EIF-based estimator under the model imposing statistical surrogacy. The corresponding EIF representations and estimator constructions are provided in Appendix \ref{app:proof}.

Without imposing statistical surrogacy, incorporating $Q$ does not improve the semiparametric efficiency bound for the ATE when the primary outcome is fully observed \citep{kallus2025role}. However, imposing statistical surrogacy can yield efficiency gains by restricting the observed-data model, as characterized by the following result.

\begin{corollary}[of Theorem \ref{thm:EIF-main} in Appendix \ref{app:proof}]\label{cor:var-main}
Suppose Condition \ref{cond:statsurr} holds and the primary outcome is fully observed, with $\mc O=(\mb X,A,Q,Y)$. Let $\mc V$ and $\widetilde{\mc V}$ denote the asymptotic variances of $\stdest$ and $\surrest$, respectively, evaluated at the same data-generating distribution. Then
\begin{align*}
\mc V-\widetilde{\mc V}
=\Ex\left\{\frac{\widetilde\pi(0\mid\mb X,Q)\widetilde\pi(1\mid\mb X,Q)}{\pi(0\mid\mb X)^2\pi(1\mid\mb X)^2}\Vx(Y\mid\mb X,Q)\right\}\geq 0.
\end{align*}
\end{corollary}

Corollary \ref{cor:var-main} quantifies the efficiency gain from imposing statistical surrogacy, rather than from incorporating surrogate information alone. The difference between the two efficiency bounds depends on the residual outcome variance $\Vx(Y\mid\mb X,Q)$ and the treatment propensity scores conditional on $\mb X$ and $(\mb X,Q)$. Notably, greater predictive strength of $Q$ for $Y$ does not necessarily translate into a larger efficiency gain from imposing statistical surrogacy. In particular, a smaller residual outcome variance may lead to a smaller difference between the two bounds. The proof is provided in Appendix \ref{app:proof}.

To illustrate this result, we conduct a simple simulation with $2{,}000$ Monte Carlo replications, each consisting of $n=1{,}000$ i.i.d. observations. We generate $X\sim\mc N(0,1)$, $A\mid X\sim\mathrm{Bernoulli}\{\mathrm{logit}^{-1}(-0.3+0.8X)\}$, $Q=0.5A+X+\varepsilon_Q$, and $Y=5+0.5Q+X+\varepsilon_Y$, where $\varepsilon_Q\sim\mc N(0,1)$ and $\varepsilon_Y\sim\mc N(0,\varepsilon^2)$ are independent, with $\varepsilon\in\{0.3,1,3,10\}$. By construction, $Y\independent A\mid X,Q$, so Condition \ref{cond:statsurr} holds, and the true ATE is $\theta=0.25$. Moreover, $\Vx(Y\mid X,Q)=\varepsilon^2$, so varying $\varepsilon$ directly controls the residual outcome variance appearing in Corollary \ref{cor:var-main}.

\begin{figure}[H]
\centering
\includegraphics[width=0.85\linewidth]{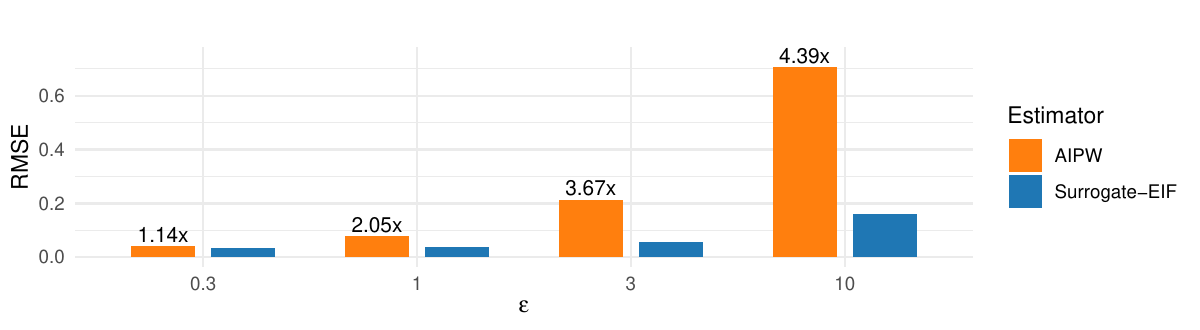}
\caption{RMSE of the standard AIPW estimator $\stdest$ and the surrogate-EIF estimator $\surrest$ across residual standard deviation $\varepsilon$. Numbers above the $\surrest$ bars report $\mathrm{RMSE}(\stdest)/\mathrm{RMSE}(\surrest)$.}
\label{fig:toy}
\end{figure}

We compare the standard AIPW estimator $\stdest$ with the surrogate-EIF estimator $\surrest$. Figure \ref{fig:toy} shows that $\surrest$ achieves a lower root mean squared error (RMSE) than $\stdest$ across all considered settings, with the difference becoming more pronounced as $\varepsilon$ increases. These results illustrate how the efficiency gain from imposing statistical surrogacy depends on the residual outcome variance, consistent with Corollary \ref{cor:var-main}.

\subsection{Outcomes missing at random}\label{subsec:missoutEIF}

\begin{figure}[ht]
    \centering
    \includegraphics[width=0.4\linewidth]{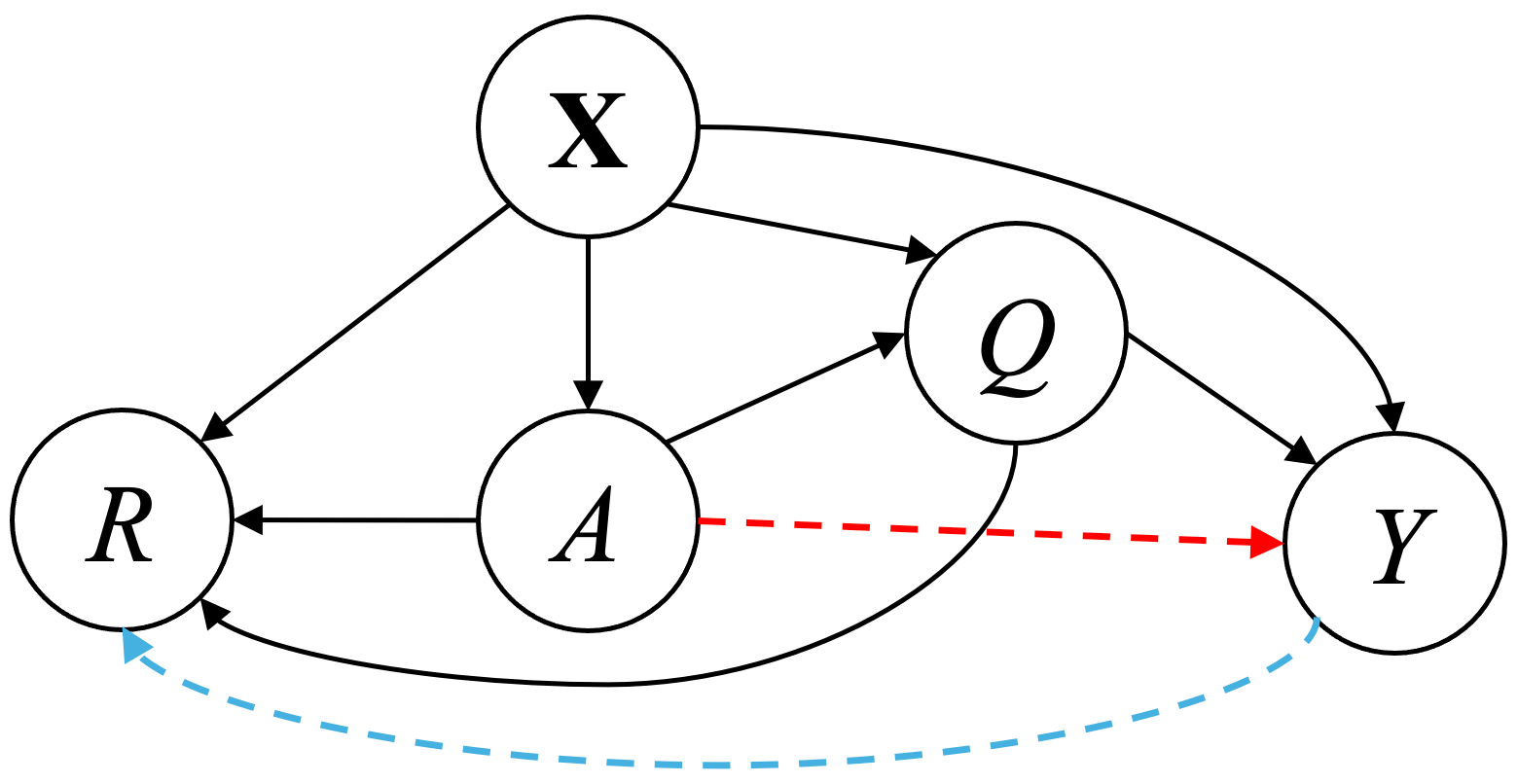}
    \caption{Directed acyclic graph for the missing-outcome setting. The unrestricted graph permits both a direct $A\to Y$ association and outcome-dependent observation through a $Y\to R$ edge, whereas the proposed model excludes these edges under statistical surrogacy and MAR, respectively.}
    \label{fig:DAG-miss}
\end{figure}

We next consider the setting where the primary outcome is subject to missingness, with observed data $\mc O=(\mb X,A,Q,R,RY)$, where $R\in\{0,1\}$ indicates whether $Y$ is observed and $Q$ is available for all participants. We impose the following MAR assumption. Figure \ref{fig:DAG-miss} further illustrates the restrictions imposed by statistical surrogacy and MAR. 

\begin{assumption}[Missing at random]\label{asp:MAR}
$Y\independent R\mid\mb X,A,Q$.
\end{assumption}

Define the pooled and treatment-specific probabilities of observing $Y$ as $\widetilde r(\mb X,Q)=\Px(R=1\mid\mb X,Q)$ and $\widetilde r_a(\mb X,Q)=\Px(R=1\mid\mb X,A=a,Q)$, respectively, for $a=0,1$. We further impose the following positivity condition.

\begin{assumption}[Positivity for outcome observation]\label{asp:positivity-r}
There exists a constant $c>0$ such that $\widetilde r(\mb X,Q)\geq c$ and $\widetilde r_a(\mb X,Q)\geq c$ for $a=0,1$, almost surely.
\end{assumption}

In this setting, let $\stdest$ denote the unrestricted EIF-based estimator under MAR \citep{kallus2025role}, which incorporates $Q$ into the treatment-specific outcome regressions and observation models without imposing statistical surrogacy. Unlike in the fully observed setting, $\stdest$ is no longer the standard complete-data AIPW estimator. Let $\surrest$ denote the EIF-based estimator under the model imposing statistical surrogacy. Together, MAR and statistical surrogacy imply $Y\independent R\mid\mb X,Q$, even when the observation probability depends on treatment (see Lemma \ref{lem:surr-MAR-imply} in Appendix \ref{app:proof}). Consequently, $\surrest$ uses a common outcome regression across treatment arms and the pooled observation probability $\widetilde r(\mb X,Q)$, whereas $\stdest$ uses treatment-specific outcome regressions and observation probabilities $\widetilde r_a(\mb X,Q)$. The corresponding EIF representations and estimator constructions are provided in Appendix \ref{app:proof}.

We next characterize the efficiency gain from imposing statistical surrogacy in the missing-outcome setting.

\begin{corollary}[of Theorem \ref{thm:EIF-miss} in Appendix \ref{app:proof}]\label{cor:var-miss}
Suppose Condition \ref{cond:statsurr}, Assumption \ref{asp:MAR}, and the relevant treatment and observation positivity conditions hold with $\mc O=(\mb X,A,Q,R,RY)$. Let $\mc V$ and $\widetilde{\mc V}$ denote the asymptotic variances of $\stdest$ and $\surrest$, respectively, evaluated at the same data-generating distribution. Then
\begin{align*}
    \mc V - \widetilde{\mc V}
    = \Ex\left\{\widetilde\omega(\mb X,Q)^2\Vx(Y\mid\mb X,Q)\right\}\geq 0,
\end{align*}
where
\begin{align*}
    \widetilde\omega(\mb X,Q) & = \frac{\pi(0\mid\mb X)\widetilde r_0(\mb X,Q) + \pi(1\mid \mb X)\widetilde r_1(\mb X,Q)}{\pi(0\mid\mb X)\pi(1\mid\mb X)}\sqrt{\frac{\widetilde\pi(0\mid\mb X, Q)\widetilde\pi(1\mid\mb X, Q)}{\widetilde r_0(\mb X,Q)\widetilde r_1(\mb X,Q)\widetilde r(\mb X,Q)}}.
\end{align*}
\end{corollary}

Corollary \ref{cor:var-miss} extends the variance comparison in Corollary \ref{cor:var-main} to the missing-outcome setting. The efficiency gain depends on the residual conditional variance of $Y$, treatment propensity scores, and outcome-observation probabilities. In particular, small observation probabilities may amplify the difference between the two efficiency bounds while also leading to unstable finite-sample estimation. When $R\equiv1$, the expression reduces to that in Corollary \ref{cor:var-main}. Importantly, both corollaries characterize the efficiency gain from imposing statistical surrogacy for a fixed information set, rather than the effect of incorporating additional surrogate information, which we investigate in Section \ref{subsec:G-B-surrogates}.

\subsection{Bias under violations of statistical surrogacy}\label{subsec:bias-by-Cond1}

The preceding efficiency comparisons assume that statistical surrogacy holds. To examine the consequences of an invalid surrogate set, define
\begin{align*}
\widetilde m(a,\mb X,Q)=\Ex(Y\mid A=a,\mb X,Q), \qquad
d(\mb X,Q) =\widetilde m(1,\mb X,Q)-\widetilde m(0,\mb X,Q).
\end{align*}
Condition \ref{cond:statsurr} implies $d(\mb X,Q)=0$ almost surely, although the converse need not hold. The following result characterizes the asymptotic bias when the surrogate-assisted estimator is constructed using nuisance estimators converging to their observed-data limits. In the missing-outcome setting, the pooled outcome regression is fitted among participants with $R=1$ and therefore converges to $\Ex(Y\mid\mb X,Q,R=1)$, which need not equal $\Ex(Y\mid\mb X,Q)$ if statistical surrogacy fails.

\begin{proposition}[Bias under violations of statistical surrogacy]\label{prop:bias-surr-main}
Suppose Assumptions \ref{asp:unconf} and \ref{asp:positivity} hold, and the nuisance estimators converge to their respective observed-data limits. Write $\pi_a(\mb X)=\pi(a\mid\mb X)$ and $\widetilde\pi_a(\mb X,Q)=\widetilde\pi(a\mid\mb X,Q)$. When $Y$ is fully observed, the asymptotic bias of $\surrest$ for the ATE is
\begin{equation}
\Delta^{\mathrm{full}}=-\Ex\left\{\frac{\widetilde\pi_0(\mb X,Q)\widetilde\pi_1(\mb X,Q)}{\pi_0(\mb X)\pi_1(\mb X)}d(\mb X,Q)\right\}.
\label{eq:bias-full-main}
\end{equation}
When $Y$ is subject to missingness, suppose additionally that Assumptions \ref{asp:MAR} and \ref{asp:positivity-r} hold, and that the pooled outcome regression is fitted among observed outcomes. Then the asymptotic bias is
\begin{equation}
\Delta^{\mathrm{mis}}=-\Ex\left[\frac{\widetilde\pi_0(\mb X,Q)\widetilde\pi_1(\mb X,Q)}{\widetilde r(\mb X,Q)}\left\{\frac{\widetilde r_0(\mb X,Q)}{\pi_1(\mb X)}+\frac{\widetilde r_1(\mb X,Q)}{\pi_0(\mb X)}\right\}d(\mb X,Q)\right].
\label{eq:bias-mis-main}
\end{equation}
\end{proposition}

Proposition \ref{prop:bias-surr-main} shows that bias depends on the residual treatment-outcome association after conditioning on $(\mb X,Q)$, weighted by treatment overlap and, in the missing-outcome setting, by the observation mechanism. Thus, zero bias is guaranteed under statistical surrogacy, whereas violations of this condition may offset the variance reduction gained by imposing the restriction. When $R\equiv1$, the missing-outcome bias expression in \eqref{eq:bias-mis-main} reduces to the fully observed expression in \eqref{eq:bias-full-main}, as expected. Detailed probability limits and proofs are provided in Appendix \ref{app:proof}.

\subsection{Choosing post-treatment information under MAR}\label{subsec:G-B-surrogates}

A separate question is whether incorporating additional post-treatment information improves estimation efficiency when outcomes are missing. To isolate this question from statistical surrogacy, we consider the unrestricted MAR model without imposing Condition \ref{cond:statsurr}. Let $\mc V_a(S)$ denote the semiparametric efficiency bound for $\theta_a$ when the post-treatment information set $S$ is used to define the MAR model.

\begin{figure}[ht]
    \centering
    \includegraphics[width=0.45\linewidth]{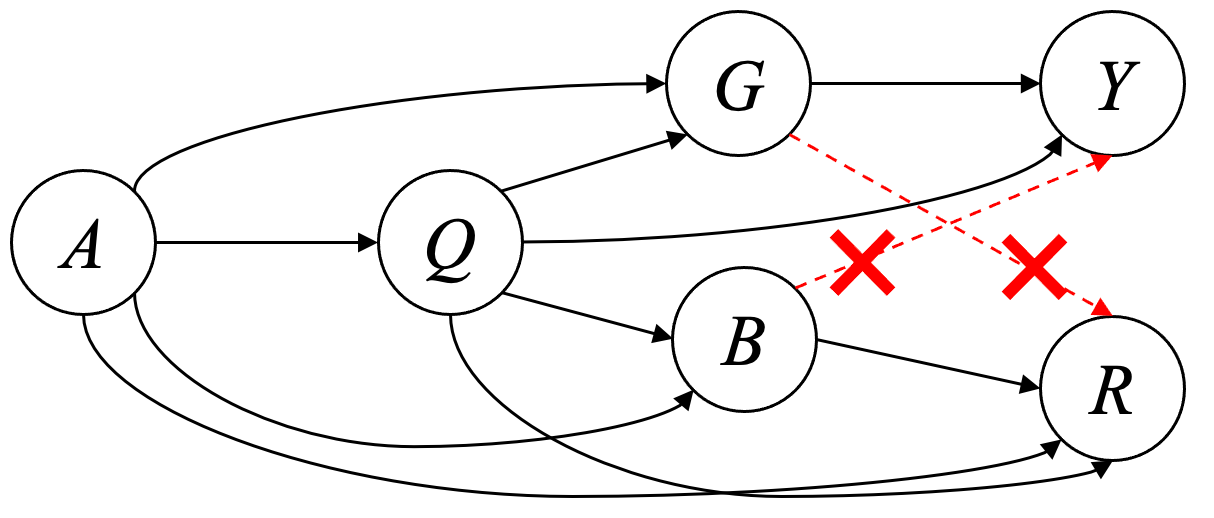}
    \caption{Illustration of additional outcome-predictive information ($G$, ``good'') and missingness-predictive information ($B$, ``bad''). Red dashed arrows indicate absent direct effects. Baseline covariates $\mb X$ and their associated arrows are omitted for clarity.}
    \label{fig:DAG-GB}
\end{figure}

Starting with an existing surrogate set $Q$, consider two additional variables $G$ and $B$, representing ``good'' and ``bad'' information, respectively. Suppose $G$ provides additional information about $Y$ but no further information about outcome missingness, whereas $B$ predicts outcome missingness but provides no additional information about $Y$. Specifically, assume
\begin{align*}
G&\independent R\mid\mb X,A,Q, \quad\qquad
G\independent R\mid\mb X,A,Q,B,\\
B&\independent Y\mid\mb X,A,Q,R, \qquad
B\independent Y\mid\mb X,A,Q,G,R,
\end{align*}
with MAR and positivity holding for all four information sets under comparison. Figure \ref{fig:DAG-GB} illustrates a possible graphical structure consistent with these conditions.

Under these conditions, the corresponding semiparametric efficiency bounds satisfy
\begin{align*}
\mc V_a(Q,G)&\leq\mc V_a(Q)\leq\mc V_a(Q,B),\\
\mc V_a(Q,G)&\leq\mc V_a(Q,G,B)\leq\mc V_a(Q,B),
\end{align*}
while $\mc V_a(Q)$ and $\mc V_a(Q,G,B)$ are not generally ordered. These comparisons follow from Theorems \ref{prop:outcome-predictive-aux} and \ref{prop:missing-predictive-aux} in Appendix \ref{app:proof}. As illustrated in Figure \ref{fig:var-order}, incorporating $G$ yields the smallest efficiency bound, whereas incorporating $B$ without $G$ yields the largest among the four information sets.

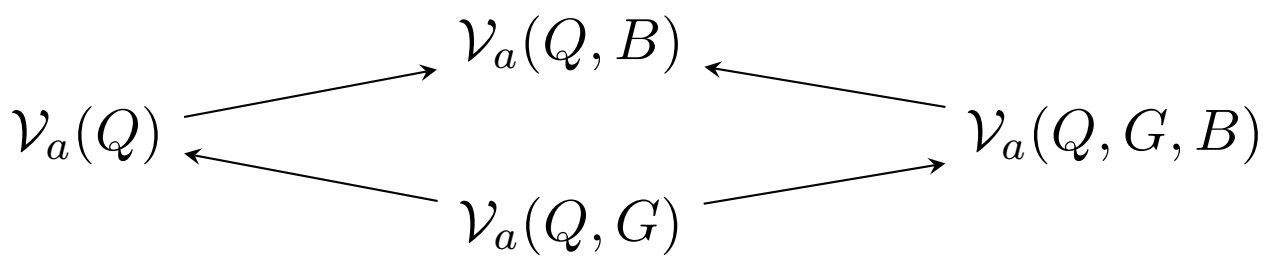
\begin{figure}[ht]
\centering
\begin{tikzpicture}[>=stealth]
    \node (top) at (0,2.8) {$\mc V_a(Q,B)$};
    \node (left) at (-3.2,2.2) {$\mc V_a(Q)$};
    \node (right) at (3.6,2.2) {$\mc V_a(Q,G,B)$};
    \node (bottom) at (0,1.6) {$\mc V_a(Q,G)$};

    \draw[->] (bottom) -- (left);
    \draw[->] (bottom) -- (right);
    \draw[->] (left) -- (top);
    \draw[->] (right) -- (top);
\end{tikzpicture}
\caption{Partial ordering of semiparametric efficiency bounds under MAR. Arrows point from smaller to larger efficiency bounds (allowing equality). The bounds based on $Q$ and $(Q,G,B)$ are not generally ordered.}
\label{fig:var-order}
\end{figure}

These results show that additional outcome-predictive information can improve efficiency, whereas information predictive only of missingness can lead to efficiency loss by increasing the variability of inverse observation weights without reducing residual outcome variation. Importantly, these comparisons concern efficiency bounds across MAR models defined by different conditioning sets, rather than the intrinsic value of observing additional variables. Together with Proposition \ref{prop:bias-surr-main}, these findings highlight the implications of surrogate-set selection for both bias and efficiency, motivating the adaptive estimator selection procedure developed in Section \ref{sec:adaptive}.

\section{Adaptive Estimator Selection Under Uncertain Statistical Surrogacy}
\label{sec:adaptive}

We develop SACES to select among a prespecified finite collection of surrogate-assisted estimators, using an unrestricted efficient estimator as a robust reference. The construction of the candidate class and the choice of reference estimator are guided by the efficiency and bias analyses in Section \ref{sec:EIF}.

Let $\widehat\theta^{(0)}=\stdest$ denote a reference estimator that does not impose Condition~\ref{cond:statsurr}. Its definition depends on whether the primary outcome is fully observed. As specified in Section \ref{sec:EIF}, with fully observed outcomes, we use the standard AIPW estimator; with missing outcomes under MAR, we use the estimator developed by \citet{kallus2025role}, which incorporates $Q$ in both the treatment-specific outcome regressions and missingness models. Thus, $\stdest$ denotes the reference estimator appropriate to each setting. Note that its consistency requires the corresponding identification and nuisance-estimation conditions, but not statistical surrogacy. Suppose we additionally consider $K$ candidate surrogate-assisted estimators, denoted by $\widehat\theta^{(1)},\ldots,\widehat\theta^{(K)}$, each constructed using a prespecified surrogate set. The candidate collection may also include prespecified weighted combinations of the reference and surrogate-assisted estimators, provided that the collection is finite and the resulting candidates satisfy the regular asymptotic linearity conditions below; this includes the weighted grids used in simulation setting II and the ACTG 175 application (Sections \ref{sec:simu} and \ref{sec:data}). 

For $k\geq1$, let $\theta^{(k)}$ denote the probability limit of $\widehat\theta^{(k)}$. If the surrogate set underlying estimator $k$ satisfies Condition \ref{cond:statsurr}, then $\theta^{(k)}=\theta$; otherwise, $\theta^{(k)}$ may differ from $\theta$. We assume that each estimator admits a regular asymptotically linear representation,
\begin{equation}
\widehat\theta^{(k)}-\theta^{(k)}
=
\frac{1}{n}\sum_{i=1}^n\varphi^{(k)}(\mc O_i;\Px)
+
o_p(n^{-1/2}),
\qquad
k=0,1,\ldots,K,
\label{eq:RAL-k}
\end{equation}
where $\theta^{(0)}=\theta$. Let $\sigma_k^2=\Ex\{\varphi^{(k)}(\mc O;\Px)^2\}$ denote the asymptotic variance of estimator $k$, and let $\tau_k^2=\Ex[\{\varphi^{(k)}(\mc O;\Px)-\varphi^{(0)}(\mc O;\Px)\}^2]$ for $k\geq1$.

SACES compares each candidate with the robust reference estimator using a modified estimated MSE criterion. Let $[x]_+=\max(x,0)$ denote the positive-part operator. For $k=1,\ldots,K$, define
\begin{equation}
\widehat{\mc R}^{\,\text{mod}}(k)
=
\left[
\left\{
\widehat\theta^{(k)}
-
\widehat\theta^{(0)}
\right\}^2
-
\frac{\widehat\tau_k^2}{n}
\right]_+
+
\frac{\widehat\sigma_k^2}{n},
\qquad
\widehat{\mc R}^{\,\text{mod}}(0)
=
\frac{\widehat\sigma_0^2}{n}.
\label{eq:riskhat-mod-k}
\end{equation}
The squared difference from the reference estimator reflects potential bias due to violations of statistical surrogacy, after accounting for its sampling variability, whereas the second term reflects the variance of the candidate estimator. The criterion therefore balances potential efficiency gains against potential bias from imposing an invalid surrogate specification.

We select
\begin{equation}
\widehat k
=
\arg\min_{k\in\{0,1,\ldots,K\}}
\widehat{\mc R}^{\,\text{mod}}(k),
\qquad
\adptest
=
\widehat\theta^{(\widehat k)}.
\label{eq:k-hat}
\end{equation}
The quantities $\widehat\sigma_k^2$ and $\widehat\tau_k^2$ are obtained from the estimated cross-fitted influence functions. Complete implementation details are provided in Appendix \ref{app:proof}. 

The following theorem establishes that SACES asymptotically avoids substantial deterioration in risk relative to the robust reference estimator. For each fixed candidate $k$, let $\mc R_n(k)=\Ex\big[\{\widehat\theta^{(k)}-\theta\}^{2}\big]$ denote its MSE for the target $\theta$. We retain the subscript $n$ to emphasize that the risk depends on the sample size.

\begin{theorem}\label{thm:SACES}
Suppose that $K$ is finite and fixed. For each $k=0,1,\ldots,K$, suppose that $\widehat\theta^{(k)}-\theta^{(k)} = n^{-1}\sum_{i=1}^n \varphi^{(k)}(\mc O_i;\Px) + r_{n,k}$, where $\Ex\{\varphi^{(k)}(\mc O;\Px)\}=0$, $\Ex\{\varphi^{(k)}(\mc O;\Px)^2\}<\infty$, and $\Ex(r_{n,k}^2)=o(n^{-1})$. Assume further that $\widehat\sigma_k^2\overset{p}{\to}\sigma_k^2$ and, for $k\geq1$, $\widehat\tau_k^2\overset{p}{\to}\tau_k^2$. Let $\theta^{(0)}=\theta$, and let $\widehat k$ minimize $\widehat{\mc R}^{\,\mathrm{mod}}(k)$ over $k=0,\ldots,K$. Then $\Px\{\theta^{(\widehat k)}=\theta\}\to 1$ and $\mc R_n(\widehat k) \leq \mc R_n(0)+o_p(n^{-1})$. Equivalently, for every $\epsilon>0$, $\Px\left\{\mc R_n(\widehat k)\leq \mc R_n(0)+{\epsilon}/{n}\right\}\to 1$.
\end{theorem}

\begin{remark}
The mean-square remainder condition $\Ex(r_{n,k}^2)=o(n^{-1})$ in Theorem \ref{thm:SACES} is stronger than the usual $o_p(n^{-1/2})$ condition for asymptotic linearity. For cross-fitted influence-function-based estimators, it can be established under suitable moment conditions and mean-square nuisance estimation rates, including strengthened product-rate conditions that control second-order bias. Sufficient conditions and a detailed justification are provided in Lemma \ref{lem:crossfit-L2-remainder} and Remark \ref{rem:nuisance-L2-rates} of Appendix \ref{app:proof}.
\end{remark}

\begin{remark}
If there exists a candidate $k^*$ such that $\theta^{(k^*)}=\theta$ and $\sigma_{k^*}^2<\sigma_0^2$, then the candidate collection contains an estimator that is asymptotically more efficient than the reference estimator. The selection procedure can therefore realize an efficiency gain when such a candidate is selected, although Theorem~\ref{thm:SACES} does not assert consistent selection of the most efficient valid candidate.

The proof of Theorem \ref{thm:SACES} is provided in Appendix \ref{app:proof}. Theorem~\ref{thm:SACES} is pointwise in the underlying data-generating distribution. In particular, it covers candidate estimators whose probability limits either equal $\theta$ or differ from $\theta$ by a fixed nonzero amount. It does not provide a uniform guarantee under local violations satisfying $\theta_n^{(k)}-\theta=O(n^{-1/2})$, for which the distinction between candidate estimators remains first-order. The intermediate-violation simulation settings (Section \ref{subsec:DGP-II}) illustrate this finite-sample difficulty.
\end{remark}

Finally, because the estimator is selected data-adaptively, inference must account for the selection step. We therefore use the post-selection confidence interval (CI) described in Proposition \ref{prop:post-select-ci} of Appendix \ref{app:proof} \citep{rothenhausler2024model}, which propagates the joint asymptotic distribution of the finite candidate vector through the selection rule. Importantly, the resulting CI targets the common causal parameter $\theta$, rather than the probability limit $\theta^{(\widehat k)}$ of the selected candidate.

\section{Simulation Studies}\label{sec:simu}

We consider three data-generating processes (DGPs) to evaluate the performance of SACES relative to several benchmark estimators. For each DGP, we generate $500$ Monte Carlo samples of size $n=1,000$.

\subsection{DGP I: Multiple candidate surrogates with differing validity}

\begin{figure}[ht]
    \centering
    \includegraphics[width=0.45\linewidth]{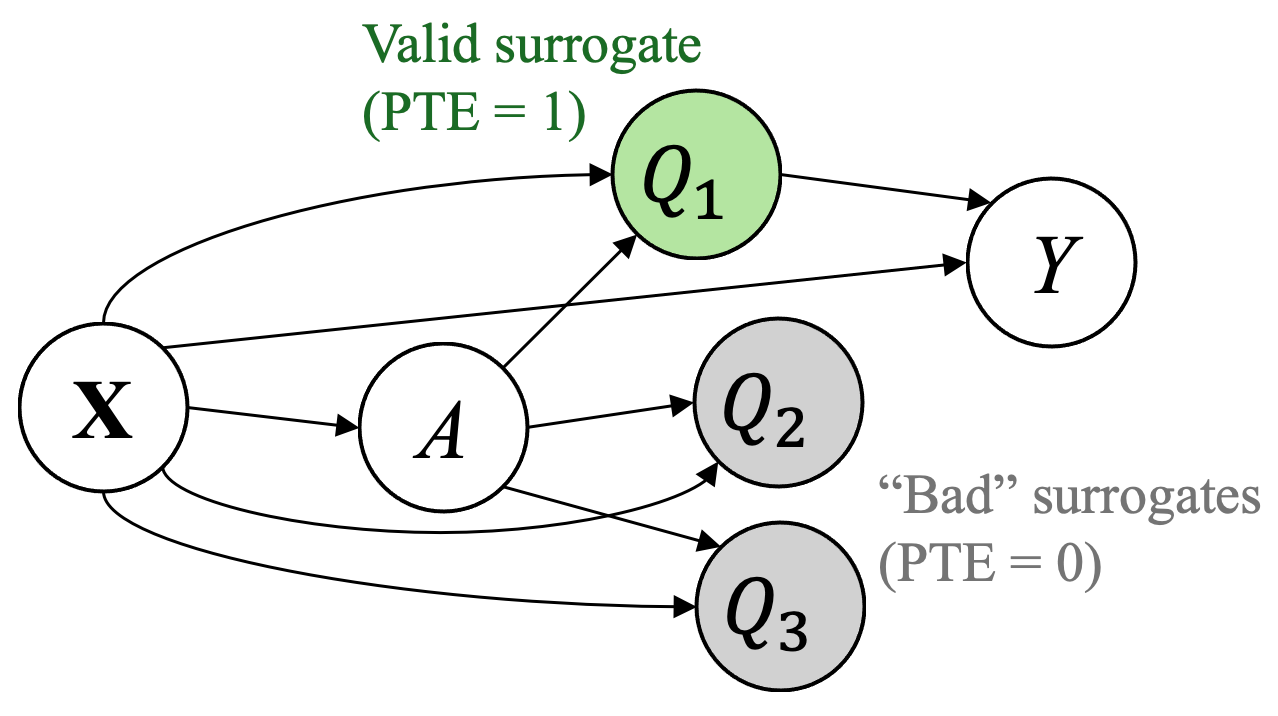}
    \caption{Causal structure of simulation DGP I.}
    \label{fig:DAG-simu1}
\end{figure}

We consider a setting with a binary treatment, three baseline covariates, and three candidate surrogate variables $(Q_1,Q_2,Q_3)$. The DGP is constructed so that $Q_1$ is the only individual candidate surrogate satisfying the statistical surrogacy condition and has a proportion of treatment effect explained (PTE; see \citet{freedman1992statistical, han2022identifying}) of one, whereas $Q_2$ and $Q_3$ may respond strongly to treatment but do not explain its effect on the primary outcome. Figure \ref{fig:DAG-simu1} illustrates the corresponding causal structure.

The true ATE is $\theta=-1$. We vary three parameters, $(\delta_2,\delta_3,\sigma)$, across simulation settings. Specifically, $(\delta_2,\delta_3)\in\{2,5,10\}^2$ control the treatment responsiveness of the invalid candidate surrogates, whereas $\sigma\in\{1,2,3\}$ controls the outcome noise level. Further details of the DGP are provided in Appendix \ref{app:add-simu}.

We compare SACES with the standard AIPW estimator $\stdest$ and three surrogate-EIF estimators: an all-surrogates adjustment using $(Q_1,Q_2,Q_3)$, a bad adjustment using $(Q_2,Q_3)$, and an oracle adjustment using $Q_1$ alone. SACES selects among the candidate surrogate subsets by minimizing the estimated modified MSE criterion.

\begin{figure}[ht]
    \centering
    \includegraphics[width=\linewidth]{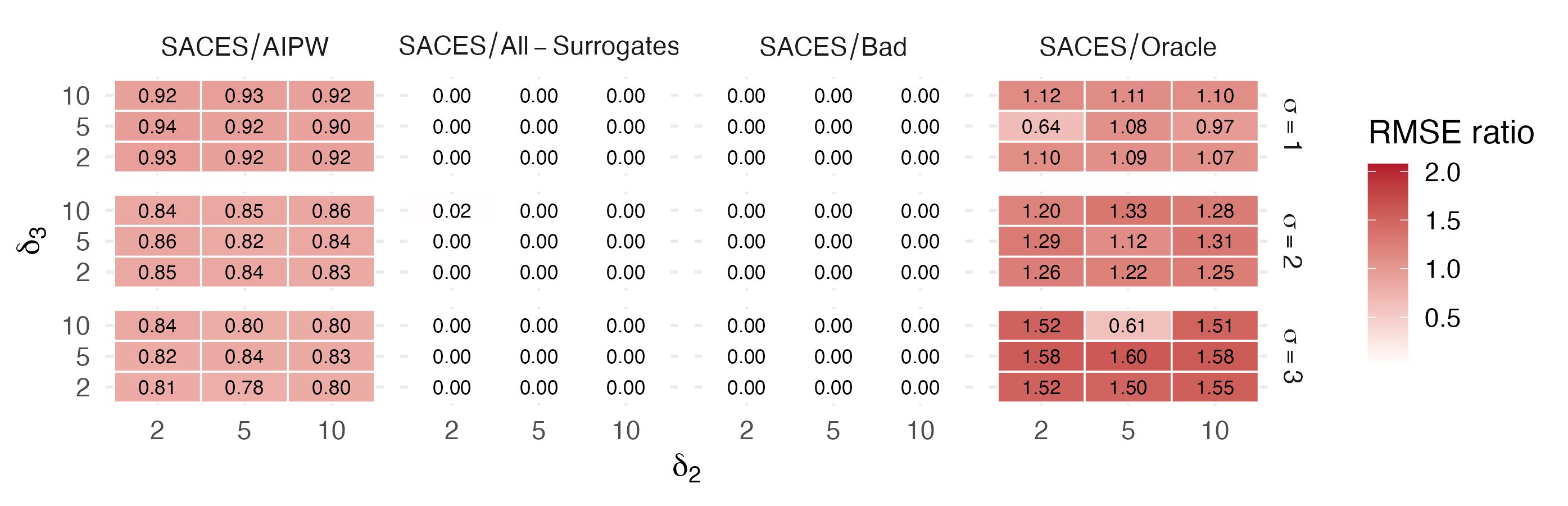}
    \includegraphics[width=\linewidth]{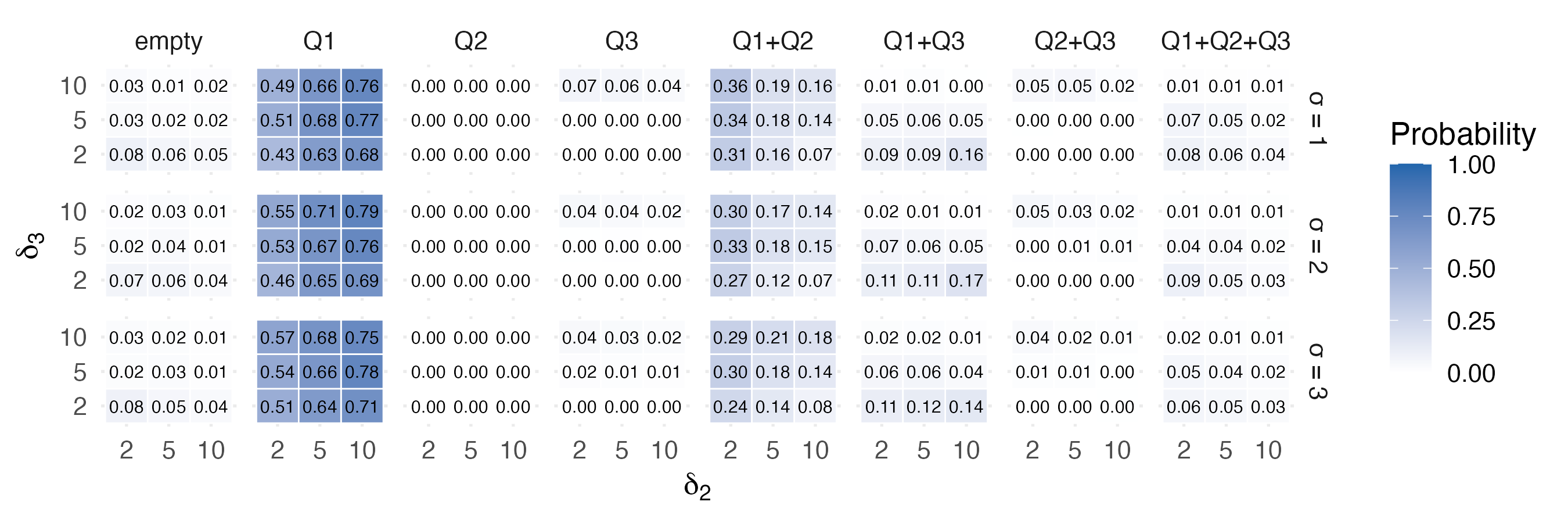}
    \caption{Main results for DGP I: RMSE ratios of $\adptest$ relative to competing estimators and probabilities of selecting each candidate surrogate subset. The lower panel shows that SACES consistently selects the oracle surrogate set $\{Q_1\}$ most frequently across all simulation scenarios.}
    \label{fig:set1-res}
\end{figure}

Figure \ref{fig:set1-res} reports RMSE ratios comparing SACES with the competing estimators and the probabilities of selecting each surrogate subset. Across the settings considered, SACES has lower RMSE than the standard AIPW, all-surrogates, and bad-adjustment estimators. Although the oracle estimator has the lowest RMSE, it requires prior knowledge of the valid surrogate specification; SACES performs close to the oracle without this information and most frequently selects the oracle subset $\{Q_1\}$. Additional results in Appendix \ref{app:add-simu} show that SACES has small bias, coverage close to the nominal 95\% level, and CI widths comparable to those of the oracle estimator.

\subsection{DGP II: A single surrogate under varying degrees of surrogacy violation}\label{subsec:DGP-II}

We next consider a single candidate surrogate $Q$ and a partially observed primary outcome, varying the degree of violation of statistical surrogacy across simulation settings. The DGP is constructed so that the true ATE remains fixed at $\theta=-1$, while $\delta\in\{0,0.2,\ldots,1.8,2\}$ controls the PTE captured by $Q$. In particular, $\delta=0$ corresponds to exact statistical surrogacy, whereas larger values of $\delta$ correspond to increasingly stronger violations. The parameter $\sigma\in\{1,2,3\}$ controls the outcome noise level. Outcome missingness is MAR conditional on $(A,X,Q)$, with a marginal missing-outcome proportion of approximately $35\%$. Further details are provided in Appendix \ref{app:add-simu}.

In addition to the Kallus--Mao unrestricted EIF-based reference estimator $\stdest$ and the surrogate-EIF estimator $\surrest$, we consider a finite class of weighted estimators
$$
\widehat{\theta}(w)
=
(1-w)\stdest+w\surrest,
\qquad
w\in\{0,0.05,\ldots,0.95,1\}.
$$
SACES selects the weight that minimizes the estimated modified MSE criterion, balancing the robustness of the unrestricted reference estimator against potential efficiency gains from imposing statistical surrogacy.

\begin{figure}[ht]
    \centering
    \includegraphics[width=\linewidth]{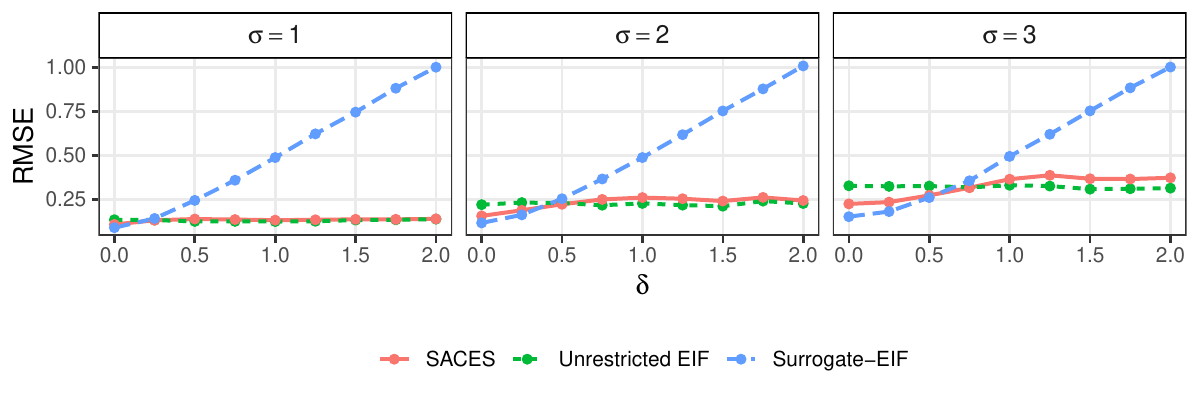}
    \includegraphics[width=\linewidth]{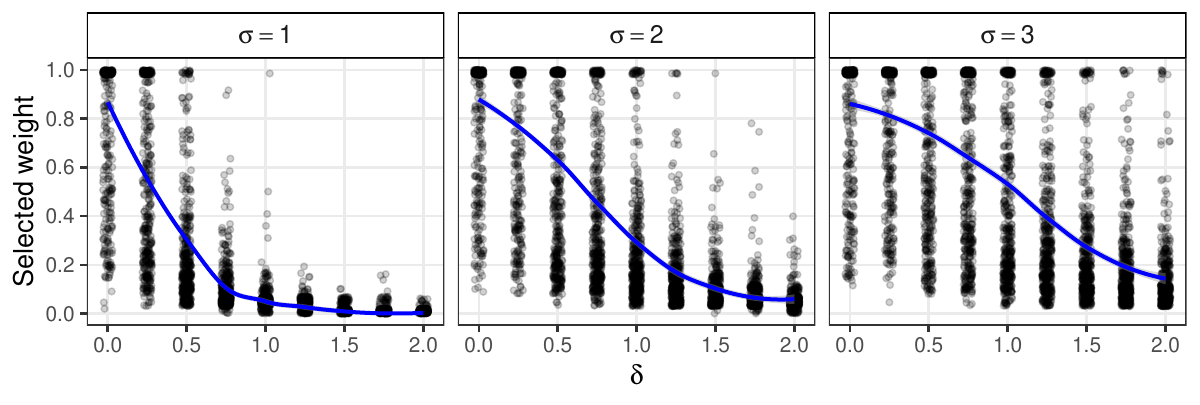}
    \caption{Main results for DGP II: RMSE and selected weights $\widehat w$ across values of $\delta$, stratified by $\sigma$. Scatterplots and average trends are shown for the selected weights.}
    \label{fig:set2-res}
\end{figure}

Figure \ref{fig:set2-res} reports the RMSE and average selected weight $\widehat w$ across values of $\delta$ and $\sigma$. Additional results on bias, coverage, and CI width are reported in Appendix \ref{app:add-simu}. 
Across the considered values of $\delta$ and $\sigma$, SACES achieves RMSE close to that of the oracle estimator. When $\delta=0$, statistical surrogacy holds and $\surrest$ is the oracle estimator. SACES accordingly assigns greater weight to $\surrest$ and captures much of its efficiency gain over $\stdest$, particularly as $\sigma$ increases. As $\delta$ increases and statistical surrogacy is more strongly violated, SACES progressively downweights $\surrest$ and moves toward $\stdest$, limiting the impact of bias in the surrogate-EIF estimator.

Additional results in Appendix \ref{app:add-simu} show that SACES generally has small bias and reasonably narrow CIs. Coverage is close to the nominal 95\% level in most settings but follows a ``U-shaped'' pattern for $\sigma=2$ and $3$ as the degree of surrogacy violation varies, similar to the pattern reported by \citet{rothenhausler2024model}. Under moderate violations, SACES may retain non-negligible weight on the surrogate-EIF estimator in finite samples, leading to residual bias and undercoverage. In these simulations, coverage is more reliable when statistical surrogacy approximately holds or when violations are sufficiently pronounced that SACES assigns little weight to the surrogate-EIF estimator. The intermediate settings warrant greater caution because the competing estimators are difficult to distinguish. In an additional experiment with $n=20,000$, undercoverage is no longer observed, consistent with the pointwise asymptotic validity of the post-selection inference procedure.

\subsection{DGP III: Additional outcome- and missingness-predictive information}\label{subsec:DGP-III}

DGP III examines the missing-outcome setting introduced in Section~\ref{subsec:G-B-surrogates}, with an existing surrogate $Q$ and two additional post-treatment variables. The variable $G$ provides additional information about the primary outcome beyond $(\mb X,Q)$, whereas $B$ primarily predicts outcome missingness and provides no additional outcome information. We vary the strength of these relationships through parameters $\gamma_G$ and $\gamma_B$. The DGP and parameter values are detailed in Appendix \ref{app:add-simu}.

We compare the robust reference estimator based on $Q$, candidate estimators based on $(Q,G)$, $(Q,B)$, and $(Q,G,B)$, and SACES. Adding $G$ generally improves efficiency as its outcome-predictive strength increases, whereas adding $B$ offers little comparable benefit and can increase estimation variability. SACES adapts to the relative usefulness of the additional variables and remains close to the reference estimator when they offer no reliable improvement. When $G$ has a nonzero coefficient in the data-generating mechanism ($\gamma_G\neq0$), SACES most frequently selects $(Q,G)$ among the candidate surrogate sets. Detailed results are reported in Appendix \ref{app:add-simu}.

\section{Application to the ACTG 175 Trial Data}\label{sec:data}

We illustrate the proposed methods using the publicly available ACTG175 dataset ($n=2,139$) from the \texttt{BART} R package, derived from AIDS Clinical Trials Group (ACTG) Study 175 \citep{hammer1996trial}. The original randomized trial enrolled adults with human immunodeficiency virus type 1 (HIV-1) infection whose baseline CD4 T-cell counts ranged from 200 to 500 cells per cubic millimeter and compared four treatment regimens: zidovudine alone, didanosine alone, zidovudine plus didanosine, and zidovudine plus zalcitabine. Following the binary treatment indicator provided in the dataset, we define the control group as participants assigned to zidovudine monotherapy ($n=532$, 25\%) and the treated group as those assigned to any of the other three regimens ($n=1,607$, 75\%). This grouping is clinically motivated by the original trial's finding that each alternative regimen reduced the incidence of its composite primary endpoint relative to zidovudine monotherapy \citep{hammer1996trial}. Accordingly, $\theta$ represents the average effect of the trial's mixture of alternative regimens relative to zidovudine monotherapy, rather than the effect of any individual regimen.

The primary outcome $Y$ is the CD4 T-cell count at week 96, and the week-20 CD4 T-cell count serves as the candidate surrogate $Q$. The week-96 CD4 count was missing for 37.3\% of participants, whereas the baseline covariates and week-20 CD4 measurements included in the analysis were fully observed. Baseline covariates $\mb X$ include age, weight, hemophilia status, homosexual activity, history of intravenous drug use, Karnofsky score, antiretroviral treatment history and duration (including prior non-zidovudine and zidovudine use), race, gender, symptomatic status, and baseline CD4 and CD8 counts. Let $R$ indicate whether the week-96 CD4 outcome is observed. The publicly available dataset does not distinguish missing measurements due to death from those due to other reasons, so $R$ represents overall observation status. The observed data are therefore $\mc O=(\mb X,A,Q,R,RY)$, corresponding to the missing-outcome setting in Section~\ref{subsec:missoutEIF}.

For conventional benchmarks, we calculated the unadjusted difference-in-means and analysis-of-covariance (ANCOVA) estimates among participants with observed week-96 CD4 counts. We include these complete-case analyses as descriptive comparisons only. We then used the full sample to compute the influence-function-based estimators. In addition to the unrestricted reference estimator $\stdest$ \citep{kallus2025role}, which incorporates $Q$ without imposing statistical surrogacy, we included an AIPW comparator that uses baseline covariates $\mb X$ but excludes $Q$ from both the outcome and missingness models. We also considered the surrogate-EIF estimator $\surrest$, which imposes statistical surrogacy, and SACES.

Following the weighted construction in DGP II (Section \ref{subsec:DGP-II}), SACES considers
$$
\widehat\theta(w) = (1-w)\stdest+w\surrest, \qquad w\in\{0,0.01,\ldots,1\},
$$
and selects the weight that minimizes the estimated modified MSE criterion. We estimate all nuisance functions using five-fold cross-fitted logistic and linear working models; further details are provided in Appendix \ref{app:data}. Table \ref{tab:actg175} reports the resulting estimates.

As shown in Table~\ref{tab:actg175}, incorporating week-20 CD4 without imposing statistical surrogacy had little effect on the estimated treatment effect. The AIPW estimator without $Q$ and the Kallus--Mao unrestricted estimator $\stdest$ yielded similar point estimates, with a slight reduction in the standard error from 9.60 to 9.21. In contrast, imposing statistical surrogacy produced a markedly different result. The surrogate-EIF estimator $\surrest$ yielded an estimate of $31.49$ CD4 cells$/\mathrm{mm}^3$, approximately half the unrestricted estimate, with a substantially smaller standard error of $3.87$. Thus, the apparent precision gain was accompanied by a large shift in the point estimate, raising concerns about the plausibility of statistical surrogacy in this application. This comparison highlights an important distinction: week-20 CD4 may provide useful predictive information without eliminating residual treatment dependence in the week-96 outcome.

\begin{table}[ht]
\centering
\caption{ATE estimates for week-96 CD4 count in the ACTG 175 trial.}
\label{tab:actg175}
\begin{threeparttable}
\begin{tabular}{lccc}
\toprule
Estimator & Estimate & Standard error & 95\% CI \\
\midrule
\multicolumn{4}{c}{\textit{Conventional benchmarks}}\\
Difference in means
& 53.83 & 10.79 & (32.69, 74.97) \\
ANCOVA
& 64.92 & 8.98 & (47.33, 82.52) \\
\midrule
\multicolumn{4}{c}{\textit{Estimators considered in this paper}}\\
AIPW without $Q$
& 61.99 & 9.60 & (43.19, 80.80) \\
$\stdest$ (Kallus--Mao unrestricted EIF)
& 62.08 & 9.21 & (44.04, 80.13) \\
$\surrest$ (surrogacy restriction)
& 31.49 & 3.87 & (23.91, 39.07) \\
$\adptest$ (SACES)
& 59.64 & -- & (43.58, 82.01) \\
\bottomrule
\end{tabular}
\begin{tablenotes}\footnotesize
\item The AIPW without $Q$ estimator is a working-model comparator that excludes $Q$ from the outcome and missingness models; it is not a complete-case estimator. A conventional standard error is not reported for SACES. Its 95\% CI is constructed using the post-selection procedure described in Appendix \ref{app:proof} and need not be centered at the point estimate. 
\end{tablenotes}
\end{threeparttable}
\end{table}

\begin{figure}[ht]
\centering
\includegraphics[width=0.9\linewidth]{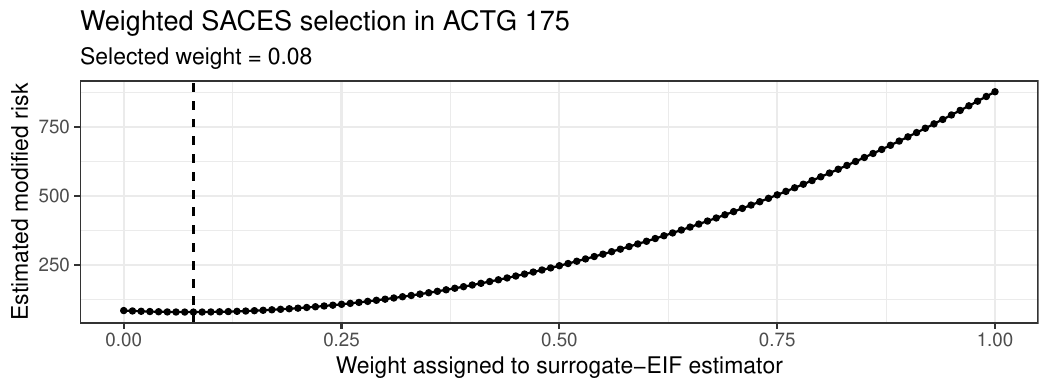}
\caption{Estimated modified risk across candidate weights in the ACTG 175 analysis. The dashed vertical line marks the selected SACES weight, $\widehat w=0.08$. A weight of zero corresponds to the unrestricted EIF estimator, whereas a weight of one corresponds to the surrogate-EIF estimator imposing statistical surrogacy.}
\label{fig:actg175-risk}
\end{figure}

SACES assigned most of the weight to the unrestricted estimator. The selected weight was $\widehat w=0.08$ (Figure \ref{fig:actg175-risk}), yielding an ATE estimate of $59.64$ CD4 cells, close to the unrestricted estimate of $62.08$. Inference uses the post-selection procedure described in Section \ref{sec:adaptive} (see also Proposition \ref{prop:post-select-ci} in Appendix \ref{app:proof}), rather than a conventional Wald interval of the form $\widehat\theta\pm1.96\,\widehat{\mathrm{SE}}$. The resulting 95\% CI for SACES was $(43.58, 82.01)$, slightly wider than that for the unrestricted estimator, consistent with accounting for uncertainty introduced by data-adaptive selection.

\subsection{Additional diagnostics for statistical surrogacy and model stability}
\label{subsec:actg-diagnostics}

We conducted additional diagnostics to investigate the discrepancy between the unrestricted Kallus--Mao and surrogate-EIF estimates. As a working diagnostic for statistical surrogacy, we examined whether the conditional mean of week-96 CD4 remained associated with treatment after adjustment for baseline covariates and week-20 CD4. Among participants with observed week-96 outcomes, we compared a linear model based on $(\mb X,Q)$ with an expanded model including treatment and its interactions with $(\mb X,Q)$. The omnibus $F$-test provided evidence of residual treatment dependence ($p=0.006$), and the average fitted conditional treatment contrast was approximately $37.0$ CD4 cells. This diagnostic assesses only a conditional-mean implication of statistical surrogacy under the specified working models; it is neither a formal test of $Y\independent A\mid\mb X,Q$ nor part of the SACES selection procedure.

We further connect the residual treatment contrast to the bias characterization in Proposition~\ref{prop:bias-surr-main}. Using the cross-fitted nuisance estimates and averaging over all participants, the plug-in estimate of the asymptotic bias of the surrogate-EIF estimator was $-31.24$ CD4 cells. This was close to the observed difference between the surrogate-EIF and unrestricted estimates, $31.49-62.08=-30.59$ CD4 cells. The estimated mean residual conditional treatment contrast from the cross-fitted nuisance models was $35.37$ CD4 cells, similar to the approximately $37.0$-cell contrast from the working-model diagnostic. Our results suggest that residual treatment dependence after conditioning on $(\mb X,Q)$ provides a plausible explanation for the large difference between the two estimators.

We also examined whether poor overlap or unstable inverse-probability weights could explain the discrepancy. The estimated treatment propensity scores indicated adequate empirical overlap for most participants. The minimum fitted probability of each participant's observed treatment was $0.0406$, and only one of the $2,139$ fitted values of $\widehat{\widetilde\pi}(1\mid\mb X,Q)$ exceeded the numerical truncation threshold of $0.99$. The fitted outcome-observation probabilities were also bounded away from zero, with minima ranging from $0.229$ to $0.247$ across the relevant models. These diagnostics offer little evidence that the discrepancy between the unrestricted and surrogate-EIF estimates is primarily driven by severe positivity violations or instability in the fitted observation models.

The ACTG 175 analysis illustrates the difference between using an intermediate outcome as auxiliary information and imposing statistical surrogacy. Incorporating week-20 CD4 without imposing statistical surrogacy produced a modest precision improvement with little change in the estimated treatment effect. Imposing statistical surrogacy yielded a substantially smaller standard error but also a markedly different point estimate, consistent with the diagnostic evidence of residual treatment dependence. SACES assigned limited weight to the surrogate-restricted estimator and produced an estimate close to the unrestricted benchmark.

\section{Concluding Remarks}\label{sec:conclude}

In this paper, we studied causal effect estimation with surrogate outcomes under uncertain statistical surrogacy. We derived the asymptotic bias arising from violations of statistical surrogacy and established that, in the unrestricted MAR model, incorporating additional post-treatment information can either improve or reduce efficiency depending on its predictive relationships with the outcome and missingness. We then developed SACES, an adaptive procedure that selects among surrogate-assisted estimators using an unrestricted efficient estimator as a robust reference. Under regularity conditions, SACES achieves an asymptotic risk no greater than that of the reference estimator up to a vanishing remainder, while allowing for efficiency gains when suitable surrogate-assisted estimators are available. Simulation studies and the ACTG 175 analysis illustrate the performance of SACES under uncertain statistical surrogacy.

There are several interesting directions for future research. First, although the post-selection inference procedure provides asymptotic coverage, finite-sample undercoverage can occur under moderate violations of statistical surrogacy. Developing inference procedures with stronger finite-sample guarantees remains an important direction, particularly when candidate estimators are difficult to distinguish. Second, our framework assumes that surrogate variables are fully observed for all participants, and extending the proposed methods to settings with partially observed surrogates is a natural extension. Other promising directions include extensions to censored survival outcomes, alternative causal estimands \citep{gao2024role,wang2025rate}, and multi-source data settings \citep{liu2025targeted,han2025federated}.

\section*{Declaration of the use of generative AI and AI-assisted technologies}

During manuscript preparation, the authors used OpenAI's ChatGPT for language editing, assistance with manuscript organization and presentation, and generating code to summarize simulation and data analysis results. All AI-assisted materials were reviewed and edited by the authors, who take full responsibility for the scientific content, analyses, and conclusions of the manuscript.

\bibliography{_refs}

\clearpage 
\appendix

\section{Technical Proofs}\label{app:proof}
\subsection{Two Equivalent Forms of Assumption \ref{asp:unconf}}\label{subsec:asp1-equi}

\begin{proposition}
The following two statements are equivalent:
\begin{enumerate}[label=(\roman*)]
    \item $(Y(a), Q(a)) \independent A \mid \mb X$.
    \item $Q(a) \independent A \mid \mb X$ and $Y(a) \independent A \mid \mb X, Q(a)$.
\end{enumerate}
\end{proposition}

\begin{proof}
\noindent\textbf{(i) $\Rightarrow$ (ii).}  
Suppose $(Y(a),Q(a))\independent A\mid\mb X$. Hence, $Q(a) \independent A\mid\mb X$. By the chain rule, 
$$
\Px(Y(a)\mid Q(a),A,\mb X) 
= \frac{\Px(Y(a), Q(a)\mid A,\mb X)}{\Px(Q(a) \mid A,\mb X)}
= \frac{\Px(Y(a), Q(a)\mid\mb X)}{\Px(Q(a) \mid \mb X)}
=\Px(Y(a) \mid Q(a), \mb X).
$$
Thus $Y(a) \independent A \mid\mb X, Q(a)$.  

\medskip
\noindent\textbf{(ii) $\Rightarrow$ (i).}  
Assume $Q(a) \independent A\mid\mb X$ and $Y(a) \independent A \mid \mb X, Q(a)$. Then
$$
\Px(Y(a), Q(a) \mid A, \mb X) 
= \Px(Y(a) \mid Q(a), A, \mb X)\, \Px(Q(a) \mid A, \mb X).
$$
By the assumed independencies,
$$
\Px(Y(a)\mid Q(a),A,\mb X) = \Px(Y(a)\mid Q(a),\mb X), \qquad
\Px(Q(a)\mid A,\mb X) = \Px(Q(a)\mid\mb X).
$$
Therefore,
$$
\Px(Y(a), Q(a) \mid A, \mb X) 
= \Px(Y(a) \mid Q(a), \mb X)\, \Px(Q(a) \mid \mb X)
= \Px(Y(a), Q(a) \mid \mb X).
$$
Hence, $(Y(a), Q(a)) \independent A \mid \mb X$.  
\end{proof}

\subsection{EIF for fully observed outcomes}\label{subsec:proof-Th1}

\begin{theorem}\label{thm:EIF-main}
With $\mc O=(\mb X,A,Q,Y)$, the EIF of $\theta=\theta_1-\theta_0$ under Assumptions \ref{asp:unconf}, \ref{asp:positivity}, and Condition \ref{cond:statsurr} is $\widetilde\varphi^*(\mc O;\Px)=\widetilde\varphi_1^*(\mc O;\Px)-\widetilde\varphi_0^*(\mc O;\Px)$, where for $a=0,1$,
\begin{align}\label{eq:eif-main}
    \widetilde\varphi_a^*(\mc O;\Px) = \frac{\Ix(A=a)}{\pi(a\mid\mb X)}\{\widetilde\nu(\mb X,Q)-\widetilde\mu(a,\mb X)\} + \frac{\widetilde\pi(a\mid\mb X,Q)}{\pi(a\mid\mb X)}\{Y-\widetilde\nu(\mb X,Q)\} + \widetilde\mu(a,\mb X)-\theta_a.
\end{align}
\end{theorem}

\begin{proof}
We follow notation in the main text. First, $\theta_a = \Ex\{Y(a)\}$ has the following identification formula with Condition \ref{cond:statsurr} and the observed data $\mc O$:
\begin{align*}
    \theta_a & = \Ex\{Y(a)\} = \Ex[\Ex\{Y(a)\mid\mb X\}] = \Ex[\Ex\{\Ex(Y(a)\mid \mb X, Q(a))\mid\mb X\}] \\ 
    & = \Ex[\Ex\{\Ex(Y(a)\mid \mb X, Q(a))\mid\mb X, A=a\}]\\
    & = \Ex[\Ex\{\Ex(Y\mid \mb X, Q)\mid\mb X, A=a\}] \\
    & = \Ex[\Ex\{\widetilde\nu(\mb X, Q)\mid\mb X, A=a\}],
\end{align*}
where $\widetilde\nu(\mb X, Q) = \Ex\{Y\mid \mb X, Q\}$. 

Then, we write down the joint likelihood of $\mc O$: 
\begin{align*}
    f(\mc O) = f(\mb X)\pi(A\mid\mb X)^A\pi(1-A\mid\mb X)^{1-A}f(Q\mid A,\mb X)f(Y\mid \mb X, Q). 
\end{align*}
The nuisance tangent space of the setting here is then characterized by
\begin{align*}
    \bar\Lambda = \bar\Lambda_{\mb X} \oplus \bar\Lambda_{A\mid\mb X} \oplus\bar\Lambda_{Q\mid A,\mb X} \oplus \bar\Lambda_{Y\mid\mb X, Q}, 
\end{align*}
where ``$\oplus$'' denotes the direct sum. Moreover, for random variables $U$, $V$ and their conditional distribution $U\mid V$, we use notation $\Lambda_{U\mid V}$ to denote set $\{l(U,V)\in L_2(U,V):\Ex\{l(U,V)\mid V\}=0\}$, where $L_2(U,V)$ is the Hilbert space containing all square-integrable functions of $(U,V)$, and then $\bar\Lambda_{U\mid V}$ denotes the mean square closure of $\Lambda_{U\mid V}$. 

Consider regular parametric submodels indexed by parameters $\epsilon\in\mathbb R$ where $\epsilon=0$ corresponds to the underlying true data distribution. We use $\Ex_\epsilon$ and $\nu_\epsilon$ to denote the expectation and outcome model $\nu$, respectively, under the submodel distribution with parameter value $\epsilon$. 

Based on identification formula above, the target parameter can be further expressed as  
\begin{align*}
    \Ex_\epsilon\{Y(a)\} & = \Ex_\epsilon[\Ex_\epsilon\{\nu_\epsilon(\mb X, Q)\mid \mb X,A=a\}]. 
\end{align*}
We then calculate the pathwise derivative for $\epsilon$. We first note that by the chain rule, 
\begin{align*}
    \frac{\partial\Ex_\epsilon\{Y(a)\}}{\partial\epsilon}~\bigg|_{\epsilon=0} & = \frac{\partial}{\partial\epsilon}\Ex_\epsilon\{\widetilde\mu(a,\mb X)\}~\bigg|_{\epsilon=0} + \Ex\left[\frac{\partial}{\partial\epsilon}\Ex_\epsilon\{\widetilde\nu(\mb X, Q)\mid\mb X,A=a\}~\bigg|_{\epsilon=0}\right] \\
    & \qquad + \Ex\left[\Ex\left\{\frac{\partial}{\partial\epsilon}\mu_\epsilon(\mb X, Q)~\bigg|_{\epsilon=0}~\bigg|~\mb X,A=a\right\}\right],
\end{align*}
where $\widetilde\mu(a,\mb X) = \Ex\{\widetilde\nu(\mb X, Q)\mid\mb X, A=a\}$. We then examine and expand these terms separately. 

First, 
\begin{align*}
    \frac{\partial}{\partial\epsilon}\Ex_\epsilon\{\widetilde\mu(a,\mb X)\}~\bigg|_{\epsilon=0} & = \Ex\{\widetilde\mu(a,\mb X)\ell(\mb X)\} = \Ex\{(\widetilde\mu(a,\mb X)-\theta_a)\ell(\mb X)\} \\
    & = \Ex\{(\widetilde\mu(a,\mb X)-\theta_a)\ell(\mc O)\}. 
\end{align*}
Second, 
\begin{align*}
    \Ex\left[\frac{\partial}{\partial\epsilon}\Ex_\epsilon\{\widetilde\nu(\mb X, Q)\mid\mb X,A=a\}~\bigg|_{\epsilon=0}\right] & = \Ex\left[\Ex\{\widetilde\nu(\mb X, Q)\ell(Q\mid\mb X, A)\mid\mb X,A=a\}\right] \\
    & = \Ex\left[\Ex\{\{\widetilde\nu(\mb X, Q)-\widetilde\mu(a,\mb X)\}\ell(Q\mid\mb X, A)\mid\mb X,A=a\}\right] \\
    & = \Ex\left[\Ex\{\{\widetilde\nu(\mb X, Q)-\widetilde\mu(a,\mb X)\}\ell(\mc O)\mid\mb X,A=a\}\right] \\
    & = \Ex\left[\frac{\Ix(A=a)}{\pi(a\mid\mb X)}\{\widetilde\nu(\mb X, Q)-\widetilde\mu(a,\mb X)\}\ell(\mc O)\right].
\end{align*}
Third, 
\begin{align*}
    \Ex\left[\Ex\left\{\frac{\partial}{\partial\epsilon}\nu_\epsilon(\mb X, Q)~\bigg|_{\epsilon=0}~\bigg|~\mb X,A=a\right\}\right] & = \Ex[\Ex\{\Ex\{Y\ell(Y\mid \mb X, Q)\mid \mb X, Q\}\mid\mb X, A=a\}] \\
    & = \Ex[\Ex\{\underbrace{\Ex\{(Y-\widetilde\nu(\mb X, Q))\ell(Y\mid \mb X, Q)\mid \mb X, Q\}}_{:= J(\mb X, Q)}\mid\mb X, A=a\}] \\
    & = \Ex\left[\frac{\Ix(A=a)}{\pi(a\mid\mb X)}J(\mb X, Q)\right] \\
    & = \Ex\left[\frac{\Ix(A=a)}{\pi(a\mid\mb X)}\Ex\{(Y-\widetilde\nu(\mb X, Q))\ell(Y\mid \mb X, Q)\mid \mb X, Q\}\right] \\
    & = \Ex\left[\frac{\Px(A=a\mid \mb X, Q)}{\pi(a\mid\mb X)}\Ex\{(Y-\widetilde\nu(\mb X, Q))\ell(Y\mid \mb X, Q)\mid \mb X, Q\}\right] \\
    & = \Ex\left[\frac{\widetilde\pi(a\mid \mb X, Q)}{\pi(a\mid\mb X)}(Y-\widetilde\nu(\mb X, Q))\ell(\mc O)\right]. 
\end{align*}
Here, the fifth ``='' used the statistical surrogacy condition $A\independent Y\mid \mb X, Q$. Putting these together, we have the EIF for $\theta_a$, denoted by $\widetilde\varphi_a^*(\mc O;\pi,\widetilde\pi,\nu,\widetilde\mu) = \widetilde\varphi_a^*(\mc O;\Px)$, derived as 
\begin{align*}
    \widetilde\varphi^*_a(\mc O;\Px) & = \frac{\Ix(A=a)}{\pi(a\mid\mb X)}\{\widetilde\nu(\mb X, Q)-\widetilde\mu(a,\mb X)\} + \frac{\widetilde\pi(a\mid \mb X, Q)}{\pi(a\mid\mb X)}\{Y-\widetilde\nu(\mb X, Q)\} + \widetilde\mu(a,\mb X) -\theta_a.
\end{align*}
Therefore, the EIF of ATE is given by $\widetilde\varphi^*(\mc O;\Px) = \widetilde\varphi^*_1(\mc O;\Px)-\widetilde\varphi^*_0(\mc O;\Px)$. 
\end{proof}

\subsection{Proof of Corollary \ref{cor:var-main}}\label{subsec:proof-cor1}

It is known by \cite{kallus2025role} that the EIF of $\theta_a$ under $\mc F = \mc O\backslash\{Q\}$ (without surrogate) or the setting of the left DAG in Figure \ref{fig:DAG-full} (surrogate plays no role with the existence of a direct effect from $A$ to $Y$) is given by 
\begin{align*}
    \varphi^*_a(\mc O;\Px) = \frac{\Ix(A=a)}{\pi(a\mid\mb X)}\{Y-m(a,\mb X)\} + m(a,\mb X) -\theta_a,
\end{align*}
where $m(a,\mb X) = \Ex\{Y(a)\mid \mb X\}$ is the outcome model under treatment $A=a$ by $\mb X$ only, instead of a conditional expectation for $\widetilde\nu(\mb X, Q)$.

Furthermore, if Condition \ref{cond:statsurr} holds, we can prove that $m(A,\mb X) = \widetilde\mu(A,\mb X)$: 
\begin{align*}
    m(A,\mb X) & = \Ex\{Y\mid\mb X, A\} = \Ex_{Q\mid\mb X, A}[\Ex\{Y\mid\mb X, A, Q\}] \\
    & = \Ex_{Q\mid\mb X, A}[\Ex\{Y\mid\mb X, Q\}]\qquad\text{(by Condition \ref{cond:statsurr})}\\
    & = \Ex[\widetilde\nu(\mb X, Q)\mid A,\mb X] = \widetilde\mu(A,\mb X). 
\end{align*}
This allows us to compare the asymptotic variance for ATE estimators defined by the above two EIFs with an explicit expression. To derive so, we first express the asymptotic variance by $\widetilde\varphi^*(\mc O;\Px)$, denoted by $\widetilde{\mc V} = \Ex\{\widetilde\varphi^*(\mc O;\Px)^2\}$. In the following, we let 
$$
w(A,\mb X) = \frac{A}{\pi(1\mid\mb X)} - \frac{1-A}{\pi(0\mid\mb X)}. 
$$
Then, we have
\begin{align*}
    \widetilde{\mc V} & = \Vx\left\{w(A,\mb X)(\widetilde\nu(\mb X, Q)-\widetilde\mu(A,\mb X))\right\} + \Vx\{\widetilde\mu(1,\mb X)-\widetilde\mu(0,\mb X)\} \\
    & \quad + \Vx\left\{\left(\frac{\widetilde\pi(1\mid \mb X, Q)}{\pi(1\mid\mb X)}-\frac{\widetilde\pi(0\mid \mb X, Q)}{\pi(0\mid\mb X)}\right)(Y-\widetilde\nu(\mb X, Q))\right\}
\end{align*}
Similarly, the asymptotic variance of the estimator defined by $\varphi^*(\mc O;\Px)$ is given by 
\begin{align*}
    \mc V & = \Vx\left\{w(A,\mb X)(Y-m(A,\mb X))\right\} + \Vx\{m(1,\mb X)-m(0,\mb X)\}. 
\end{align*}
Hence,
\begin{align*}
    \mc V - \widetilde{\mc V} & = \Vx\left\{w(A,\mb X)(Y-m(A,\mb X))\right\} - \Vx\left\{w(A,\mb X)(\widetilde\nu(\mb X, Q)-\widetilde\mu(A,\mb X))\right\} \\
    & \quad - \Vx\left\{\left(\frac{\widetilde\pi(1\mid \mb X, Q)}{\pi(1\mid\mb X)}-\frac{\widetilde\pi(0\mid \mb X, Q)}{\pi(0\mid\mb X)}\right)(Y-\widetilde\nu(\mb X, Q))\right\}\\
    & = \Ex[w(A,\mb X)^2\{\Vx(Y\mid A,\mb X) - \Vx(\widetilde\nu(\mb X, Q)\mid A,\mb X)\}]\\
    & \quad - \Ex\left\{\left(\frac{\widetilde\pi(1\mid \mb X, Q)}{\pi(1\mid\mb X)}-\frac{\widetilde\pi(0\mid \mb X, Q)}{\pi(0\mid\mb X)}\right)^2\Vx(Y\mid \mb X, Q)\right\}. 
\end{align*}
Note that by law of total variance,
\begin{align*}
    \Vx(Y\mid A,\mb X) & = \Ex\{\Vx(Y\mid Q,A,\mb X)\mid A,\mb X\} + \Vx[\Ex\{Y\mid Q, A,\mb X\}\mid A,\mb X] \\
    & = \Ex\{\Vx(Y\mid \mb X, Q)\mid A,\mb X\} + \Vx\{\widetilde\nu(\mb X, Q)\mid A,\mb X\}. 
\end{align*}
Therefore, 
\begin{align*}
    \mc V - \widetilde{\mc V}
    & = \Ex[w(A,\mb X)^2\Ex\{\Vx(Y\mid \mb X, Q)\mid A,\mb X\}] - \Ex\left\{\left(\frac{\widetilde\pi(1\mid \mb X, Q)}{\pi(1\mid\mb X)}-\frac{\widetilde\pi(0\mid \mb X, Q)}{\pi(0\mid\mb X)}\right)^2\Vx(Y\mid \mb X, Q)\right\} \\
    & = \Ex\left\{\left(\frac{A}{\pi(1\mid\mb X)^2}+\frac{1-A}{\pi(0\mid\mb X)^2}\right)\Vx(Y\mid \mb X, Q)\right\}- \Ex\left\{\left(\frac{\widetilde\pi(1\mid \mb X, Q)}{\pi(1\mid\mb X)}-\frac{\widetilde\pi(0\mid \mb X, Q)}{\pi(0\mid\mb X)}\right)^2\Vx(Y\mid \mb X, Q)\right\} \\
    & = \Ex\left[\left\{\Ex\left(\frac{A}{\pi(1\mid\mb X)^2}+\frac{1-A}{\pi(0\mid\mb X)^2}\bigg|~\mb X, Q\right) - \left(\frac{\widetilde\pi(1\mid \mb X, Q)}{\pi(1\mid\mb X)}-\frac{\widetilde\pi(0\mid \mb X, Q)}{\pi(0\mid\mb X)}\right)^2\right\}\Vx(Y\mid \mb X, Q)\right] \\
    & = \Ex\left\{\widetilde\pi(1\mid \mb X, Q)\widetilde\pi(0\mid \mb X, Q)\left(\frac{1}{\pi(1\mid\mb X)^2}+\frac{1}{\pi(0\mid\mb X)^2} + \frac{2}{\pi(1\mid\mb X)\pi(0\mid\mb X)}\right)\Vx(Y\mid \mb X, Q)\right\} \\
    & = \Ex\left\{\widetilde\pi(1\mid \mb X, Q)\widetilde\pi(0\mid \mb X, Q)\left(\frac{1}{\pi(1\mid\mb X)}+\frac{1}{\pi(0\mid\mb X)}\right)^2\Vx(Y\mid \mb X, Q)\right\} \\ 
    & = \Ex\left\{\frac{\widetilde\pi(0\mid \mb X, Q)\widetilde\pi(1\mid \mb X, Q)}{\pi(0\mid\mb X)^2\pi(1\mid\mb X)^2}\Vx(Y\mid \mb X, Q)\right\} \geq 0. 
\end{align*}
which quantifies the efficiency gain by knowing the absence of the direct effect from $A$ to $Y$, i.e., knowing Condition \ref{cond:statsurr} holds. The proof is completed.  

\subsection{EIF for outcomes missing at random}\label{subsec:proof-Th2}

\begin{lemma}\label{lem:surr-MAR-imply}
    Under Condition \ref{cond:statsurr} and Assumption \ref{asp:MAR}, $Y\independent R\mid\mb X,Q$. 
\end{lemma}
\begin{proof}
By Assumption \ref{asp:MAR},
$$
Y\independent R\mid\mb X,A,Q.
$$
Together with Condition \ref{cond:statsurr}, which states that
$Y\independent A\mid\mb X,Q$, this implies
$$
Y\independent (A,R)\mid\mb X,Q
$$
by the contraction property of conditional independence.
Marginalizing over $A$ yields
$Y\independent R\mid\mb X,Q$.
\end{proof}

\begin{theorem}\label{thm:EIF-miss}
With $\mc O=(\mb X,A,Q,R,RY)$, the EIF of $\theta$ under Assumptions \ref{asp:unconf}--\ref{asp:positivity-r}, and Condition \ref{cond:statsurr} is $\widetilde\psi^*(\mc O;\Px)=\widetilde\psi_1^*(\mc O;\Px)-\widetilde\psi_0^*(\mc O;\Px)$, where for $a=0,1$,
\begin{align}\label{eq:eif-miss}
    \widetilde\psi_a^*(\mc O;\Px) = \frac{\widetilde\pi(a\mid\mb X,Q)R}{\pi(a\mid\mb X)\widetilde r(\mb X,Q)}\{Y-\widetilde\nu(\mb X,Q)\} + \frac{\Ix(A=a)}{\pi(a\mid\mb X)}\{\widetilde\nu(\mb X,Q)-\widetilde\mu(a,\mb X)\} + \widetilde\mu(a,\mb X)-\theta_a.
\end{align}
\end{theorem}
\begin{proof}
We first write the identification formula below:
\begin{align*}
    \theta_a & = \Ex\{Y(a)\} = \Ex[\Ex\{Y(a)\mid\mb X\}] = \Ex[\Ex\{\Ex(Y(a)\mid \mb X, Q(a))\mid\mb X\}] \nonumber \\ 
    & = \Ex[\Ex\{\Ex(Y(a)\mid \mb X, Q(a))\mid\mb X, A=a\}] \nonumber \\
    & = \Ex[\Ex\{\Ex(Y\mid \mb X, Q)\mid\mb X, A=a\}], \nonumber \\
    & = \Ex[\Ex\{\Ex(Y\mid \mb X, Q, R=1)\mid\mb X, A=a\}]\qquad\text{(Assumption \ref{asp:MAR})}. 
\end{align*}
The joint likelihood, based on Assumptions \ref{cond:statsurr} and \ref{asp:MAR}, is written as
\begin{align*}
    f(\mc O) = f(\mb X)f(A\mid\mb X)f(Q\mid\mb X, A)f(R\mid\mb X,Q,A)f(Y\mid\mb X, Q, R=1). 
\end{align*}

Following a similar derivation process in Appendix \ref{subsec:proof-Th1}, we found that to get the pathwise derivative of $\theta$ evaluated at $\epsilon=0$, we only need to change the most inner derivative, where
\begin{align*}
    & \Ex\left[\Ex\left\{\frac{\partial}{\partial\epsilon}\Ex_\epsilon(Y\mid \mb X, Q, R=1)~\bigg|_{\epsilon=0}~\bigg|~\mb X,A=a\right\}\right] \\
    & = \Ex[\Ex\{\Ex\{Y\ell(Y\mid \mb X, Q, R)\mid \mb X, Q, R=1\}\mid\mb X, A=a\}] \\
    & = \Ex\left[\Ex\left\{\Ex\left\{\frac{R}{\widetilde r(\mb X, Q)}(Y-\widetilde\nu(\mb X, Q, R))\ell(Y\mid \mb X, Q, R)\mid \mb X, Q\right\}~\bigg|~\mb X, A=a\right\}\right] \\
    & = \Ex\left[\frac{\Ix(A=a)}{\pi(a\mid\mb X)}\Ex\left\{\frac{R}{\widetilde r(\mb X, Q)}(Y-\widetilde\nu(\mb X, Q, R))\ell(Y\mid \mb X, Q, R)\mid \mb X, Q\right\}\right] \\
    & = \Ex\left[\frac{\Px(A=a\mid\mb X,Q)}{\pi(a\mid\mb X)}\Ex\left\{\frac{R}{\widetilde r(\mb X, Q)}(Y-\widetilde\nu(\mb X, Q, R))\ell(Y\mid \mb X, Q, R)\mid \mb X, Q\right\}\right] \\
    & = \Ex\left[\frac{\widetilde\pi(a\mid\mb X,Q)}{\pi(a\mid\mb X)}\frac{R}{\widetilde r(\mb X, Q)}(Y-\widetilde\nu(\mb X, Q, R))\ell(Y\mid \mb X, Q, R)\right] \\
    & = \Ex\left[\frac{\widetilde\pi(a\mid\mb X,Q)R}{\pi(a\mid\mb X)\widetilde r(\mb X,Q)}(Y-\widetilde\nu(\mb X, Q))\ell(\mc O)\right]. 
\end{align*}
Here, $\widetilde r(\mb X, Q)=\Px(R=1\mid \mb X, Q)$, and $\widetilde\nu(\mb X, Q,r)=\Ex\{Y\mid \mb X, Q, R=r\}$ for $r=0,1$. Under Condition \ref{cond:statsurr} and Assumption \ref{asp:MAR} (or Lemma \ref{lem:surr-MAR-imply}), we have $\widetilde\nu(\mb X, Q,r)=\Ex\{Y\mid \mb X, Q\}=\widetilde\nu(\mb X, Q)$, which justifies the last ``$=$.''  
Hence, the EIF of $\theta_a$ is given as
\begin{align*}
    \widetilde\psi_a^*(\mc O;\Px) = \frac{\widetilde\pi(a\mid\mb X,Q)R}{\pi(a\mid\mb X)\widetilde r(\mb X,Q)}\{Y-\widetilde\nu(\mb X, Q)\} + \frac{\Ix(A=a)}{\pi(a\mid\mb X)}\{\widetilde\nu(\mb X, Q)-\widetilde\mu(a,\mb X)\} + \widetilde\mu(a,\mb X) - \theta_a,
\end{align*}
and so the EIF for $\theta$ is $\widetilde\psi^*(\mc O;\Px) = \widetilde\psi_1^*(\mc O;\Px)-\widetilde\psi_0^*(\mc O;\Px)$.
\end{proof}

\begin{remark}\label{rmk:another-EIF-form}
    An alternative form of the EIF incorporates a missingness model explicitly involving $A$ in the predictors. By Bayes's rule, we have 
    \begin{align*}
        \widetilde r(\mb X,Q) & = \Px(R=1\mid\mb X,Q) = \frac{\Px(R=1\mid\mb X,Q,A=a)\Px(A=a\mid\mb X,Q)}{\Px(A=a\mid\mb X,Q,R=1)}. 
    \end{align*}
    A more explicit explanation for the above equation is $\Px(R=1\mid\mb X,Q)\Px(A=a\mid\mb X,Q,R=1) = \Px(R=1,A=a\mid\mb X,Q) = \Px(A=a\mid\mb X,Q)\Px(R=1\mid\mb X,Q,A=a)$. Therefore, 
    \begin{align*}
        \frac{\widetilde\pi(a\mid\mb X,Q)}{\widetilde r(\mb X,Q)}  = \frac{\Px(A=a\mid\mb X,Q)}{\Px(R=1\mid\mb X,Q)} = \frac{\Px(A=a\mid\mb X,Q,R=1)}{\Px(R=1\mid\mb X,Q,A=a)}.
    \end{align*}
    Letting $\widetilde\pi_{R=1}(a\mid\mb X,Q)=\Px(A=a\mid\mb X,Q,R=1)$ and $\widetilde r_a(\mb X,Q) = \Px(R=1\mid\mb X,Q,A=a)$, one could re-write the EIF of $\theta_a$ \eqref{eq:eif-miss} as 
    \begin{align*}
        \widetilde\psi_a^*(\mc O;\Px) = \frac{\widetilde\pi_{R=1}(a\mid\mb X,Q)R}{\pi(a\mid\mb X)\widetilde r_a(\mb X,Q)}\{Y-\widetilde\nu(\mb X, Q)\} + \frac{\Ix(A=a)}{\pi(a\mid\mb X)}\{\widetilde\nu(\mb X, Q)-\widetilde\mu(a,\mb X)\} + \widetilde\mu(a,\mb X) - \theta_a. 
    \end{align*}
    Therefore, only the weight term before $Y-\widetilde\nu(\mb X,Q)$ is modified. The two EIF forms are conceptually equivalent, differing only in implementation details. This remark provides an alternative interpretation: although incorporating the missingness model $\widetilde r(\mb X,Q)$ may seem ``counter-intuitive,'' our formulation reveals that the missingness can be explicitly adjusted by treatment $A$, yet the contribution of this adjustment ultimately cancels out. 
\end{remark}

\subsection{Proof of Corollary \ref{cor:var-miss}}\label{subsec:proof-cor2}

Recall that the unrestricted MAR EIF of $\theta_a$ from \citet{kallus2025role}, without imposing Condition \ref{cond:statsurr}, is given by 
\begin{align*}
\psi_a^*(\mc O;\Px) = \frac{\Ix(A=a)R}{\pi(a\mid\mb X)\widetilde r_a(\mb X, Q)}\{Y-\widetilde m(a,\mb X, Q)\} + \frac{\Ix(A=a)}{\pi(a\mid\mb X)}\{\widetilde m(a,\mb X, Q)-m(a,\mb X)\} + m(a,\mb X) - \theta_a, 
\end{align*}
so the EIF of $\theta$ in this case is given by $\psi^*(\mc O;\Px) = \psi_1^*(\mc O;\Px)-\psi_0^*(\mc O;\Px)$. 

We then calculate the asymptotic efficiency difference by $\psi^*(\mc O;\Px)$ and $\widetilde\psi^*(\mc O;\Px)$, denoted by $\mc V = \Ex\{\psi^*(\mc O;\Px)^2\}$ and $\widetilde{\mc V} = \Ex\{\widetilde\psi^*(\mc O;\Px)^2\}$, respectively. Similar to the no missing data case, we first note that $m(a,\mb X) = \widetilde\mu(a,\mb X)$ and $\widetilde m(a,\mb X,Q) = \Ex\{Y\mid \mb X, Q, A=a\} = \Ex\{Y\mid \mb X, Q\} = \widetilde\nu(\mb X, Q)$ under Condition \ref{cond:statsurr}. 

In the following, let 
$$
w(A,\mb X) = \frac{A}{\pi(1\mid\mb X)} - \frac{1-A}{\pi(0\mid\mb X)}. 
$$
Then,  
\begin{align*}
    \widetilde{\mc V} & = \Vx\left\{w(A,\mb X)(\widetilde\nu(\mb X, Q)-\widetilde\mu(A,\mb X))\right\} + \Vx\{\widetilde\mu(1,\mb X)-\widetilde\mu(0,\mb X)\} \\
    & \quad + \Vx\left\{\left(\frac{\widetilde\pi(1\mid \mb X, Q)}{\pi(1\mid\mb X)}-\frac{\widetilde\pi(0\mid \mb X, Q)}{\pi(0\mid\mb X)}\right)\frac{R}{\widetilde r(\mb X, Q)}(Y-\widetilde\nu(\mb X, Q))\right\},
\end{align*}
and
\begin{align*}
    \mc V & = \Vx\left\{w(A,\mb X)(\widetilde m(A, \mb X, Q)-m(A,\mb X))\right\} + \Vx\{m(1,\mb X)-m(0,\mb X)\} \\
    & \quad + \Vx\left\{w(A,\mb X)\frac{R}{\widetilde r_A(\mb X, Q)}(Y-\widetilde m(A, \mb X, Q))\right\}.
\end{align*}
Recall $\widetilde\nu(\mb X, Q)=\widetilde m(A, \mb X, Q)$ and $m(A,\mb X) = \widetilde\mu(A,\mb X)$. 
Therefore, we could express
\begin{align*}
    \mc V - \widetilde{\mc V} & = \Vx\left\{w(A,\mb X)\frac{R}{\widetilde r_A(\mb X, Q)}(Y-\widetilde\nu(\mb X, Q))\right\} \\
    & \qquad - \Vx\left\{\left(\frac{\widetilde\pi(1\mid \mb X, Q)}{\pi(1\mid\mb X)}-\frac{\widetilde\pi(0\mid \mb X, Q)}{\pi(0\mid\mb X)}\right)\frac{R}{\widetilde r(\mb X, Q)}(Y-\widetilde\nu(\mb X, Q))\right\}\\
    & = \Ex\left\{\frac{w(A,\mb X)^2}{\widetilde r_A(\mb X, Q)}(Y-\widetilde\nu(\mb X,Q))^2\right\} \\
    & \qquad - \Ex\left\{\left(\frac{\widetilde\pi(1\mid \mb X, Q)}{\pi(1\mid\mb X)} -\frac{\widetilde\pi(0\mid \mb X, Q)}{\pi(0\mid\mb X)}\right)^2\frac1{\widetilde r(\mb X,Q)}(Y-\nu(\mb X,Q))^2\right\} \\
    & = \Ex\left[\frac{w(A,\mb X)^2}{\widetilde r_A(\mb X, Q)}\Ex\{\Vx(Y\mid \mb X, Q)\mid A,\mb X, Q\}\right] \\
    & \qquad - \Ex\left\{\left(\frac{\widetilde\pi(1\mid \mb X, Q)}{\pi(1\mid\mb X)} -\frac{\widetilde\pi(0\mid \mb X, Q)}{\pi(0\mid\mb X)}\right)^2\frac1{\widetilde r(\mb X,Q)}\Vx(Y\mid \mb X, Q)\right\} \\
    & = \Ex\left\{\frac{w(A,\mb X)^2}{\widetilde r_A(\mb X, Q)}\Vx(Y\mid \mb X, Q)\right\} - \Ex\left\{\left(\frac{\widetilde\pi(1\mid \mb X, Q)}{\pi(1\mid\mb X)} -\frac{\widetilde\pi(0\mid \mb X, Q)}{\pi(0\mid\mb X)}\right)^2\frac1{\widetilde r(\mb X,Q)}\Vx(Y\mid \mb X, Q)\right\}
\end{align*}
Note that for the first term above, we could further express it as
\begin{align*}
    \Ex\left\{\frac{w(A,\mb X)^2}{\widetilde r_A(\mb X, Q)}\Vx(Y\mid \mb X, Q)\right\} & = \Ex\left[\left\{\frac{A}{\pi(1\mid\mb X)^2\widetilde r_1(\mb X, Q)} + \frac{1-A}{\pi(0\mid\mb X)^2\widetilde r_0(\mb X, Q)}\right\}\Vx(Y\mid \mb X, Q)\right] \\
    & = \Ex\left[\Ex\left\{\frac{A}{\pi(1\mid\mb X)^2\widetilde r_1(\mb X, Q)} + \frac{1-A}{\pi(0\mid\mb X)^2\widetilde r_0(\mb X, Q)}~\bigg|~\mb X,Q\right\}\Vx(Y\mid \mb X, Q)\right] \\
    & = \Ex\left[\left\{\frac{\widetilde\pi(1\mid\mb X,Q)}{\pi(1\mid\mb X)^2\widetilde r_1(\mb X, Q)} + \frac{\widetilde\pi(0\mid\mb X,Q)}{\pi(0\mid\mb X)^2\widetilde r_0(\mb X, Q)}\right\}\Vx(Y\mid \mb X, Q)\right]. 
\end{align*}
We further note that by law of total probability, 
\begin{align*}
    \widetilde r(\mb X, Q) = \sum_{a=0}^1\widetilde r_a(\mb X, Q)\Px(A=a\mid\mb X, Q) = \sum_{a=0}^1\widetilde r_a(\mb X, Q)\widetilde\pi(a\mid\mb X,Q). 
\end{align*}
Based on these two facts, we continue the expression on $\mc V-\widetilde{\mc V}$. For simplicity, let us now focus on the terms multiplied to $\Vx(Y\mid\mb X,Q)$ in the expectations, i.e., 
\begin{align}\label{eq:diff-term-in-varcomp}
    \left\{\frac{\widetilde\pi(1\mid\mb X,Q)}{\pi(1\mid\mb X)^2\widetilde r_1(\mb X, Q)} + \frac{\widetilde\pi(0\mid\mb X,Q)}{\pi(0\mid\mb X)^2\widetilde r_0(\mb X, Q)}\right\} - \left(\frac{\widetilde\pi(1\mid \mb X, Q)}{\pi(1\mid\mb X)} -\frac{\widetilde\pi(0\mid \mb X, Q)}{\pi(0\mid\mb X)}\right)^2\frac1{\widetilde r(\mb X,Q)},
\end{align}
since we only need to show it is non-negative. In the following, for succinct, denote $\pi_a=\pi(a\mid\mb X)$, $\widetilde\pi_a=\pi(a\mid\mb X,Q)$, $\widetilde r=\widetilde r(\mb X,Q)$, and $\widetilde r_a=\widetilde r(a,\mb X,Q)$ for $a=0,1$. Therefore $\widetilde r= \widetilde\pi_0\widetilde r_0 + \widetilde\pi_1\widetilde r_1$. 

Therefore, 
\begin{align*}
    \eqref{eq:diff-term-in-varcomp} & = \frac{\widetilde\pi_1}{\pi_1^2\widetilde r_1}+\frac{\widetilde\pi_0}{\pi_0^2\widetilde r_0} - \frac{(\widetilde\pi_1\pi_0 - \widetilde\pi_0\pi_1)^2}{\pi_0^2\pi_1^2}\frac{1}{\widetilde\pi_0\widetilde r_0 + \widetilde\pi_1\widetilde r_1} \\
    & = \frac{\widetilde\pi_1\pi_0^2\widetilde r_0 + \widetilde\pi_0\pi_1^2\widetilde r_1}{\pi_0^2\pi_1^2\widetilde r_0\widetilde r_1} - \frac{(\widetilde\pi_1\pi_0 - \widetilde\pi_0\pi_1)^2}{\pi_0^2\pi_1^2}\frac{1}{\widetilde\pi_0\widetilde r_0 + \widetilde\pi_1\widetilde r_1} \\
    & = \frac{1}{\pi_0^2\pi_1^2}\left(\frac{\widetilde\pi_1\pi_0^2\widetilde r_0 + \widetilde\pi_0\pi_1^2\widetilde r_1}{\widetilde r_0\widetilde r_1} - \frac{(\widetilde\pi_1\pi_0-\widetilde\pi_0\pi_1)^2}{\widetilde\pi_0\widetilde r_0 + \widetilde\pi_1\widetilde r_1}\right) \\
    & = \frac{1}{\pi_0^2\pi_1^2\widetilde r_0\widetilde r_1(\widetilde\pi_0\widetilde r_0 + \widetilde\pi_1\widetilde r_1)}\left\{(\widetilde\pi_0\widetilde r_0 + \widetilde\pi_1\widetilde r_1)(\widetilde\pi_1\pi_0^2\widetilde r_0 + \widetilde\pi_0\pi_1^2\widetilde r_1) - \widetilde r_0\widetilde r_1(\widetilde\pi_1^2\pi_0^2-2\widetilde\pi_0\pi_0\widetilde\pi_1\pi_1+\widetilde\pi_0^2\pi_1^2)\right\} \\
    & = \frac{1}{\pi_0^2\pi_1^2\widetilde r_0\widetilde r_1(\widetilde\pi_0\widetilde r_0 + \widetilde\pi_1\widetilde r_1)}\big\{\widetilde\pi_0\widetilde r_0^2\widetilde\pi_1\pi_0^2 + \widetilde\pi_0^2\widetilde r_0\pi_1^2\widetilde r_1^2 + \widetilde\pi_1^2\widetilde r_1\widetilde r_0\pi_0^2 + \widetilde\pi_1\pi_1^2\widetilde\pi_0\widetilde r_1^2 \\
    & \qquad\qquad\qquad\qquad\qquad\qquad-\widetilde r_0\widetilde r_1\widetilde\pi_1^2\pi_0^2+2\widetilde r_0\widetilde r_1\pi_0\widetilde\pi_0\pi_1\widetilde\pi_1-\widetilde r_0\widetilde r_1\widetilde\pi_0^2\pi_1^2\big\} \\
    & = \frac{\widetilde\pi_0\widetilde\pi_1(\pi_0^2\widetilde r_0^2+2\pi_0\widetilde r_0\pi_1\widetilde r_1+\pi_1^2\widetilde r_1^2)}{\pi_0^2\pi_1^2\widetilde r_0\widetilde r_1(\widetilde\pi_0\widetilde r_0 + \widetilde\pi_1\widetilde r_1)} \\
    & = \frac{\widetilde\pi_0\widetilde\pi_1(\pi_0\widetilde r_0 + \pi_1\widetilde r_1)^2}{\pi_0^2\pi_1^2\widetilde r_0\widetilde r_1\widetilde r} \geq 0.  
\end{align*}
Therefore, 
\begin{align*}
    \eqref{eq:diff-term-in-varcomp} & = \frac{\widetilde\pi(0\mid\mb X, Q)\widetilde\pi(1\mid\mb X, Q)\{\pi(0\mid\mb X)\widetilde r_0(\mb X,Q) + \pi(1\mid \mb X)\widetilde r_1(\mb X,Q)\}^2}{\pi(0\mid\mb X)^2\pi(1\mid\mb X)^2\widetilde r_0(\mb X,Q)\widetilde r_1(\mb X,Q)\widetilde r(\mb X,Q)} = \widetilde\omega(\mb X, Q)^2,
\end{align*}
where $\widetilde\omega(\mb X, Q)$ was defined in Section \ref{subsec:missoutEIF}. 

\subsection{Asymptotic properties of the surrogate-EIF estimators}
\label{subsec:asy-surr}

We establish the regular and asymptotically linear (RAL) properties of $\surrest$. For generality, we first consider the missing-outcome setting. The fully observed setting follows as a special case by taking $R\equiv 1$ and $\widetilde r(\mb X,Q)\equiv 1$. Based on the EIF in Theorem \ref{thm:EIF-miss}, define
$$
\widehat\theta_a^{\mathrm{surr}}
=
\Px_n
\left[
\frac{
\widehat{\widetilde\pi}(a\mid\mb X,Q)R}
{\widehat\pi(a\mid\mb X)
 \widehat{\widetilde r}(\mb X,Q)}
\{Y-\widehat{\widetilde\nu}(\mb X,Q)\}
+
\frac{\Ix(A=a)}
{\widehat\pi(a\mid\mb X)}
\{
\widehat{\widetilde\nu}(\mb X,Q)
-
\widehat{\widetilde\mu}(a,\mb X)
\}
+
\widehat{\widetilde\mu}(a,\mb X)
\right],
$$
for $a\in\{0,1\}$, and let
$$
\surrest
=
\widehat\theta_1^{\mathrm{surr}}
-
\widehat\theta_0^{\mathrm{surr}}.
$$

For brevity, write
$$
\pi_a(\mb X)=\pi(a\mid\mb X), \qquad
\widetilde\pi_a(\mb X,Q)=\widetilde\pi(a\mid\mb X,Q), \qquad
\widetilde\mu_a(\mb X)=\widetilde\mu(a,\mb X).
$$

\begin{theorem}[RAL in the missing-outcome setting]
\label{thm:asy-miss-supp}
Suppose Assumptions \ref{asp:unconf}--\ref{asp:positivity-r} and Condition \ref{cond:statsurr} hold. Assume that all nuisance functions are estimated using cross-fitting and are consistent in $L_2(\Px)$. Suppose that the estimated treatment and non-missingness propensity scores are bounded away from zero with probability tending to one and, for $a=0,1$,
$$
\|\widehat{\widetilde\nu}-\widetilde\nu\|_2
\left\{
\|\widehat{\widetilde\pi}_a-\widetilde\pi_a\|_2
+
\|\widehat{\widetilde r}-\widetilde r\|_2
\right\}
+
\|\widehat\pi_a-\pi_a\|_2
\|\widehat{\widetilde\mu}_a-\widetilde\mu_a\|_2
=
o_p(n^{-1/2}),
$$
where $\|\cdot\|_2$ denotes the $L_2(\Px)$ norm. Then
$$
\sqrt n
\left(
\widehat\theta_a^{\mathrm{surr}}
-
\theta_a
\right)
=
\frac{1}{\sqrt n}
\sum_{i=1}^n
\widetilde\psi_a^*(\mc O_i;\Px)
+
o_p(1),
$$
and therefore
$$
\sqrt n(\surrest-\theta)
=
\frac{1}{\sqrt n}
\sum_{i=1}^n
\widetilde\psi^*(\mc O_i;\Px)
+
o_p(1).
$$
If
$\Ex\{\widetilde\psi^*(\mc O;\Px)^2\}<\infty$, then
$$
\sqrt n(\surrest-\theta)
\overset{d}{\longrightarrow}
\mc N(0,\widetilde{\mc V}),
\qquad
\widetilde{\mc V}
=
\Ex\{\widetilde\psi^*(\mc O;\Px)^2\}.
$$
Thus, $\surrest$ attains the semiparametric efficiency bound under the statistical surrogacy and MAR conditions.
\end{theorem}

\begin{proof}
Consider a fixed treatment level $a$. Let
$$
\widehat\eta
=
\left(
\widehat\pi_a,
\widehat{\widetilde\pi}_a,
\widehat{\widetilde r},
\widehat{\widetilde\nu},
\widehat{\widetilde\mu}_a
\right),
$$
and let $\eta$ denote the corresponding true nuisance functions. We decompose
$$
\widehat\theta_a^{\mathrm{surr}}-\theta_a
=
(\Px_n-\Px)
\widetilde\psi_a^*(\mc O;\eta)
+
(\Px_n-\Px)
\left\{
\widetilde\psi_a^*(\mc O;\widehat\eta)
-
\widetilde\psi_a^*(\mc O;\eta)
\right\}
+
\text{Rem}^{\mathrm{mis}}_a(\widehat\eta,\eta).
$$
By cross-fitting and $L_2(\Px)$ consistency of the nuisance estimators, the second empirical-process term is
$o_p(n^{-1/2})$.

The remaining bias term can be written as
\begin{align}
\text{Rem}^{\mathrm{mis}}_a(\widehat\eta,\eta)
&=
\Ex\left[
\frac{
\widehat{\widetilde\pi}_a(\mb X,Q)
\widetilde r(\mb X,Q)
-
\widetilde\pi_a(\mb X,Q)
\widehat{\widetilde r}(\mb X,Q)}
{\widehat\pi_a(\mb X)
 \widehat{\widetilde r}(\mb X,Q)}
\{
\widetilde\nu(\mb X,Q)
-
\widehat{\widetilde\nu}(\mb X,Q)
\}
\right]
\nonumber\\
&\quad+
\Ex\left[
\frac{
\pi_a(\mb X)-\widehat\pi_a(\mb X)}
{\widehat\pi_a(\mb X)}
\{
\widetilde\mu_a(\mb X)
-
\widehat{\widetilde\mu}_a(\mb X)
\}
\right].
\label{eq:rem-miss}
\end{align}

Since
$$
\widehat{\widetilde\pi}_a\widetilde r
-
\widetilde\pi_a\widehat{\widetilde r}
=
\widetilde r
(\widehat{\widetilde\pi}_a-\widetilde\pi_a)
+
\widetilde\pi_a
(\widetilde r-\widehat{\widetilde r}),
$$
the positivity conditions and the Cauchy--Schwarz inequality imply
\begin{align*}
|\text{Rem}^{\mathrm{mis}}_a(\widehat\eta,\eta)|
\lesssim\;&
\|\widehat{\widetilde\nu}-\widetilde\nu\|_2
\left\{
\|\widehat{\widetilde\pi}_a-\widetilde\pi_a\|_2
+
\|\widehat{\widetilde r}-\widetilde r\|_2
\right\}\\
&+
\|\widehat\pi_a-\pi_a\|_2
\|\widehat{\widetilde\mu}_a-\widetilde\mu_a\|_2,
\end{align*}
which is $o_p(n^{-1/2})$ by assumption. Therefore,
$$
\widehat\theta_a^{\mathrm{surr}}-\theta_a
=
(\Px_n-\Px)
\widetilde\psi_a^*(\mc O;\Px)
+
o_p(n^{-1/2}).
$$
Taking the difference between $a=1$ and $a=0$ gives the expansion for $\surrest$. The asymptotic normality follows from the central limit theorem. Since $\widetilde\psi^*$ is the EIF established in Theorem \ref{thm:EIF-miss}, $\surrest$ attains the semiparametric efficiency bound.
\end{proof}

\begin{corollary}[RAL in the full-outcome setting]
\label{cor:asy-full-supp}
Suppose Assumptions \ref{asp:unconf} and \ref{asp:positivity} and Condition \ref{cond:statsurr} hold, and suppose that $Y$ is fully observed. Define
$$
\widehat\theta_a^{\mathrm{surr}}
=
\Px_n
\left[
\frac{\Ix(A=a)}{\widehat\pi(a\mid\mb X)}
\{
\widehat{\widetilde\nu}(\mb X,Q)
-
\widehat{\widetilde\mu}(a,\mb X)
\}
+
\frac{\widehat{\widetilde\pi}(a\mid\mb X,Q)}
{\widehat\pi(a\mid\mb X)}
\{
Y-\widehat{\widetilde\nu}(\mb X,Q)
\}
+
\widehat{\widetilde\mu}(a,\mb X)
\right],
$$
for $a\in\{0,1\}$, and let
$$
\surrest
=
\widehat\theta_1^{\mathrm{surr}}
-
\widehat\theta_0^{\mathrm{surr}}.
$$

Assume that the nuisance functions are estimated using cross-fitting and are consistent in $L_2(\Px)$, and that the estimated treatment propensity scores are bounded away from zero with probability tending to one. If, for $a=0,1$,
$$
\|\widehat{\widetilde\pi}_a-\widetilde\pi_a\|_2
\|\widehat{\widetilde\nu}-\widetilde\nu\|_2
+
\|\widehat\pi_a-\pi_a\|_2
\|\widehat{\widetilde\mu}_a-\widetilde\mu_a\|_2
=
o_p(n^{-1/2}),
$$
then
$$
\sqrt n
\left(
\widehat\theta_a^{\mathrm{surr}}
-
\theta_a
\right)
=
\frac{1}{\sqrt n}
\sum_{i=1}^n
\widetilde\varphi_a^*(\mc O_i;\Px)
+
o_p(1),
$$
and therefore
$$
\sqrt n(\surrest-\theta)
=
\frac{1}{\sqrt n}
\sum_{i=1}^n
\widetilde\varphi^*(\mc O_i;\Px)
+
o_p(1).
$$
If
$\Ex\{\widetilde\varphi^*(\mc O;\Px)^2\}<\infty$, then
$$
\sqrt n(\surrest-\theta)
\overset{d}{\longrightarrow}
\mc N(0,\widetilde{\mc V}),
\qquad
\widetilde{\mc V}
=
\Ex\{\widetilde\varphi^*(\mc O;\Px)^2\}.
$$
Hence, $\surrest$ attains the semiparametric efficiency bound under Condition \ref{cond:statsurr}.
\end{corollary}

\begin{proof}
The result follows from Theorem \ref{thm:asy-miss-supp} by taking
$$
R\equiv 1,
\qquad
\widetilde r(\mb X,Q)\equiv 1.
$$
Under this specialization, the EIF $\widetilde\psi_a^*$ reduces to $\widetilde\varphi_a^*$ in Theorem \ref{thm:EIF-main}. Moreover, the remainder term \eqref{eq:rem-miss} reduces to
\begin{align}
\text{Rem}_a(\widehat\eta,\eta)
&=
\Ex\left[
\frac{
\widehat{\widetilde\pi}_a(\mb X,Q)
-
\widetilde\pi_a(\mb X,Q)}
{\widehat\pi_a(\mb X)}
\{
\widetilde\nu(\mb X,Q)
-
\widehat{\widetilde\nu}(\mb X,Q)
\}
\right]
\nonumber\\
&\quad+
\Ex\left[
\frac{
\pi_a(\mb X)-\widehat\pi_a(\mb X)}
{\widehat\pi_a(\mb X)}
\{
\widetilde\mu_a(\mb X)
-
\widehat{\widetilde\mu}_a(\mb X)
\}
\right].
\label{eq:rem-full}
\end{align}
By the positivity condition and the Cauchy--Schwarz inequality,
$$
|\text{Rem}_a(\widehat\eta,\eta)|
\lesssim
\|\widehat{\widetilde\pi}_a-\widetilde\pi_a\|_2
\|\widehat{\widetilde\nu}-\widetilde\nu\|_2
+
\|\widehat\pi_a-\pi_a\|_2
\|\widehat{\widetilde\mu}_a-\widetilde\mu_a\|_2,
$$
which is $o_p(n^{-1/2})$ under the stated product-rate condition.
\end{proof}

\subsection{Multiple robustness of the surrogate-EIF estimators}
\label{subsec:multiple-robust}

We next characterize the robustness of $\surrest$ to misspecification of the nuisance functions. Unlike the RAL results in Section \ref{subsec:asy-surr}, which require all nuisance estimators to converge to their true counterparts at appropriate rates, consistency of $\surrest$ can hold under substantially weaker conditions. In particular, the second-order remainder expressions \eqref{eq:rem-miss} and \eqref{eq:rem-full} reveal a multiple robustness property.

We first consider the missing-outcome setting. Suppose that the nuisance estimators converge in $L_2(\Px)$ to possibly misspecified limits
$$
\widehat\pi_a \longrightarrow \bar\pi_a,\qquad
\widehat{\widetilde\pi}_a \longrightarrow \bar{\widetilde\pi}_a,\qquad
\widehat{\widetilde r} \longrightarrow \bar{\widetilde r},\qquad
\widehat{\widetilde\nu} \longrightarrow \bar{\widetilde\nu},\qquad
\widehat{\widetilde\mu}_a \longrightarrow \bar{\widetilde\mu}_a,
$$
for $a=0,1$, where the limiting propensity-score and missingness functions are bounded away from zero.

\begin{theorem}[Multiple robustness with missing outcomes]
\label{thm:mr-miss}
Suppose Assumptions \ref{asp:unconf}--\ref{asp:positivity-r} and Condition \ref{cond:statsurr} hold. Then $\surrest$ is consistent for $\theta$ if, for each $a=0,1$, both of the following conditions are satisfied:
$$
\bar{\widetilde\nu}=\widetilde\nu
\qquad\text{or}\qquad
\left(
\bar{\widetilde\pi}_a=\widetilde\pi_a
\ \text{and}\
\bar{\widetilde r}=\widetilde r
\right),
$$
and
$$
\bar\pi_a=\pi_a
\qquad\text{or}\qquad
\bar{\widetilde\mu}_a=\widetilde\mu_a.
$$
Equivalently, consistency is guaranteed under any of the following four combinations of correctly specified nuisance-function blocks:
\begin{enumerate}[label=(\roman*)]
    \item $\widetilde\nu$ and $\pi_a$;
    \item $\widetilde\nu$ and $\widetilde\mu_a$;
    \item $(\widetilde\pi_a,\widetilde r)$ and $\pi_a$;
    \item $(\widetilde\pi_a,\widetilde r)$ and $\widetilde\mu_a$.
\end{enumerate}
\end{theorem}

\begin{proof}
From the remainder expression in \eqref{eq:rem-miss}, the probability limit of the bias term is
\begin{align*}
\overline{\text{Rem}}^{\mathrm{mis}}_a
&=
\Ex\left[
\frac{
\bar{\widetilde\pi}_a(\mb X,Q)
\widetilde r(\mb X,Q)
-
\widetilde\pi_a(\mb X,Q)
\bar{\widetilde r}(\mb X,Q)}
{\bar\pi_a(\mb X)
 \bar{\widetilde r}(\mb X,Q)}
\{
\widetilde\nu(\mb X,Q)
-
\bar{\widetilde\nu}(\mb X,Q)
\}
\right]\\
&\quad+
\Ex\left[
\frac{
\pi_a(\mb X)-\bar\pi_a(\mb X)}
{\bar\pi_a(\mb X)}
\{
\widetilde\mu_a(\mb X)
-
\bar{\widetilde\mu}_a(\mb X)
\}
\right].
\end{align*}
The first term is zero if $\bar{\widetilde\nu}=\widetilde\nu$. Alternatively, it is zero if both
$\bar{\widetilde\pi}_a=\widetilde\pi_a$ and
$\bar{\widetilde r}=\widetilde r$. The second term is zero if either
$\bar\pi_a=\pi_a$ or
$\bar{\widetilde\mu}_a=\widetilde\mu_a$.
Therefore, under any combination satisfying one condition from each pair,
$$
\overline{\text{Rem}}^{\mathrm{mis}}_a=0.
$$
It follows that
$$
\widehat\theta_a^{\mathrm{surr}}
\overset{p}{\longrightarrow}
\theta_a,
$$
for $a=0,1$, and hence
$$
\surrest
\overset{p}{\longrightarrow}
\theta.
$$
\end{proof}

The fully observed setting yields a simpler robustness structure because the missingness nuisance function is absent.

\begin{corollary}[Multiple robustness with fully observed outcomes]
\label{cor:mr-full}
Suppose Assumptions \ref{asp:unconf} and \ref{asp:positivity} and Condition \ref{cond:statsurr} hold, and suppose that $Y$ is fully observed. Then $\surrest$ is consistent for $\theta$ if, for each $a=0,1$,
$$
\bar{\widetilde\nu}=\widetilde\nu
\qquad\text{or}\qquad
\bar{\widetilde\pi}_a=\widetilde\pi_a,
$$
and
$$
\bar\pi_a=\pi_a
\qquad\text{or}\qquad
\bar{\widetilde\mu}_a=\widetilde\mu_a.
$$
Equivalently, consistency is guaranteed under any of the following four combinations:
\begin{enumerate}[label=(\roman*)]
    \item $\widetilde\nu$ and $\pi_a$;
    \item $\widetilde\nu$ and $\widetilde\mu_a$;
    \item $\widetilde\pi_a$ and $\pi_a$;
    \item $\widetilde\pi_a$ and $\widetilde\mu_a$.
\end{enumerate}
\end{corollary}

\begin{proof}
By \eqref{eq:rem-full}, the limiting bias term is
\begin{align*}
\overline{\text{Rem}}_a
=
\Ex\left[
\frac{
\bar{\widetilde\pi}_a(\mb X,Q)
-
\widetilde\pi_a(\mb X,Q)}
{\bar\pi_a(\mb X)}
\{
\widetilde\nu(\mb X,Q)
-
\bar{\widetilde\nu}(\mb X,Q)
\}
\right] +
\Ex\left[
\frac{
\pi_a(\mb X)-\bar\pi_a(\mb X)}
{\bar\pi_a(\mb X)}
\{
\widetilde\mu_a(\mb X)
-
\bar{\widetilde\mu}_a(\mb X)
\}
\right].
\end{align*}
The first term vanishes if either
$\bar{\widetilde\nu}=\widetilde\nu$ or
$\bar{\widetilde\pi}_a=\widetilde\pi_a$, while the second vanishes if either
$\bar\pi_a=\pi_a$ or
$\bar{\widetilde\mu}_a=\widetilde\mu_a$.
Thus, under any combination satisfying one condition from each pair,
$$
\overline{\text{Rem}}_a=0,
$$
which implies consistency of $\widehat\theta_a^{\mathrm{surr}}$ and hence of $\surrest$.
\end{proof}

These results show that consistency of the surrogate-EIF estimator does not require simultaneous correct specification of all nuisance functions.

\subsection{Bias under violations of statistical surrogacy}
\label{subapp:bias}

We next characterize the asymptotic bias of the surrogate-EIF estimator stated in Section \ref{subsec:bias-by-Cond1} when Condition \ref{cond:statsurr} generally fails. For convenience, we recall the notation
$$
\pi_a(\mb X)=\Px(A=a\mid\mb X), \qquad \widetilde\pi_a(\mb X,Q)=\Px(A=a\mid\mb X,Q),
$$
and let
$$
\widetilde m(a,\mb X,Q)=\Ex(Y\mid\mb X,A=a,Q).
$$
Define the residual conditional treatment contrast
\begin{equation}
d(\mb X,Q)=\widetilde m(1,\mb X,Q)-\widetilde m(0,\mb X,Q).
\label{eq:surrogacy-violation}
\end{equation}
Condition \ref{cond:statsurr} implies $d(\mb X,Q)=0$ almost surely, although the latter is only a conditional-mean restriction and is therefore weaker than statistical surrogacy.

\subsubsection{Fully observed outcomes}

Suppose first that $Y$ is fully observed. Recall
$$
\widetilde\nu(\mb X,Q)=\Ex(Y\mid\mb X,Q), \qquad \widetilde\mu_a(\mb X)=\Ex\{\widetilde\nu(\mb X,Q)\mid\mb X,A=a\}.
$$
Without imposing Condition \ref{cond:statsurr} and by law of total iteration, the pooled outcome regression satisfies
\begin{equation}
\widetilde\nu(\mb X,Q)=\widetilde\pi_1(\mb X,Q)\widetilde m(1,\mb X,Q)+\widetilde\pi_0(\mb X,Q)\widetilde m(0,\mb X,Q).
\label{eq:pooled-outcome-full-bias}
\end{equation}

Consider the surrogate-EIF estimator $\widehat\theta_a^{\mathrm{surr}}$ defined in Section \ref{subsec:asy-surr}, with all nuisance functions consistently estimated but without assuming Condition \ref{cond:statsurr}.

\begin{proposition}[Bias under violations of statistical surrogacy with fully observed outcomes]
\label{prop:bias-full-surr}
Suppose Assumptions \ref{asp:unconf} and \ref{asp:positivity} hold and $Y$ is fully observed. If the nuisance functions in $\widehat\theta_a^{\mathrm{surr}}$ are consistently estimated, then
\begin{equation}
\widehat\theta_a^{\mathrm{surr}}\overset{p}{\longrightarrow}\theta_a^{\dagger,\mathrm{full}}
=\Ex\left[\Ex\{\widetilde\nu(\mb X,Q)\mid\mb X,A=a\}\right].
\label{eq:pseudotarget-full}
\end{equation}
Let $\Delta_a^{\mathrm{full}}=\theta_a^{\dagger,\mathrm{full}}-\theta_a$. Then
\begin{equation}
\Delta_1^{\mathrm{full}}=-\Ex\left\{\frac{\widetilde\pi_0(\mb X,Q)\widetilde\pi_1(\mb X,Q)}{\pi_1(\mb X)}d(\mb X,Q)\right\},
\label{eq:bias-full-a1}
\end{equation}
and
\begin{equation}
\Delta_0^{\mathrm{full}}=\Ex\left\{\frac{\widetilde\pi_0(\mb X,Q)\widetilde\pi_1(\mb X,Q)}{\pi_0(\mb X)}d(\mb X,Q)\right\}.
\label{eq:bias-full-a0}
\end{equation}
Consequently, the asymptotic bias of $\surrest$ for the ATE is
\begin{equation}
\Delta^{\mathrm{full}}=\Delta_1^{\mathrm{full}}-\Delta_0^{\mathrm{full}}
=-\Ex\left\{\frac{\widetilde\pi_0(\mb X,Q)\widetilde\pi_1(\mb X,Q)}{\pi_0(\mb X)\pi_1(\mb X)}d(\mb X,Q)\right\}.
\label{eq:bias-full-ATE}
\end{equation}
\end{proposition}

\begin{proof}
For a fixed treatment level $a$, the probability limit of $\widehat\theta_a^{\mathrm{surr}}$ is the expectation of
$$
\frac{\Ix(A=a)}{\pi_a(\mb X)}\{\widetilde\nu(\mb X,Q)-\widetilde\mu_a(\mb X)\}
+\frac{\widetilde\pi_a(\mb X,Q)}{\pi_a(\mb X)}\{Y-\widetilde\nu(\mb X,Q)\}
+\widetilde\mu_a(\mb X).
$$
The first term has expectation zero by the definition of $\widetilde\mu_a$, while the second has expectation zero because $\Ex\{Y-\widetilde\nu(\mb X,Q)\mid\mb X,Q\}=0$. Hence,
$$
\theta_a^{\dagger,\mathrm{full}}=\Ex\{\widetilde\mu_a(\mb X)\}
=\Ex\left[\Ex\{\widetilde\nu(\mb X,Q)\mid\mb X,A=a\}\right].
$$
Under Assumptions \ref{asp:unconf} and \ref{asp:positivity},
$$
\theta_a=\Ex\left[\Ex\{\widetilde m(a,\mb X,Q)\mid\mb X,A=a\}\right].
$$
Therefore,
\begin{equation}
\Delta_a^{\mathrm{full}}
=\Ex\left[\Ex\{\widetilde\nu(\mb X,Q)-\widetilde m(a,\mb X,Q)\mid\mb X,A=a\}\right].
\label{eq:bias-full-intermediate}
\end{equation}
By \eqref{eq:surrogacy-violation} and \eqref{eq:pooled-outcome-full-bias},
$$
\widetilde\nu(\mb X,Q)-\widetilde m(1,\mb X,Q)=-\widetilde\pi_0(\mb X,Q)d(\mb X,Q),
$$
and
$$
\widetilde\nu(\mb X,Q)-\widetilde m(0,\mb X,Q)=\widetilde\pi_1(\mb X,Q)d(\mb X,Q).
$$
For any integrable function $h(\mb X,Q)$,
\begin{equation}
\Ex\left[\Ex\{h(\mb X,Q)\mid\mb X,A=a\}\right]
=\Ex\left\{\frac{\widetilde\pi_a(\mb X,Q)}{\pi_a(\mb X)}h(\mb X,Q)\right\}.
\label{eq:change-measure-Q}
\end{equation}
Applying \eqref{eq:change-measure-Q} to \eqref{eq:bias-full-intermediate} gives \eqref{eq:bias-full-a1} and \eqref{eq:bias-full-a0}. Finally,
$$
\Delta^{\mathrm{full}}
=-\Ex\left[\widetilde\pi_0(\mb X,Q)\widetilde\pi_1(\mb X,Q)
\left\{\frac{1}{\pi_1(\mb X)}+\frac{1}{\pi_0(\mb X)}\right\}d(\mb X,Q)\right],
$$
which reduces to \eqref{eq:bias-full-ATE} because $\pi_0(\mb X)+\pi_1(\mb X)=1$.
\end{proof}

\subsubsection{Missing outcomes}

We next consider the setting in which $Y$ is subject to missingness. We continue to assume Assumption \ref{asp:MAR},
$$
Y\independent R\mid\mb X,A,Q,
$$
but allow Condition \ref{cond:statsurr} to fail. Recall
$$
\widetilde r_a(\mb X,Q)=\Px(R=1\mid\mb X,Q,A=a), \qquad
\widetilde r(\mb X,Q)=\Px(R=1\mid\mb X,Q).
$$
As in Remark \ref{rmk:another-EIF-form}, define
$$
\widetilde\pi_{R=1}(a\mid\mb X,Q)=\Px(A=a\mid\mb X,Q,R=1).
$$
By Bayes's rule,
\begin{equation}
\widetilde\pi_{R=1}(a\mid\mb X,Q)
=\frac{\widetilde\pi_a(\mb X,Q)\widetilde r_a(\mb X,Q)}{\widetilde r(\mb X,Q)}.
\label{eq:pi-r1-bias}
\end{equation}

When Condition \ref{cond:statsurr} holds, Assumption \ref{asp:MAR} implies that the pooled outcome regression among observations with $R=1$ equals $\widetilde\nu(\mb X,Q)$. This equality need not hold when statistical surrogacy fails. Accordingly, write
$$
\widetilde\nu(\mb X,Q,1)=\Ex(Y\mid\mb X,Q,R=1).
$$
Under Assumption \ref{asp:MAR},
$$
\Ex(Y\mid\mb X,A=a,Q,R=1)=\widetilde m(a,\mb X,Q),
$$
and hence
\begin{equation}
\widetilde\nu(\mb X,Q,1)=\widetilde\pi_{R=1}(1\mid\mb X,Q)\widetilde m(1,\mb X,Q)+\widetilde\pi_{R=1}(0\mid\mb X,Q)\widetilde m(0,\mb X,Q).
\label{eq:pooled-outcome-miss-bias}
\end{equation}

We consider the natural implementation of the missing-outcome surrogate-EIF estimator in which $\widehat{\widetilde\nu}$ is fitted using observations with $R=1$. Under violation of statistical surrogacy, its probability limit is therefore $\widetilde\nu(\mb X,Q,1)$. Correspondingly, the probability limit of $\widehat{\widetilde\mu}_a(\mb X)$ is
$$
\Ex\{\widetilde\nu(\mb X,Q,1)\mid\mb X,A=a\}.
$$

\begin{proposition}[Bias under violations of statistical surrogacy with missing outcomes]
\label{prop:bias-miss-surr}
Suppose Assumptions \ref{asp:unconf}--\ref{asp:positivity-r} hold, except that Condition \ref{cond:statsurr} is not necessarily satisfied. Suppose further that the nuisance estimators converge to the observed-data limits described above. Then
\begin{equation}
\widehat\theta_a^{\mathrm{surr}}\overset{p}{\longrightarrow}\theta_a^{\dagger,\mathrm{mis}}
=\Ex\left[\Ex\{\widetilde\nu(\mb X,Q,1)\mid\mb X,A=a\}\right].
\label{eq:pseudotarget-mis}
\end{equation}
Let $\Delta_a^{\mathrm{mis}}=\theta_a^{\dagger,\mathrm{mis}}-\theta_a$. Then
\begin{equation}
\Delta_1^{\mathrm{mis}}
=-\Ex\left\{\frac{\widetilde\pi_1(\mb X,Q)\widetilde\pi_{R=1}(0\mid\mb X,Q)}{\pi_1(\mb X)}d(\mb X,Q)\right\},
\label{eq:bias-mis-a1}
\end{equation}
and
\begin{equation}
\Delta_0^{\mathrm{mis}}
=\Ex\left\{\frac{\widetilde\pi_0(\mb X,Q)\widetilde\pi_{R=1}(1\mid\mb X,Q)}{\pi_0(\mb X)}d(\mb X,Q)\right\}.
\label{eq:bias-mis-a0}
\end{equation}
Consequently, the asymptotic bias of $\surrest$ for the ATE is
\begin{equation}
\Delta^{\mathrm{mis}}
=-\Ex\left[\frac{\widetilde\pi_0(\mb X,Q)\widetilde\pi_1(\mb X,Q)}{\widetilde r(\mb X,Q)}
\left\{\frac{\widetilde r_0(\mb X,Q)}{\pi_1(\mb X)}+\frac{\widetilde r_1(\mb X,Q)}{\pi_0(\mb X)}\right\}d(\mb X,Q)\right].
\label{eq:bias-mis-ATE}
\end{equation}
\end{proposition}

\begin{proof}
Under the stated nuisance limits, the probability limit of $\widehat\theta_a^{\mathrm{surr}}$ is the expectation of
$$
\frac{\widetilde\pi_a(\mb X,Q)R}{\pi_a(\mb X)\widetilde r(\mb X,Q)}
\{Y-\widetilde\nu(\mb X,Q,1)\}
+\frac{\Ix(A=a)}{\pi_a(\mb X)}
\left[\widetilde\nu(\mb X,Q,1)-\Ex\{\widetilde\nu(\mb X,Q,1)\mid\mb X,A=a\}\right]
+\Ex\{\widetilde\nu(\mb X,Q,1)\mid\mb X,A=a\}.
$$
The first term has expectation zero because
$$
\Ex\left[R\{Y-\widetilde\nu(\mb X,Q,1)\}\mid\mb X,Q\right]
=\widetilde r(\mb X,Q)\Ex\{Y-\widetilde\nu(\mb X,Q,1)\mid\mb X,Q,R=1\}
=0.
$$
The second term also has expectation zero by iterated expectation. Therefore,
$$
\theta_a^{\dagger,\mathrm{mis}}
=\Ex\left[\Ex\{\widetilde\nu(\mb X,Q,1)\mid\mb X,A=a\}\right].
$$
Using Assumption \ref{asp:MAR},
\begin{equation}
\Delta_a^{\mathrm{mis}}
=\Ex\left[\Ex\{\widetilde\nu(\mb X,Q,1)-\widetilde m(a,\mb X,Q)\mid\mb X,A=a\}\right].
\label{eq:bias-mis-intermediate}
\end{equation}
By \eqref{eq:pooled-outcome-miss-bias},
$$
\widetilde\nu(\mb X,Q,1)-\widetilde m(1,\mb X,Q)
=-\widetilde\pi_{R=1}(0\mid\mb X,Q)d(\mb X,Q),
$$
and
$$
\widetilde\nu(\mb X,Q,1)-\widetilde m(0,\mb X,Q)
=\widetilde\pi_{R=1}(1\mid\mb X,Q)d(\mb X,Q).
$$
Applying \eqref{eq:change-measure-Q} gives \eqref{eq:bias-mis-a1} and \eqref{eq:bias-mis-a0}. Finally, substituting \eqref{eq:pi-r1-bias} yields
$$
\Delta^{\mathrm{mis}}
=-\Ex\left[\frac{\widetilde\pi_0(\mb X,Q)\widetilde\pi_1(\mb X,Q)}{\widetilde r(\mb X,Q)}
\left\{\frac{\widetilde r_0(\mb X,Q)}{\pi_1(\mb X)}+\frac{\widetilde r_1(\mb X,Q)}{\pi_0(\mb X)}\right\}d(\mb X,Q)\right],
$$
which proves \eqref{eq:bias-mis-ATE}.
\end{proof}

\begin{corollary}
\label{cor:bias-miss-full}
Under the conditions of Proposition \ref{prop:bias-miss-surr}, if additionally
$$
R \independent A\mid\mb X,Q,
$$
then $\widetilde r_0(\mb X,Q)=\widetilde r_1(\mb X,Q)=\widetilde r(\mb X,Q)$ and
\begin{equation}
\Delta^{\mathrm{mis}}=\Delta^{\mathrm{full}}
=-\Ex\left\{\frac{\widetilde\pi_0(\mb X,Q)\widetilde\pi_1(\mb X,Q)}{\pi_0(\mb X)\pi_1(\mb X)}d(\mb X,Q)\right\}.
\label{eq:bias-mis-full-equal}
\end{equation}
\end{corollary}

\begin{proof}
The result follows immediately from \eqref{eq:bias-mis-ATE} after substituting $\widetilde r_0=\widetilde r_1=\widetilde r$ and using $\pi_0(\mb X)+\pi_1(\mb X)=1$.
\end{proof}

These results show explicitly how violations of statistical surrogacy induce bias in the surrogate-EIF estimator. In both outcome-observation regimes, the bias is driven by the residual conditional treatment contrast $d(\mb X,Q)$ and is weighted by treatment overlap after conditioning on $(\mb X,Q)$. Condition \ref{cond:statsurr} therefore guarantees zero bias because it implies $d(\mb X,Q)=0$ almost surely. However, the converse need not hold: the ATE bias can vanish through weighted cancellation even when the full statistical surrogacy condition fails. With missing outcomes, the bias additionally depends on treatment-specific missingness probabilities through the treatment composition among observations with $R=1$.

\subsection{Variance ordering with additional surrogate information under outcome missingness}
\label{subsec:var-order}

Recall that Section \ref{subsec:G-B-surrogates} also characterizes the efficiency implications of incorporating additional surrogate information when the primary outcome is subject to missingness. An additional variable may provide further information about the outcome, the missingness mechanism, or both. Under the standard MAR assumption, we consider augmenting an existing surrogate $Q$ with an additional variable $G$ or $B$.

For a generic surrogate set $S$, define
$$
\widetilde\mu_S(a,\mb X,S)=\Ex(Y\mid\mb X,A=a,S,R=1),
$$
$$
\widetilde\nu_S(a,\mb X)=\Ex\{\widetilde\mu_S(a,\mb X,S)\mid\mb X,A=a\},
$$
and
$$
\widetilde r_S(a,\mb X,S)=\Px(R=1\mid\mb X,A=a,S).
$$
Under Assumptions \ref{asp:unconf} and \ref{asp:positivity}, together with MAR and the corresponding missingness-positivity condition given $S$, the nonparametric EIF for $\theta_a$ is
\begin{align}
\widetilde\varphi_a^{S}(\mc O;\Px)
=
\frac{\Ix(A=a)}{\pi(a\mid\mb X)}
\{\widetilde\mu_S(a,\mb X,S)-\widetilde\nu_S(a,\mb X)\}
+
\frac{\Ix(A=a)R}{\pi(a\mid\mb X)\widetilde r_S(a,\mb X,S)}
\{Y-\widetilde\mu_S(a,\mb X,S)\}
+
\widetilde\nu_S(a,\mb X)-\theta_a.
\label{eq:eif-generic-surr}
\end{align}
This is the standard EIF under MAR without imposing Condition \ref{cond:statsurr}. We use it to compare the efficiency consequences of two qualitatively different types of additional information.

\subsubsection{Outcome-predictive additional information}

Suppose an additional variable $G$ provides information about the outcome beyond $(\mb X,A,Q)$ but does not further explain the missingness mechanism. Specifically, assume
$$
G\independent R\mid\mb X,A,Q,
$$
and that MAR continues to hold after augmenting the surrogate set:
$$
Y\independent R\mid\mb X,A,Q,G.
$$
Write
$$
\widetilde\mu_Q(a,\mb X,Q)=\Ex(Y\mid\mb X,A=a,Q,R=1),
$$
and
$$
\widetilde\mu_{Q,G}(a,\mb X,Q,G)=\Ex(Y\mid\mb X,A=a,Q,G,R=1).
$$
Also define
$$
\widetilde r_Q(a,\mb X,Q)=\Px(R=1\mid\mb X,A=a,Q).
$$
Because $G\independent R\mid\mb X,A,Q$,
$$
\widetilde r_{Q,G}(a,\mb X,Q,G)=\widetilde r_Q(a,\mb X,Q).
$$

\begin{theorem}[Outcome-predictive-only variable]
\label{prop:outcome-predictive-aux}
Under the conditions above,
$$
\Vx\{\widetilde\varphi_a^{Q}(\mc O;\Px)\}
-
\Vx\{\widetilde\varphi_a^{Q,G}(\mc O;\Px)\}
$$
is equal to
$$
\Ex\left[
\frac{\Ix(A=a)}{\pi(a\mid\mb X)^2}
\{\widetilde\mu_{Q,G}(a,\mb X,Q,G)-\widetilde\mu_Q(a,\mb X,Q)\}^2
\frac{1-\widetilde r_Q(a,\mb X,Q)}
{\widetilde r_Q(a,\mb X,Q)}
\right]
\geq 0.
$$
Therefore, augmenting $Q$ with outcome-predictive information $G$ that provides no additional information about missingness cannot increase the semiparametric efficiency bound.
\end{theorem}

\begin{proof}
By the law of iterated expectations and $G\independent R\mid\mb X,A,Q$,
$$
\Ex\{\widetilde\mu_{Q,G}(a,\mb X,Q,G)\mid\mb X,A=a,Q\}
=
\widetilde\mu_Q(a,\mb X,Q),
$$
and hence
$$
\widetilde\nu_{Q,G}(a,\mb X)=\widetilde\nu_Q(a,\mb X).
$$
Moreover,
$$
\widetilde r_{Q,G}(a,\mb X,Q,G)=\widetilde r_Q(a,\mb X,Q).
$$
Using \eqref{eq:eif-generic-surr}, direct algebra yields
\begin{align}
\widetilde\varphi_a^{Q}(\mc O;\Px)
=
\widetilde\varphi_a^{Q,G}(\mc O;\Px)
+
\frac{\Ix(A=a)}{\pi(a\mid\mb X)}
\left\{
\frac{R}{\widetilde r_Q(a,\mb X,Q)}-1
\right\}
\{\widetilde\mu_{Q,G}(a,\mb X,Q,G)-\widetilde\mu_Q(a,\mb X,Q)\}.
\label{eq:eif-G-decomp}
\end{align}
Let
$$
D_G
=
\frac{\Ix(A=a)}{\pi(a\mid\mb X)}
\left\{
\frac{R}{\widetilde r_Q(a,\mb X,Q)}-1
\right\}
\{\widetilde\mu_{Q,G}(a,\mb X,Q,G)-\widetilde\mu_Q(a,\mb X,Q)\}.
$$
Since $G\independent R\mid\mb X,A,Q$, we have
$$
\Ex(D_G\mid\mb X,A,Q,G)=0.
$$
The components of $\widetilde\varphi_a^{Q,G}$ that are measurable with respect to $(\mb X,A,Q,G)$ are therefore uncorrelated with $D_G$. For the remaining outcome-residual component, MAR implies
$$
\Ex\{Y-\widetilde\mu_{Q,G}(a,\mb X,Q,G)\mid\mb X,A=a,Q,G,R=1\}=0,
$$
which gives
$$
\text{Cov}\{\widetilde\varphi_a^{Q,G}(\mc O;\Px),D_G\}=0.
$$
It follows that
$$
\Vx\{\widetilde\varphi_a^{Q}(\mc O;\Px)\}
=
\Vx\{\widetilde\varphi_a^{Q,G}(\mc O;\Px)\}
+
\Vx(D_G).
$$
Finally,
\begin{align*}
\Vx(D_G)
&=
\Ex\left[
\frac{\Ix(A=a)}{\pi(a\mid\mb X)^2}
\{\widetilde\mu_{Q,G}(a,\mb X,Q,G)-\widetilde\mu_Q(a,\mb X,Q)\}^2
\Vx\left\{
\left.
\frac{R}{\widetilde r_Q(a,\mb X,Q)}-1
\right|
\mb X,A,Q,G
\right\}
\right] \\
&=
\Ex\left[
\frac{\Ix(A=a)}{\pi(a\mid\mb X)^2}
\{\widetilde\mu_{Q,G}(a,\mb X,Q,G)-\widetilde\mu_Q(a,\mb X,Q)\}^2
\frac{1-\widetilde r_Q(a,\mb X,Q)}
{\widetilde r_Q(a,\mb X,Q)}
\right]
\geq 0.
\end{align*}
This proves the result.
\end{proof}

\subsubsection{Missingness-predictive additional information}

We next consider the opposite case. Suppose an additional variable $B$ provides information about the missingness mechanism but no additional information about the outcome beyond $(\mb X,A,Q,R)$. Specifically, assume
$$
B\independent Y\mid\mb X,A,Q,R,
$$
and that MAR continues to hold after augmenting the surrogate set:
$$
Y\independent R\mid\mb X,A,Q,B.
$$
Define
$$
\widetilde r_{Q,B}(a,\mb X,Q,B)
=
\Px(R=1\mid\mb X,A=a,Q,B).
$$
Unlike the preceding case, $\widetilde r_{Q,B}$ may depend on $B$.

\begin{theorem}[Missingness-predictive-only variable]
\label{prop:missing-predictive-aux}
Under the conditions above,
$$
\Vx\{\widetilde\varphi_a^{Q}(\mc O;\Px)\}
-
\Vx\{\widetilde\varphi_a^{Q,B}(\mc O;\Px)\}
\leq 0.
$$
More explicitly,
\begin{align}
& \Vx\{\widetilde\varphi_a^{Q}(\mc O;\Px)\}
-
\Vx\{\widetilde\varphi_a^{Q,B}(\mc O;\Px)\} \nonumber \\
 & 
=
-\Ex\left[
\frac{\Ix(A=a)R}{\pi(a\mid\mb X)^2}
\{Y-\widetilde\mu_Q(a,\mb X,Q)\}^2
\Vx\left\{
\left.
\frac{1}{\widetilde r_{Q,B}(a,\mb X,Q,B)}
\right|
\mb X,A,Q,R=1
\right\}
\right]
\leq 0.
\label{eq:var-bad-aux}
\end{align}
Thus, augmenting $Q$ with information that refines the missingness mechanism but provides no additional outcome information cannot reduce the semiparametric efficiency bound.
\end{theorem}

\begin{proof}
The assumption
$$
B\independent Y\mid\mb X,A,Q,R
$$
implies that, among observations with $R=1$,
$$
\widetilde\mu_{Q,B}(a,\mb X,Q,B)
=
\Ex(Y\mid\mb X,A=a,Q,B,R=1)
=
\widetilde\mu_Q(a,\mb X,Q).
$$
Consequently,
$$
\widetilde\nu_{Q,B}(a,\mb X)=\widetilde\nu_Q(a,\mb X).
$$

Now consider the EIF based on $(Q,B)$. Conditional on $(\mb X,A,Q,R,Y)$, the only remaining random component is the inverse-missingness weight involving $B$. By Bayes's rule,
$$
\Ex\left\{
\left.
\frac{1}{\widetilde r_{Q,B}(a,\mb X,Q,B)}
\right|
\mb X,A=a,Q,R=1
\right\}
=
\frac{1}{\widetilde r_Q(a,\mb X,Q)}.
$$
It follows that
$$
\Ex\{\widetilde\varphi_a^{Q,B}(\mc O;\Px)\mid\mb X,A,Q,R,Y\}
=
\widetilde\varphi_a^{Q}(\mc O;\Px).
$$
Therefore, by the law of total variance,
\begin{align*}
\Vx\{\widetilde\varphi_a^{Q,B}(\mc O;\Px)\}
&=
\Vx\left[
\Ex\{\widetilde\varphi_a^{Q,B}(\mc O;\Px)\mid\mb X,A,Q,R,Y\}
\right]
+
\Ex\left[
\Vx\{\widetilde\varphi_a^{Q,B}(\mc O;\Px)\mid\mb X,A,Q,R,Y\}
\right] \\
&=
\Vx\{\widetilde\varphi_a^{Q}(\mc O;\Px)\}
+
\Ex\left[
\Vx\{\widetilde\varphi_a^{Q,B}(\mc O;\Px)\mid\mb X,A,Q,R,Y\}
\right].
\end{align*}
The remaining conditional variance is
$$
\frac{\Ix(A=a)R}{\pi(a\mid\mb X)^2}
\{Y-\widetilde\mu_Q(a,\mb X,Q)\}^2
\Vx\left\{
\left.
\frac{1}{\widetilde r_{Q,B}(a,\mb X,Q,B)}
\right|
\mb X,A,Q,R=1
\right\}.
$$
Substituting this expression into the preceding variance decomposition gives \eqref{eq:var-bad-aux} and completes the proof.
\end{proof}

\subsection{Implementation of SACES}
\label{subapp:SACES-algorithm}

We provide the implementation details of SACES. Suppose that, in addition to the reference estimator $\widehat\theta^{(0)}=\stdest$, we consider $K$ candidate surrogate-assisted estimators $\widehat\theta^{(1)},\ldots,\widehat\theta^{(K)}$, each constructed using a prespecified surrogate set. Let $\widehat\varphi^{(k)}(\mc O_i)$ denote the estimated cross-fitted influence-function value corresponding to estimator $k$, for $k=0,1,\ldots,K$.

We estimate the asymptotic variance of each estimator by
\begin{equation}
\widehat\sigma_k^2
=
\Px_n\left[\widehat\varphi^{(k)}(\mc O)^2\right],
\qquad k=0,1,\ldots,K,
\label{eq:sigmahat-k}
\end{equation}
and, for each candidate $k=1,\ldots,K$, estimate the variability of its difference from the reference estimator by
\begin{equation}
\widehat\tau_k^2
=
\Px_n\left[
\left\{
\widehat\varphi^{(k)}(\mc O)
-
\widehat\varphi^{(0)}(\mc O)
\right\}^2
\right].
\label{eq:tauhat-k}
\end{equation}

For the reference estimator, define
\begin{equation}
\widehat{\mc R}^{\,\mathrm{mod}}(0)
=
\frac{\widehat\sigma_0^2}{n}.
\label{eq:riskhat-mod-0-supp}
\end{equation}
For $k=1,\ldots,K$, define
\begin{equation}
\widehat{\mc R}^{\,\mathrm{mod}}(k)
=
\left[
\left\{
\widehat\theta^{(k)}
-
\widehat\theta^{(0)}
\right\}^2
-
\frac{\widehat\tau_k^2}{n}
\right]_+
+
\frac{\widehat\sigma_k^2}{n}.
\label{eq:riskhat-mod-k-supp}
\end{equation}
The selected index and the resulting SACES estimator are
\begin{equation}
\widehat k
=
\arg\min_{k\in\{0,1,\ldots,K\}}
\widehat{\mc R}^{\,\mathrm{mod}}(k),
\qquad
\adptest
=
\widehat\theta^{(\widehat k)}.
\label{eq:k-hat-supp}
\end{equation}

Algorithm \ref{alg:SACES} summarizes the complete procedure.

\begin{algorithm}[H]
\caption{SACES: Surrogate-Assisted Adaptive Causal Effect Estimation}
\label{alg:SACES}
\begin{algorithmic}[1]

\REQUIRE Observed data $\{\mc O_i\}_{i=1}^n$ and $K$ prespecified candidate surrogate sets.

\STATE Construct the reference estimator $\widehat\theta^{(0)}=\stdest$ using cross-fitted nuisance-function estimators, and obtain the corresponding estimated influence-function values $\widehat\varphi^{(0)}(\mc O_i)$.

\FOR{$k=1,\ldots,K$}

\STATE Construct the surrogate-EIF estimator $\widehat\theta^{(k)}$ using candidate surrogate set $k$ and the same cross-fitting scheme.

\STATE Obtain the corresponding estimated influence-function values $\widehat\varphi^{(k)}(\mc O_i)$.

\ENDFOR

\STATE Compute $\widehat\sigma_k^2$ for $k=0,1,\ldots,K$ using \eqref{eq:sigmahat-k}.

\STATE Compute $\widehat\tau_k^2$ for $k=1,\ldots,K$ using \eqref{eq:tauhat-k}.

\STATE Compute $\widehat{\mc R}^{\,\mathrm{mod}}(k)$ for $k=0,1,\ldots,K$ using \eqref{eq:riskhat-mod-0-supp} and \eqref{eq:riskhat-mod-k-supp}.

\STATE Select $\widehat k$ according to \eqref{eq:k-hat-supp}.

\ENSURE The SACES estimator $\adptest=\widehat\theta^{(\widehat k)}$.

\end{algorithmic}
\end{algorithm}

The candidate surrogate sets can consist of individual surrogate variables or prespecified combinations of multiple surrogates. More generally, the same procedure applies to any fixed finite collection of surrogate-assisted estimators for which the corresponding estimated influence functions are available.

\subsection{Proof of Theorem \ref{thm:SACES}}\label{subapp:proof-SACES}

We first provide sufficient conditions for the mean-square remainder assumption in Theorem \ref{thm:SACES}. These conditions strengthen the usual probability-based remainder control for asymptotic linearity.

\begin{lemma}[Mean-square control of cross-fitted remainders]
\label{lem:crossfit-L2-remainder}
Suppose that, for each $k=0,\ldots,K$, the estimator $\widehat\theta^{(k)}$ is constructed using a fixed number $V$ of cross-fitting folds $\mc I_1,\ldots,\mc I_V$, with fold sizes $n_v$ satisfying $n_v/n$ bounded away from zero.

Let $g_k$ denote a population estimating function satisfying
$$
\Px(g_k)=\theta^{(k)},
\qquad
\varphi^{(k)}(\mc O;\Px)=g_k(\mc O)-\theta^{(k)}.
$$
For each fold $v$, let $\widehat g_{k,v}$ be estimated using observations outside $\mc I_v$, and suppose that
$$
\widehat\theta^{(k)}
=
\sum_{v=1}^V\frac{n_v}{n}
\Px_{n,v}(\widehat g_{k,v}),
$$
where $\Px_{n,v}$ denotes the empirical measure of fold $v$.

Assume that, uniformly over the fixed finite collection of candidates and folds,
\begin{align}
\Ex\left[
\|\widehat g_{k,v}-g_k\|_2^2
\right]
&=o(1),
\label{eq:crossfit-score-L2}\\
\Ex\left[
\left\{
\Px(\widehat g_{k,v}-g_k)
\right\}^2
\right]
&=o(n^{-1}).
\label{eq:crossfit-bias-L2}
\end{align}
Then
$$
\widehat\theta^{(k)}-\theta^{(k)}
=
\frac1n\sum_{i=1}^n
\varphi^{(k)}(\mc O_i;\Px)
+r_{n,k},
$$
where
$$
\max_{0\leq k\leq K}
\Ex(r_{n,k}^2)
=o(n^{-1}).
$$
\end{lemma}

\begin{proof}
Fix a candidate $k$. By adding and subtracting $g_k$, we obtain
\begin{align*}
\widehat\theta^{(k)}-\theta^{(k)}
=
(\Px_n-\Px)(g_k)+
\sum_{v=1}^V\frac{n_v}{n}
(\Px_{n,v}-\Px)
(\widehat g_{k,v}-g_k)+
\sum_{v=1}^V\frac{n_v}{n}
\Px(\widehat g_{k,v}-g_k).
\end{align*}
Since $\Px (g_k)=\theta^{(k)}$, the first term equals
$$
(\Px_n-\Px)(g_k)
=
\frac1n\sum_{i=1}^n
\varphi^{(k)}(\mc O_i;\Px).
$$

Define the empirical-process and bias remainders as
\begin{align*}
D_{n,k}
&=
\sum_{v=1}^V\frac{n_v}{n}
(\Px_{n,v}-\Px)
(\widehat g_{k,v}-g_k),\\
B_{n,k}
&=
\sum_{v=1}^V\frac{n_v}{n}
\Px(\widehat g_{k,v}-g_k).
\end{align*}
Thus,
$$
r_{n,k}=D_{n,k}+B_{n,k}.
$$

We first control $D_{n,k}$. Conditional on the training observations outside fold $v$, the function $\widehat g_{k,v}-g_k$ is fixed and the validation observations are independent of the fitted function. Therefore,
\begin{align*}
\Ex\left[
\left\{
(\Px_{n,v}-\Px)
(\widehat g_{k,v}-g_k)
\right\}^2
\;\middle|\;
\mc D_{-v}
\right]=
\frac1{n_v}
\Vx\left(
\widehat g_{k,v}(\mc O)-g_k(\mc O)
\;\middle|\;
\mc D_{-v}
\right)\leq
\frac1{n_v}
\|\widehat g_{k,v}-g_k\|_2^2,
\end{align*}
where $\mc D_{-v}$ denotes the training data excluding fold $v$.

By Jensen's inequality,
\begin{align*}
\Ex(D_{n,k}^2)
&\leq
\sum_{v=1}^V\frac{n_v}{n}
\Ex\left[
\left\{
(\Px_{n,v}-\Px)
(\widehat g_{k,v}-g_k)
\right\}^2
\right]\\
&\leq
\frac1n\sum_{v=1}^V
\Ex\left[
\|\widehat g_{k,v}-g_k\|_2^2
\right]\\
&=o(n^{-1}),
\end{align*}
where the final equality follows from \eqref{eq:crossfit-score-L2} and the fixed number of folds.

Similarly, another application of Jensen's inequality gives
\begin{align*}
\Ex(B_{n,k}^2)\leq
\sum_{v=1}^V\frac{n_v}{n}
\Ex\left[
\left\{
\Px(\widehat g_{k,v}-g_k)
\right\}^2
\right]=o(n^{-1})
\end{align*}
by \eqref{eq:crossfit-bias-L2}.

Consequently,
\begin{align*}
\Ex(r_{n,k}^2)=
\Ex\{(D_{n,k}+B_{n,k})^2\}\leq
2\Ex(D_{n,k}^2)
+ 2\Ex(B_{n,k}^2)=o(n^{-1}).
\end{align*}
Since $K$ is fixed and finite, this result holds uniformly over $k=0,\ldots,K$.
\end{proof}

\begin{remark}[Sufficient nuisance estimation rates]
\label{rem:nuisance-L2-rates}
The conditions of Lemma \ref{lem:crossfit-L2-remainder} can be verified using nuisance estimation rates.

For the surrogate-EIF estimator in the missing-outcome setting, the second-order remainder derived in \eqref{eq:rem-miss} satisfies, under uniformly bounded inverse propensity weights,
\begin{align*}
|\text{Rem}^{\mathrm{mis}}_a|
\lesssim\;
\|\widehat{\widetilde\nu}-\widetilde\nu\|_2
\left\{
\|\widehat{\widetilde\pi}_a-\widetilde\pi_a\|_2
+
\|\widehat{\widetilde r}-\widetilde r\|_2
\right\} +
\|\widehat\pi_a-\pi_a\|_2
\|\widehat{\widetilde\mu}_a-\widetilde\mu_a\|_2.
\end{align*}
Here, the true nuisance functions may be understood as the relevant observed-data probability limits when statistical surrogacy does not hold.

A sufficient condition for mean-square control of this bias is
\begin{align}
\Ex\left[
\left\{\|\widehat{\widetilde\nu}-\widetilde\nu\|_2
\left(
\|\widehat{\widetilde\pi}_a-\widetilde\pi_a\|_2
+
\|\widehat{\widetilde r}-\widetilde r\|_2
\right) +
\|\widehat\pi_a-\pi_a\|_2
\|\widehat{\widetilde\mu}_a-\widetilde\mu_a\|_2
\right\}^2
\right]
=o(n^{-1}).
\label{eq:mean-square-product-rate}
\end{align}

For example, let $\eta$ denote any nuisance function in the collection
$\{\pi_a,\widetilde\pi_a,\widetilde r,\widetilde\nu,\widetilde\mu_a\}$, $a=0,1$,
with $\widehat\eta$ denoting its corresponding estimator.
By the Cauchy--Schwarz inequality, condition \eqref{eq:mean-square-product-rate} follows if, for each $\eta$,
$$
\left[
\Ex\left\{
\|\widehat\eta-\eta\|_2^4
\right\}
\right]^{1/4}
=o(n^{-1/4}).
$$
These rates are sufficient but not necessary; slower rates for some nuisance functions are permitted when compensated by faster rates for others.

Under suitable outcome moment conditions, mean-square consistency of the nuisance estimators and uniformly bounded inverse propensity weights also ensure
\eqref{eq:crossfit-score-L2}. In particular, the boundedness requirement can be enforced for estimated propensity scores through truncation.

Together, these conditions establish the mean-square remainder requirement for the cross-fitted surrogate-EIF estimator. The fully observed case follows by setting $R\equiv1$ and $\widetilde r\equiv1$. Analogous product-rate conditions apply to the unrestricted reference estimator, using its treatment-specific outcome and observation models. The same argument extends to any fixed finite collection of candidate estimators satisfying the corresponding conditions.
\end{remark}

We now prove Theorem \ref{thm:SACES}.

\begin{proof}
Let
$$
\Delta_k=\theta^{(k)}-\theta^{(0)},
\qquad k=0,\ldots,K.
$$
By assumption, $\theta^{(0)}=\theta$.

First consider a candidate $k$ with $\Delta_k\neq0$. The assumed asymptotic linear representations imply
\begin{align*}
\widehat\theta^{(k)}
-\widehat\theta^{(0)}=
\Delta_k+
\frac1n\sum_{i=1}^n
\left\{
\varphi^{(k)}(\mc O_i;\Px)-\varphi^{(0)}(\mc O_i;\Px)
\right\}+r_{n,k}-r_{n,0}.
\end{align*}
Since the influence functions have finite second moments, the empirical average is $O_p(n^{-1/2})$. Moreover, the mean-square remainder condition implies
$$
r_{n,k}-r_{n,0}=o_p(n^{-1/2}).
$$
Consequently,
$$
\widehat\theta^{(k)}
-\widehat\theta^{(0)}
=
\Delta_k+O_p(n^{-1/2}).
$$
Squaring both sides gives
$$
\left\{
\widehat\theta^{(k)}
-\widehat\theta^{(0)}
\right\}^2
=
\Delta_k^2+O_p(n^{-1/2}).
$$

Since $\widehat\tau_k^2=O_p(1)$,
$$
\frac{\widehat\tau_k^2}{n}=O_p(n^{-1}).
$$
For fixed $\Delta_k\neq0$, the quantity inside the positive-part operator in $\widehat{\mc R}^{\,\mathrm{mod}}(k)$ is therefore strictly positive with probability tending to one. Hence,
$$
\widehat{\mc R}^{\,\mathrm{mod}}(k)
=
\Delta_k^2+O_p(n^{-1/2}).
$$
On the other hand, the reference criterion satisfies
$$
\widehat{\mc R}^{\,\mathrm{mod}}(0)
=
\frac{\widehat\sigma_0^2}{n}
=
O_p(n^{-1}).
$$
Because $\Delta_k^2>0$,
$$
\Px\left\{
\widehat{\mc R}^{\,\mathrm{mod}}(k)
>
\widehat{\mc R}^{\,\mathrm{mod}}(0)
\right\}
\to1.
$$
Thus,
$$
\Px(\widehat k=k)\to0
$$
for every candidate with $\Delta_k\neq0$. Since the candidate collection is finite,
$$
\Px\{\theta^{(\widehat k)}=\theta\}\to1.
$$

Next consider a candidate $k$ satisfying $\Delta_k=0$. By the definition of the modified estimated risk,
$$
\widehat{\mc R}^{\,\mathrm{mod}}(k)
\geq
\frac{\widehat\sigma_k^2}{n}
=
\frac{\sigma_k^2}{n}
+
o_p(n^{-1}),
$$
whereas
$$
\widehat{\mc R}^{\,\mathrm{mod}}(0)
=
\frac{\sigma_0^2}{n}
+
o_p(n^{-1}).
$$
If $\sigma_k^2>\sigma_0^2$, then
\begin{align*}
n\left\{
\widehat{\mc R}^{\,\mathrm{mod}}(k)
-
\widehat{\mc R}^{\,\mathrm{mod}}(0)
\right\}
&\geq
\sigma_k^2-\sigma_0^2+o_p(1).
\end{align*}
Since $\sigma_k^2-\sigma_0^2>0$, it follows that
$$
\Px\left\{
\widehat{\mc R}^{\,\mathrm{mod}}(k)
>
\widehat{\mc R}^{\,\mathrm{mod}}(0)
\right\}
\to1.
$$

Combining the two cases, define
$$
\mc K_0
=
\left\{
k:
\theta^{(k)}=\theta,\;
\sigma_k^2\leq\sigma_0^2
\right\}.
$$
Because the reference estimator is included in the candidate collection, $\mc K_0$ is nonempty. Since $K$ is finite,
$$
\Px(\widehat k\in\mc K_0)\to1.
$$

We now establish the risk comparison. For any fixed $k\in\mc K_0$,
$$
\widehat\theta^{(k)}-\theta
=
\overline\varphi_k+r_{n,k},
$$
where
$$
\overline\varphi_k
=
\frac1n\sum_{i=1}^n
\varphi^{(k)}(\mc O_i;\Px).
$$
Therefore,
\begin{align*}
\mc R_n(k)=
\Ex\left[
\{\widehat\theta^{(k)}-\theta\}^2
\right]=
\Ex(\overline\varphi_k^2)
+
2\Ex(\overline\varphi_k r_{n,k})
+
\Ex(r_{n,k}^2).
\end{align*}
Since the observations are independent and the influence function has mean zero,
$$
\Ex(\overline\varphi_k^2)
=
\frac{\sigma_k^2}{n}.
$$
By the Cauchy--Schwarz inequality,
\begin{align*}
\left|
\Ex(\overline\varphi_k r_{n,k})
\right|
\leq
\{\Ex(\overline\varphi_k^2)\}^{1/2}
\{\Ex(r_{n,k}^2)\}^{1/2}=
\left(\frac{\sigma_k^2}{n}\right)^{1/2}
o(n^{-1/2})=o(n^{-1}).
\end{align*}
Together with $\Ex(r_{n,k}^2)=o(n^{-1})$, this yields
$$
\mc R_n(k)
=
\frac{\sigma_k^2}{n}
+
o(n^{-1}).
$$

Because $\mc K_0$ is finite, the expansion is uniform over $k\in\mc K_0$, namely,
$$
\max_{k\in\mc K_0}
\left|
\mc R_n(k)
-
\frac{\sigma_k^2}{n}
\right|
=
o(n^{-1}).
$$

On the event $\{\widehat k\in\mc K_0\}$,
\begin{align*}
\mc R_n(\widehat k)=
\frac{\sigma_{\widehat k}^2}{n}
+
o(n^{-1})\leq
\frac{\sigma_0^2}{n}
+
o(n^{-1})=
\mc R_n(0)+o(n^{-1}).
\end{align*}
Since
$$
\Px(\widehat k\in\mc K_0)\to1,
$$
we conclude that
$$
\mc R_n(\widehat k)
\leq
\mc R_n(0)+o_p(n^{-1}).
$$
Equivalently, for every $\epsilon>0$,
$$
\Px\left\{
\mc R_n(\widehat k)
\leq
\mc R_n(0)+\frac{\epsilon}{n}
\right\}
\to1.
$$
This completes the proof.
\end{proof}

\subsection{Inference after estimator selection}\label{subapp:infer-G}

After selecting $\widehat k$, na\"ive confidence intervals (CIs) based on influence function of the selected estimator can fail to account for the uncertainty induced by the selection step. Following \cite{rothenhausler2024model}, we view the procedure as estimator selection over a finite set of candidate estimators indexed by $k\in\{0,1,\dots,K\}$. With a fixed finite number of candidates, asymptotically valid post-selection inference can be constructed by propagating the joint asymptotic distribution of
$\big(\widehat\theta^{(0)},\widehat\theta^{(1)},\dots,\widehat\theta^{(G)}\big)$ through the MSE selection rule \eqref{eq:riskhat-mod-k}. 

\begin{proposition}[Post-selection inference]
\label{prop:post-select-ci}
Let
$$
\widehat{\bd\theta}
=
\big(\widehat\theta^{(0)},\widehat\theta^{(1)},\dots,\widehat\theta^{(K)}\big)^\top \text{ and }
\bd\theta
= \big(\theta^{(0)},\theta^{(1)},\dots,\theta^{(K)}\big)^\top,
$$
where $\theta^{(0)}=\theta$ is the target ATE and $\widehat\theta^{(0)}=\stdest$ is the reference estimator. Assume that, for fixed $K$,
\begin{align*}
\sqrt{n}(\widehat{\bd\theta}-\bd\theta)\overset{d}{\longrightarrow} \mc N(\mathbf 0,\Sigma),
\end{align*}
for some positive definite covariance matrix $\Sigma\in\mathbb R^{(K+1)\times(K+1)}$. Let $\widehat\Sigma$ be a consistent estimator of $\Sigma$, and let $\widehat\sigma_k$ and $\widehat\tau_k$ be consistent estimators of the asymptotic standard deviations of $\sqrt n(\widehat\theta^{(k)}-\theta^{(k)})$ and $\sqrt n\{(\widehat\theta^{(k)}-\widehat\theta^{(0)})-(\theta^{(k)}-\theta^{(0)})\}$, respectively, for $g=0,1,\dots,G$. Recall the selected estimator index in \eqref{eq:riskhat-mod-k}. Conditionally on the observed data, draw
$$
\bd Z=(Z_0,Z_1,\dots,Z_G)^\top \sim \mathcal N(\sqrt n\,\widehat{\bd\theta},\,\widehat\Sigma).
$$
For each $k\in\{0,1,\dots,K\}$, define the selection event
\begin{align*}
A_k^{\text{sel}}
& =
\bigg\{
\left[(Z_k-Z_0)^2-\widehat\tau_k^2\right]_+ + \widehat\sigma_k^2
<
\min_{k'\neq k}
\left(
\left[(Z_{k'}-Z_0)^2-\widehat\tau_{k'}^2\right]_+ + \widehat\sigma_{k'}^2
\right)
\bigg\}. 
\end{align*}
For any $\beta\in\mathbb R$, define
\begin{align*}
b_{\mc D}(\beta)
=
\sum_{k=0}^K
\Px\left(
\left\{Z_k-\sqrt n\,\widehat\theta^{(0)}\le \beta\right\}
\cap A_k^{\text{sel}}~\bigg|~\mc D
\right),
\end{align*}
where $\mc D=\{\mc O_1,\dots,\mc O_n\}$ denotes the observed data. Then, for any fixed $\alpha\in(0,1)$, with probability tending to one the inverse map $b_{\mc D}^{-1}:(0,1)\to\mathbb R$ is well-defined, and the interval
\begin{align*}
\left[
\widehat\theta^{(\widehat k)}-\frac{b_{\mc D}^{-1}(1-\alpha/2)}{\sqrt n},
\ 
\widehat\theta^{(\widehat k)}-\frac{b_{\mc D}^{-1}(\alpha/2)}{\sqrt n}
\right]
\end{align*}
is an asymptotically valid $100(1-\alpha)\%$ CI for $\theta^{(0)}=\theta$, that is,
$$
\Px\left(
\theta \in
\left[
\widehat\theta^{(\widehat k)}-\frac{b_{\mc D}^{-1}(1-\alpha/2)}{\sqrt n},
\ 
\widehat\theta^{(\widehat k)}-\frac{b_{\mc D}^{-1}(\alpha/2)}{\sqrt n}
\right]
\right)
\longrightarrow 1-\alpha.
$$
\end{proposition}

\newpage

\section{Additional Simulation Details}\label{app:add-simu}

We include additional simulation details and results that are not presented in the main text in this section. 

\subsection{Data generation details of DGP I}\label{subapp:DGP-I}

We consider a study with a binary treatment $A\in\{0,1\}$, baseline covariates
$\mb X=(X_1,X_2,X_3)^\top$ with
$\mb X\sim\mc N(0,I_3)$, and three candidate surrogate variables
$(Q_1,Q_2,Q_3)$.

Treatment assignment depends on all three baseline covariates according to
$$
\Px(A=1\mid\mb X)
=
\mathrm{logit}^{-1}
\left(
0.6X_1-0.4X_2+0.3X_3
\right).
$$

The surrogate variables are generated as
\begin{align*}
Q_1
&=
10+ AX_1^2+X_1+\varepsilon_{Q1},\\
Q_2
&=
-5+\delta_2 A 
+
\sin(X_3) + X_2^2
+
\varepsilon_{Q2},\\
Q_3
&=
5+\delta_3 A
+
\Ix(X_2>0)
+
X_3^2
+
\varepsilon_{Q3},
\end{align*}
where
$$
\varepsilon_{Q1}\sim\mc N(0,1),\qquad
\varepsilon_{Q2}\sim\mc N(0,1),\qquad
\varepsilon_{Q3}\sim\mc N(0,1),
$$
and the three error terms are mutually independent and independent of
$(A,\mb X)$.

The primary outcome is generated according to
$$
Y
=
-Q_1+X_2+X_3+\varepsilon_Y,
\qquad
\varepsilon_Y\sim\mc N(0,\sigma^2),
$$
where $\varepsilon_Y$ is independent of
$(A,\mb X,Q_1,Q_2,Q_3)$.

Under this DGP, $Q_1$ is the only candidate surrogate through which treatment affects the primary outcome. In particular, treatment changes $Q_1$ by $\Ex(X_1^2)=1$, while the coefficient of $Q_1$ in the outcome model is $-1$, yielding a treatment effect through $Q_1$ of $-1$. This equals the total ATE, so under the structural interpretation of this DGP, $Q_1$ has a proportion of treatment effect explained (PTE) equal to one and satisfies the statistical surrogacy condition when used alone. 

In contrast, although $Q_2$ and $Q_3$ may be strongly affected by treatment, neither enters the outcome model conditional on $(\mb X,Q_1)$. Their corresponding PTEs are therefore zero. We refer to $Q_2$ and $Q_3$ as ``bad'' candidate surrogates in the sense that their treatment responsiveness may make them appear plausible as surrogate variables even though they do not explain the treatment effect on the primary outcome. When used without $Q_1$, they fail to satisfy the statistical surrogacy condition.

We vary
$$
(\delta_2,\delta_3)\in\{2,5,10\}^2
$$
to change the degree to which the invalid candidate surrogates respond to treatment, without changing their relevance for the primary outcome. We additionally consider
$$
\sigma\in\{1,2,3\}
$$
to vary the residual variability of the primary outcome.

For each simulated data set, all candidate estimators were computed using five-fold cross-fitting. For the standard AIPW estimator without surrogate information, the treatment propensity score was fitted by logistic regression of $A$ on $(X_1,X_2,X_3)$, and the outcome regressions were fitted separately within each treatment group using linear regression of $Y$ on the baseline covariates. 

For a candidate surrogate set $Q_S\subseteq\{Q_1,Q_2,Q_3\}$, the surrogate-EIF estimator used four working nuisance models. First, the treatment propensity $\Px(A=1\mid\mb X)$ was estimated by logistic regression
on $\mb X$. Second, the augmented treatment propensity
$\Px(A=1\mid\mb X,Q_S)$ was estimated by logistic regression on the baseline covariates and $Q_S$. Third, the conditional outcome mean $\Ex(Y\mid\mb X,Q_S)$ was estimated by a linear regression of $Y$ on $(\mb X,Q_S)$. Finally, the fitted conditional outcome means were regressed separately within the two treatment groups on $\mb X$. All nuisance
models were fitted in the training folds and evaluated in the held-out fold. The candidate collection consisted of the empty surrogate set, each singleton, each pair, and the full set $(Q_1,Q_2,Q_3)$. SACES used the resulting
cross-fitted influence-function estimates for candidate selection and post-selection inference.

\subsection{Additional simulation results of DGP I}\label{subapp:add-simu-I}

\begin{figure}[H]
    \centering
    \includegraphics[width=\linewidth]{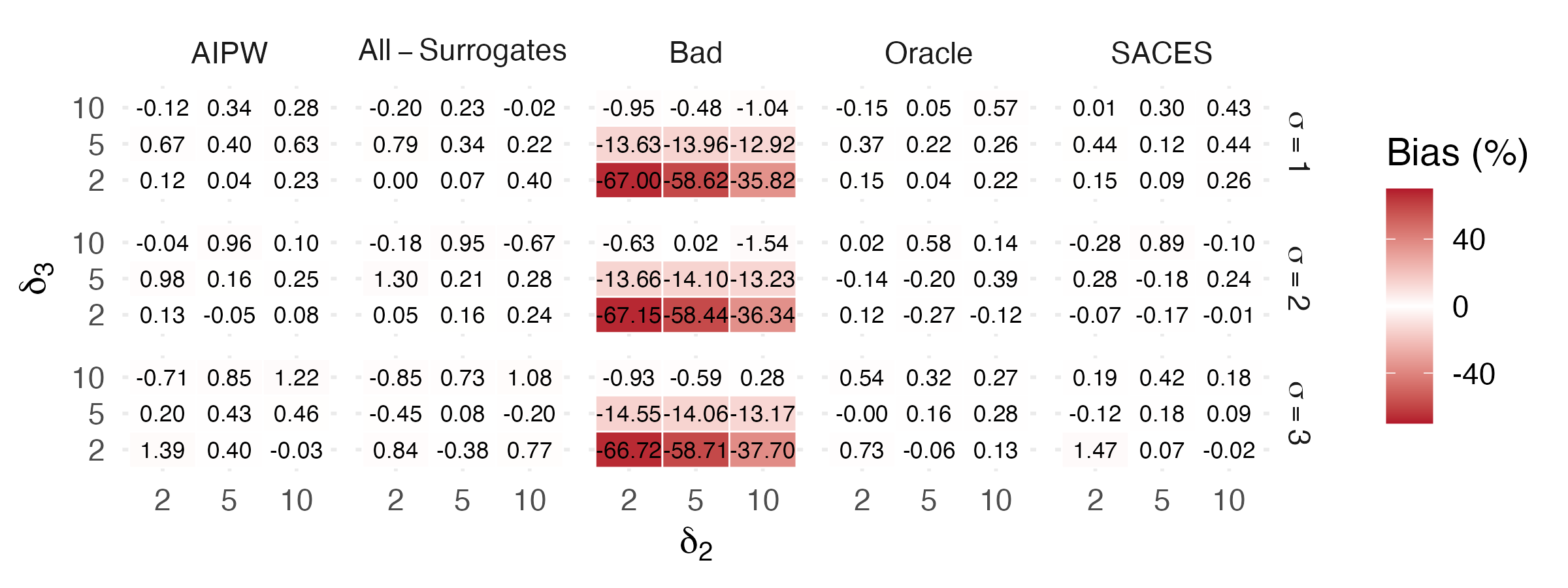}
    \caption{Percent relative estimation bias of all methods, average over 500 simulations. The proposed SACES estimator has small bias.}
\end{figure}

\begin{figure}[H]
    \centering
    \includegraphics[width=\linewidth]{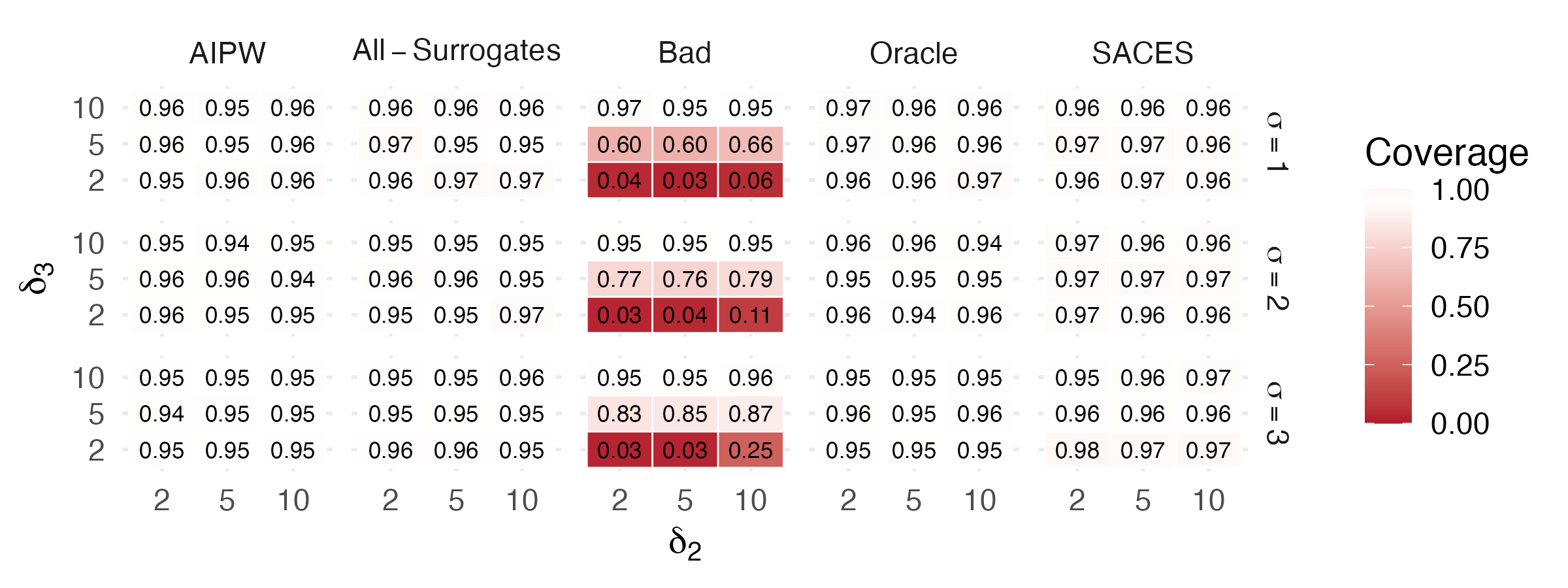}
    \caption{Coverage probabilities of 95\% CIs by all methods. The proposed SACES estimator provides valid inference, while the bad adjustment has notable issue in covering the true ATE.}
\end{figure}

\begin{figure}[H]
    \centering
    \includegraphics[width=\linewidth]{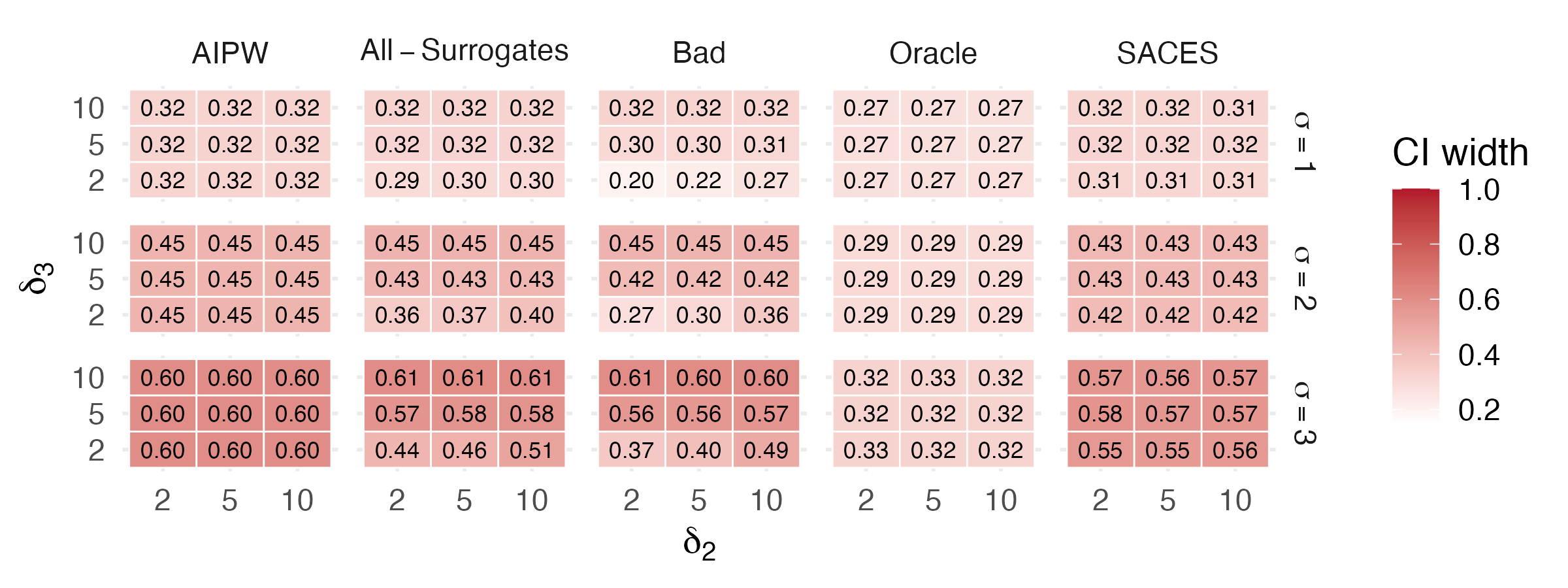}
    \caption{Interval width of 95\% CIs by all methods, average over 500 simulations. The proposed SACES estimator provides reasonable CI widths that are overall narrower than AIPW, all-surrogates and bad adjustments. The oracle method provides the shortest CIs.}
\end{figure}

\subsection{Data generation details of DGP II}\label{subapp:DGP-II}

We consider a setting with a single baseline covariate $X\sim\mc N(0,1)$, a binary treatment $A$, a candidate surrogate $Q$, and a primary outcome $Y$ subject to missingness. The treatment assignment follows
$$
A\mid X\sim\mathrm{Bernoulli}\{\mathrm{expit}(1.2X)\}.
$$
The surrogate is generated according to
$$
Q=0.5(2-\delta)A+3X+\varepsilon_Q,
$$
where $\varepsilon_Q\sim\mc N(0,1)$. The full primary outcome is generated as
$$
Y=-Q-0.5\delta A+\varepsilon_Y,
$$
where $\varepsilon_Y\sim\mc N(0,\sigma^2)$ independently of $(X,A,\varepsilon_Q)$.

This DGP is constructed to vary the degree to which the treatment effect is captured by the surrogate while keeping the target ATE fixed. Specifically, treatment changes $Q$ by $0.5(2-\delta)$, and the coefficient of $Q$ in the outcome model is $-1$. Therefore, the treatment effect operating through $Q$ is
$$
-0.5(2-\delta),
$$
while the remaining direct treatment effect on $Y$ is $-0.5\delta$. The total treatment effect is consequently
$$
-0.5(2-\delta)-0.5\delta=-1,
$$
so the true ATE is $\theta=-1$ for all values of $\delta$.

Under the structural interpretation of this DGP, the PTE by $Q$ is
$$
\mathrm{PTE}
=
\frac{-0.5(2-\delta)}{-1}
=
1-\frac{\delta}{2}.
$$
Thus, when $\delta=0$, $Q$ captures the entire treatment effect, with $\mathrm{PTE}=1$, and the statistical surrogacy condition holds. As $\delta$ increases, an increasing proportion of the treatment effect acts directly through $A$ rather than through $Q$, corresponding to progressively stronger violations of statistical surrogacy. When $\delta=2$, treatment no longer affects $Q$, so $\mathrm{PTE}=0$ and the entire treatment effect is direct.

To introduce missing primary outcomes, let $R$ denote the outcome-observation indicator. We generate
$$
R\mid A,X,Q
\sim
\mathrm{Bernoulli}\left[
\mathrm{expit}\left\{
\alpha_\delta+0.2A+0.2X+0.6(Q-3X)
\right\}
\right],
$$
where $\alpha_\delta$ is calibrated separately for each value of $\delta$ so that the marginal missing-outcome proportion is approximately $35\%$. The observed outcome is $Y$ when $R=1$ and is otherwise missing. By construction, the missingness mechanism satisfies $Y\independent R\mid A,X,Q$.

We consider
$$
\delta\in\{0,0.25,\ldots,1.75,2\},
$$
which spans settings ranging from exact statistical surrogacy to complete absence of treatment-effect explanation by the surrogate. We additionally vary
$$
\sigma\in\{1,2,3\}
$$
to control the residual outcome variability.

For each simulated dataset, we consider the unrestricted EIF-based reference estimator $\stdest$, the surrogate-EIF estimator $\surrest$, and a finite collection of weighted estimators
$$
\widehat\theta(w)
=
(1-w)\stdest+w\surrest,
\qquad
w\in\{0,0.05,\ldots,0.95,1\}.
$$
The SACES procedure treats these weighted estimators as candidate estimators and selects the value of $w$ that minimizes the estimated MSE criterion. Thus, $w=0$ corresponds to the unrestricted reference estimator, $w=1$ corresponds to the surrogate-EIF estimator, and intermediate values allow partial incorporation of the statistical surrogacy restriction.

For each simulated dataset, all nuisance quantities are estimated using five-fold cross-fitting. The treatment propensity score $\pi(1\mid X)=\Px(A=1\mid X)$ is estimated by logistic regression of $A$ on $X$, while $\widetilde\pi(1\mid X,Q)=\Px(A=1\mid X,Q)$ is estimated by logistic regression of $A$ on $(X,Q)$. The treatment-specific outcome-observation probabilities $r(a,X,Q)=\Px(R=1\mid A=a,X,Q)$ are estimated by logistic regression of $R$ on $(A,X,Q)$. The marginal observation probability $r(X,Q)=\Px(R=1\mid X,Q)$ is then obtained by averaging the fitted treatment-specific probabilities over $\widetilde\pi(A\mid X,Q)$.

For the unrestricted reference estimator, the treatment-specific outcome regressions $\widetilde\mu(a,X,Q)=\Ex(Y\mid A=a,X,Q)$ are fitted separately among observed outcomes using linear regression of $Y$ on $(X,Q)$. Their corresponding $X$-specific regressions are obtained by projecting the fitted values onto $X$ separately within each treatment group. For the surrogate-EIF estimator, we fit the pooled regression $\nu(X,Q)=\Ex(Y\mid X,Q)$ among participants with observed outcomes and then project its fitted values onto $X$ separately within each treatment group. All nuisance models are fitted in the training folds and evaluated in the held-out fold.

\subsection{Additional simulation results of DGP II}\label{subapp:add-simu-II}

\begin{figure}[H]
    \centering
    \includegraphics[width=\linewidth]{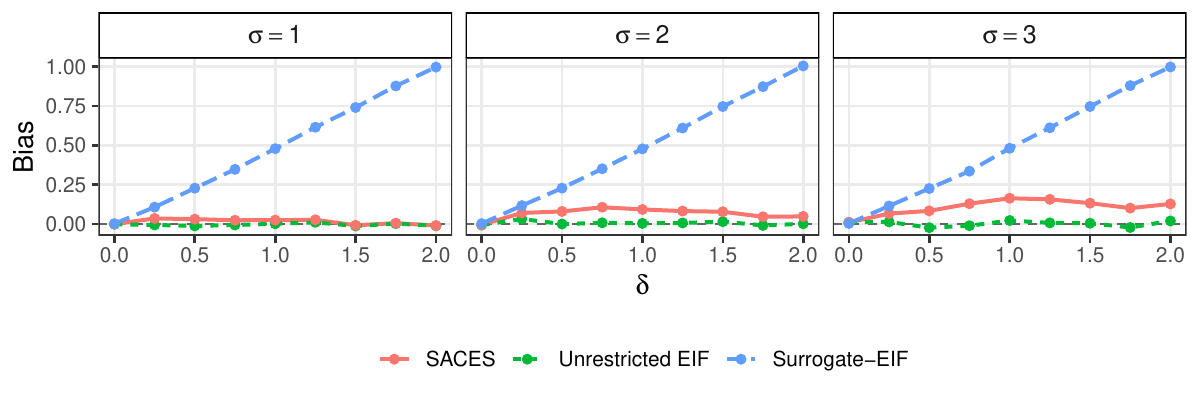}
    \caption{Bias vs. $\delta$ in simulation under three $\sigma$ values, average over 500 simulations. }
\end{figure}

\begin{figure}[H]
    \centering
    \includegraphics[width=\linewidth]{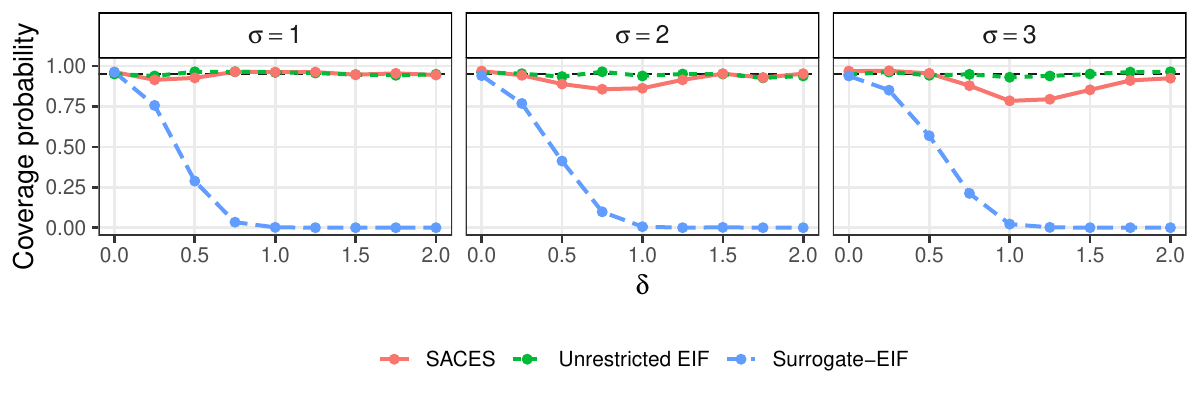}
    \caption{Coverage probability vs. $\delta$ in simulation under three $\sigma$ values. }
\end{figure}

\begin{figure}[H]
    \centering
    \includegraphics[width=\linewidth]{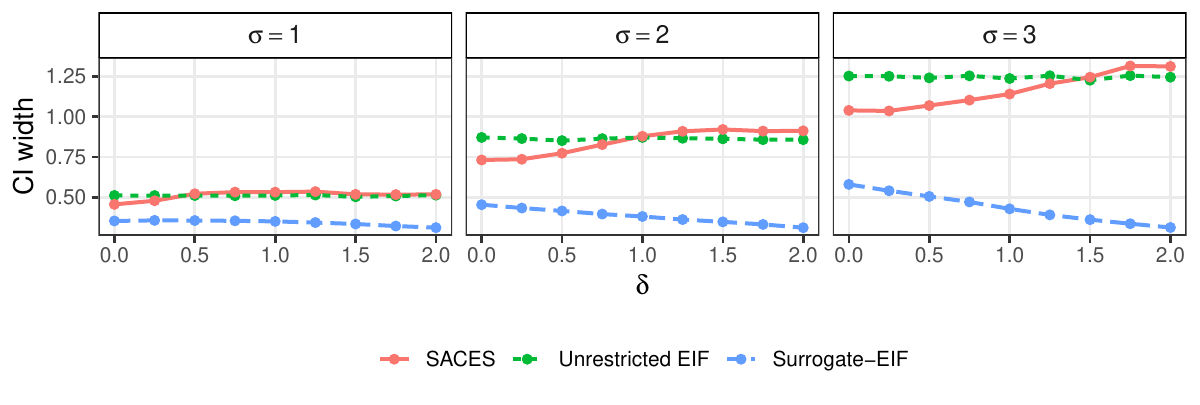}
    \caption{CI width vs. $\delta$ in simulation under three $\sigma$ values, average over 500 simulations. }
\end{figure}

\begin{figure}[H]
    \centering
    \includegraphics[width=\linewidth]{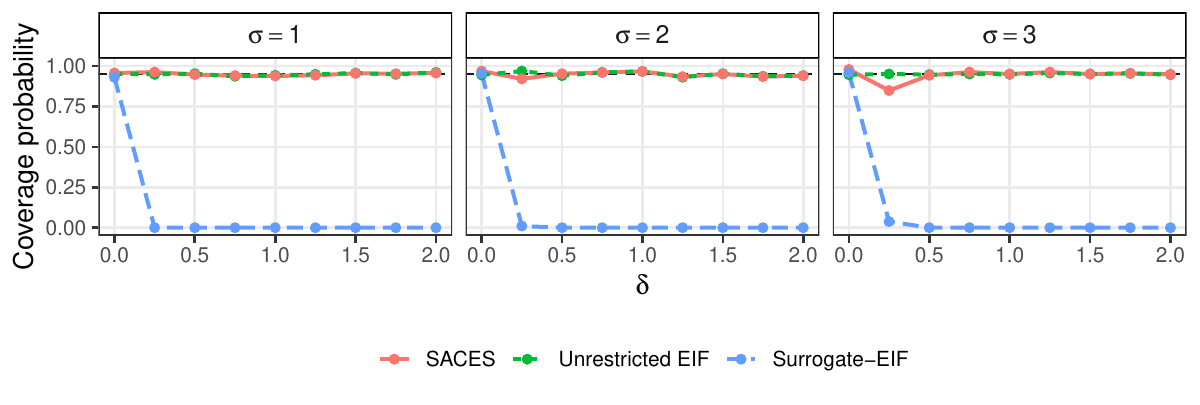}
    \caption{Coverage probability vs. $\delta$ in simulation under three $\sigma$ values and a larger sample size $n=20,000$. This results justify that the inference procedure is asymptotically valid with sufficiently large $n$. }
\end{figure}

\subsection{Data generation details of DGP III}\label{subapp:DGP-III}

For each Monte Carlo replication, we generated $n=1,000$ observations according to
\[
X\sim\mc N(0,1),
\qquad
A\mid X\sim\mathrm{Bernoulli}
\{\operatorname{expit}(-0.3+0.8X)\},
\]
and
\[
Q=0.5A+X+\varepsilon_Q,
\qquad
G=1.5X+\varepsilon_G,
\qquad
B=\varepsilon_B,
\]
where $\varepsilon_Q,\varepsilon_G,\varepsilon_B$ are mutually independent standard normal variables, independent of $(X,A)$. The primary outcome was generated as
\[
Y=5+0.5Q+\gamma_GG+\varepsilon_Y,
\qquad
\varepsilon_Y\sim\mc N(0,1),
\]
with $\varepsilon_Y$ independent of all other variables. Thus, $\gamma_G$ controls the additional outcome information provided by $G$. Since treatment affects $Y$ only through $Q$, the true ATE is $\theta=0.5\times0.5=0.25$ throughout, and statistical surrogacy holds. However, none of the estimators compared in this experiment imposes statistical surrogacy.

Let $R$ indicate that $Y$ is observed. We generated
\[
R\mid A,X,Q,G,B,Y
\sim\mathrm{Bernoulli}
\left[
\operatorname{expit}
\{\alpha_a+0.25X+0.15Q+\gamma_BB+0.35a\}
\right],
\qquad A=a.
\]
Within each generated sample, the intercept $\alpha_a$ was calibrated separately for each treatment group so that its average observation probability was $0.70$. The realized missingness proportions therefore fluctuate around $30\%$. The parameter $\gamma_B$ controls the additional missingness information provided by $B$.

By construction, $G$ provides outcome information but no additional information about missingness given $(X,A,Q)$, whereas $B$ provides missingness information but no additional outcome information given $(X,A,Q,R)$. MAR holds for each information set considered. We varied
\[
\gamma_G\in\{0,0.8,0.6,2.4\},
\qquad
\gamma_B\in\{0,0.5,1.0,1.5\},
\]
yielding 16 scenarios, each with 500 Monte Carlo replications. When either coefficient is zero, the corresponding variable provides no additional information of the specified type.

We compared the unrestricted MAR influence-function estimators using $Q$, $(Q,G)$, $(Q,B)$, and $(Q,G,B)$, together with SACES. All estimators used the same five cross-fitting folds within each replication. The baseline treatment propensity was fitted using logistic regression on $X$. For each surrogate set $S$, treatment-specific observation probabilities were fitted using logistic regression on $(X,S)$, and treatment-specific outcome regressions were fitted using linear regression on $(X,S)$ among participants with observed outcomes. The resulting outcome predictions were then regressed on $X$ within each treatment group to estimate the corresponding baseline-conditional means. 

SACES used the Kallus--Mao's unrestricted-EIF estimator using $Q$ as the only surrogate as its reference and selected among all four unrestricted-EIF estimators by minimizing the modified estimated risk criterion. Its variance and covariance inputs were computed jointly from their cross-fitted influence-function scores. For the four fixed estimators, we constructed Wald 95\% CIs using influence-function-based standard errors. For SACES, we used the Gaussian post-selection procedure with 1,000 draws per replication.

\subsection{Additional simulation results of DGP III}\label{subapp:add-simu-III}

\begin{figure}[H]
    \centering
    \includegraphics[width=\linewidth]{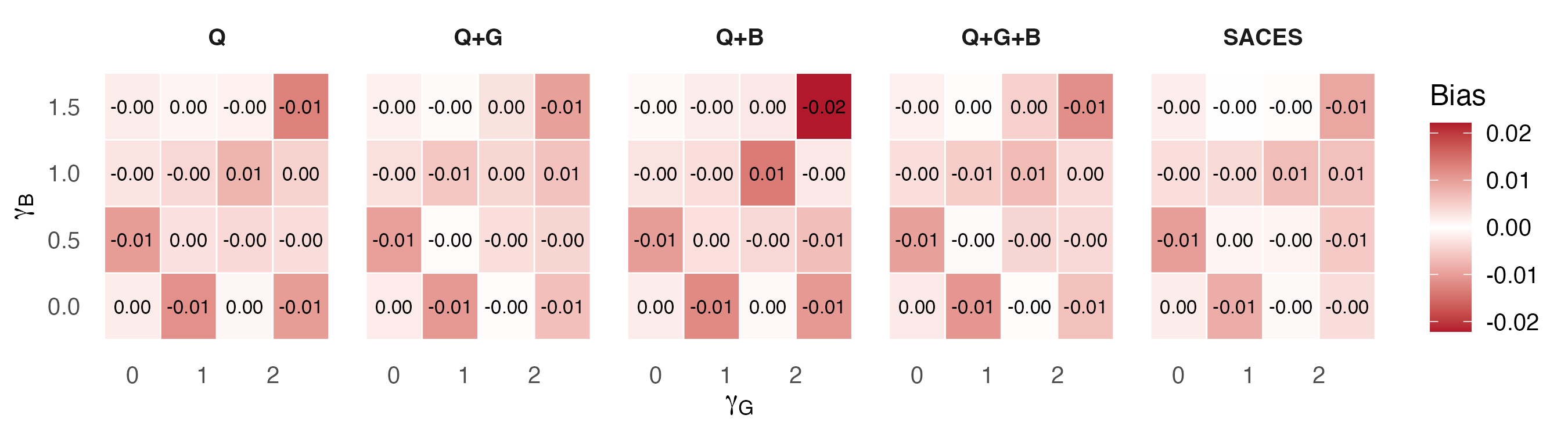}
    \caption{Bias of the candidate surrogate-assisted estimators and SACES
    across the considered values of $\gamma_G$ and $\gamma_B$ in DGP~III.}
    \label{fig:qgb-bias}
\end{figure}

\begin{figure}[H]
    \centering
    \includegraphics[width=\linewidth]{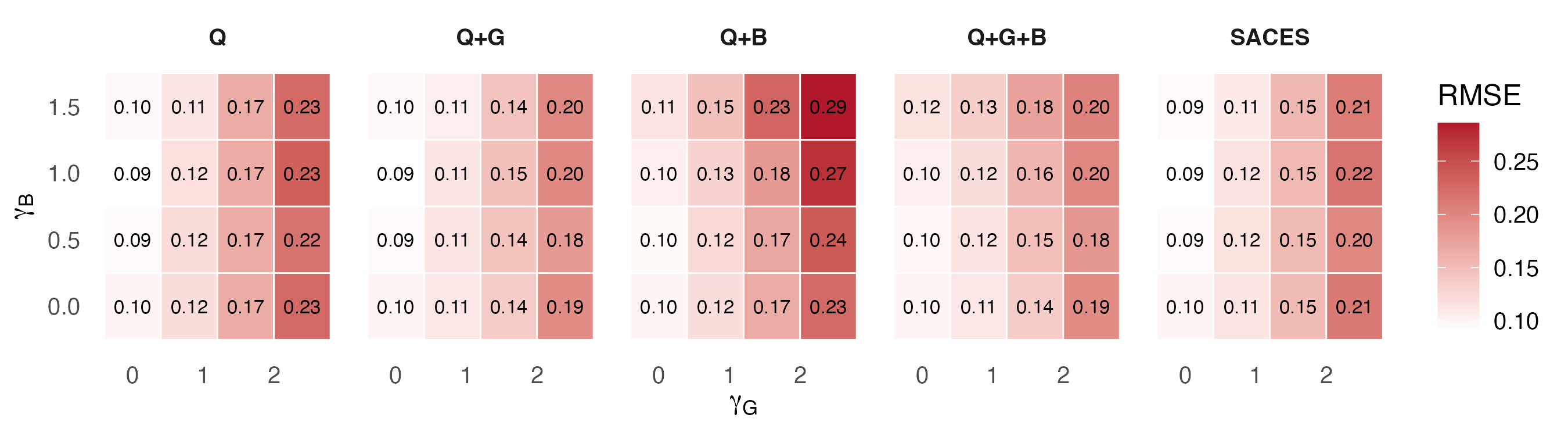}
    \caption{RMSE of the candidate surrogate-assisted estimators and SACES
    across the considered values of $\gamma_G$ and $\gamma_B$ in DGP~III.}
    \label{fig:qgb-rmse}
\end{figure}

\begin{figure}[H]
    \centering
    \includegraphics[width=\linewidth]{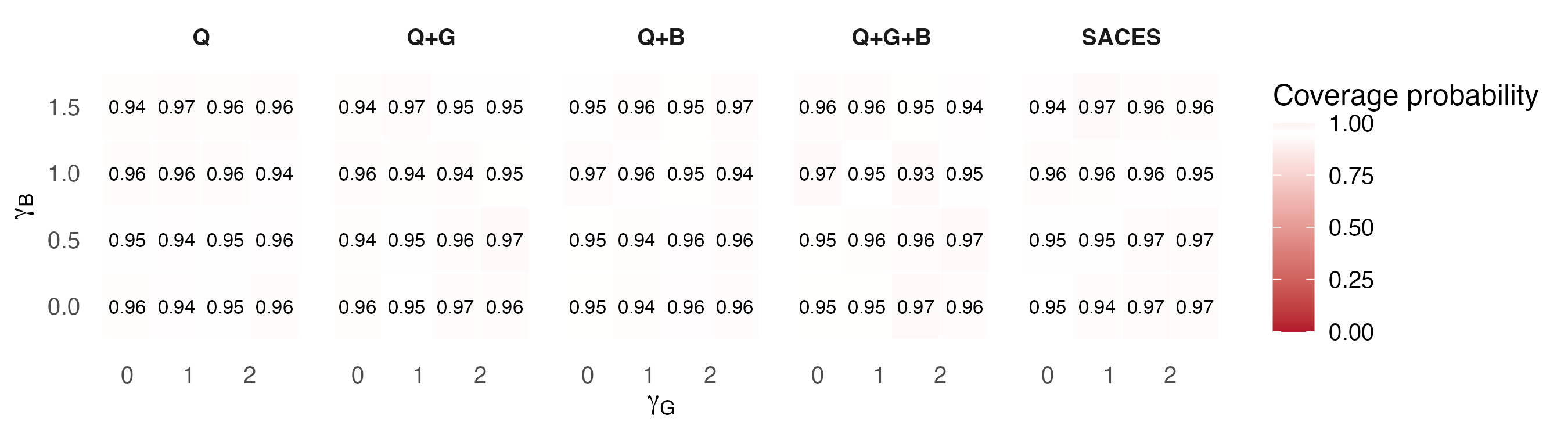}
    \caption{Coverage probabilities of the candidate surrogate-assisted
    estimators and SACES across the considered values of $\gamma_G$ and $\gamma_B$ in DGP~III.}
    \label{fig:qgb-coverage}
\end{figure}

\begin{figure}[H]
    \centering
    \includegraphics[width=\linewidth]{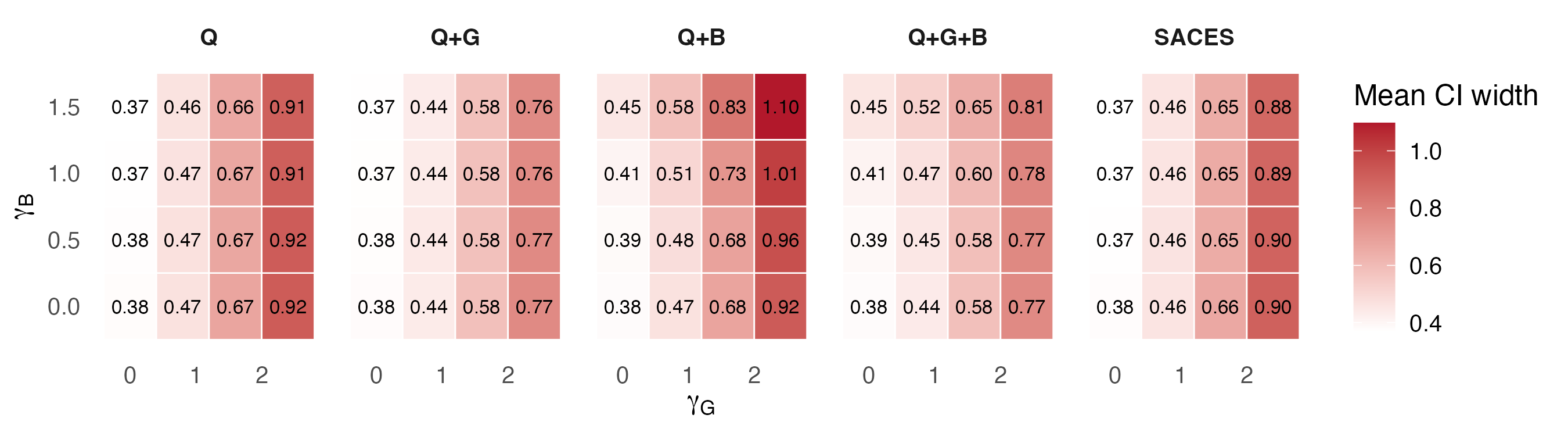}
    \caption{CI widths of the candidate surrogate-assisted
    estimators and SACES across the considered values of $\gamma_G$ and $\gamma_B$ in DGP~III.}
    \label{fig:qgb-ciw}
\end{figure}

\begin{figure}[H]
    \centering
    \includegraphics[width=\linewidth]{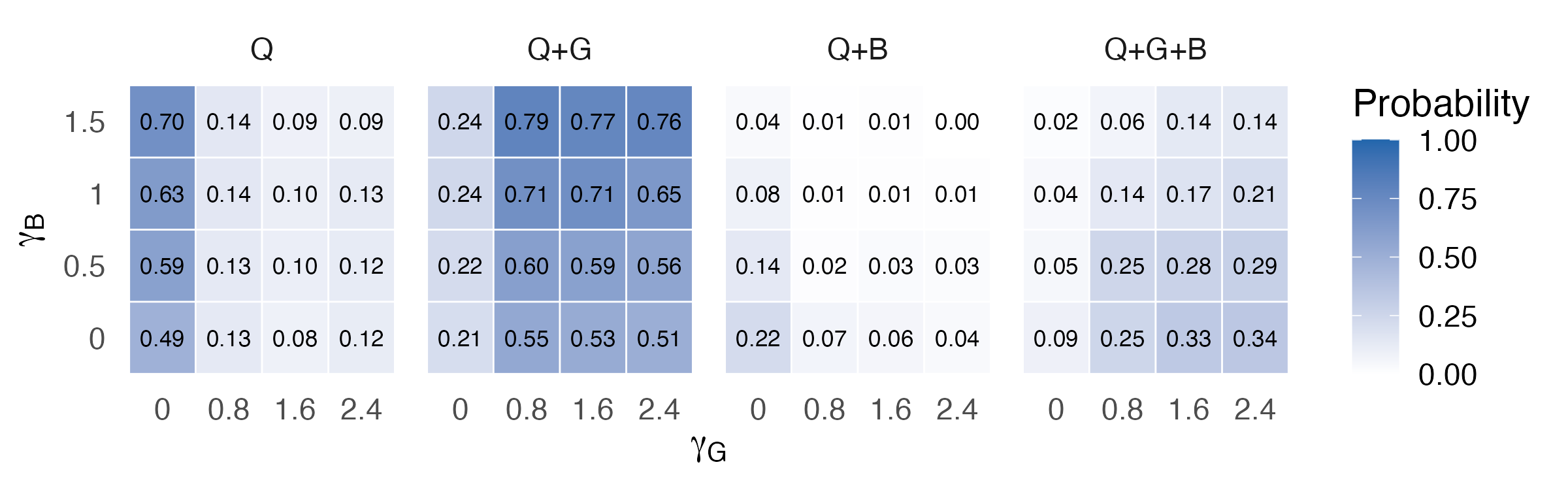}
    \caption{Selection frequencies of the candidate surrogate sets across the considered values of $\gamma_G$ and $\gamma_B$ in DGP~III.}
    \label{fig:qgb-selection}
\end{figure}

\newpage

\section{Additional Data Analysis Details}\label{app:data}

\subsection{Nuisance-model fitting}
\label{subapp:nuisance-details}

All nuisance functions in the ACTG~175 analysis were estimated using
five-fold cross-fitting. The treatment propensity score
$\pi(1\mid\mb X)=\Px(A=1\mid\mb X)$ and the surrogate-adjusted treatment propensity score $\widetilde\pi(1\mid\mb X,Q)=\Px(A=1\mid\mb X,Q)$ were fitted using logistic regression. The treatment-specific missingness probabilities $\widetilde r_a(\mb X,Q)=\Px(R=1\mid A=a,\mb X,Q)$ and the pooled missingness probability $\widetilde r(\mb X,Q)=\Px(R=1\mid\mb X,Q)$ were also fitted using logistic regression.

Among participants with observed week-96 CD4 counts, the treatment-specific outcome regressions
$\widetilde m(a,\mb X,Q)=\Ex(Y\mid A=a,\mb X,Q,R=1)$ were fitted using ordinary least-squares linear regression. The corresponding baseline regressions $m(a,\mb X)$ were obtained by regressing the fitted $\widetilde m(a,\mb X,Q)$ values on $\mb X$ within each treatment group. For the surrogate-EIF estimator, the pooled regression $\widetilde\nu(\mb X,Q)=\Ex(Y\mid\mb X,Q,R=1)$ was fitted using ordinary least squares, and $\widetilde\mu(a,\mb X)$ was obtained by regressing the fitted $\widetilde\nu(\mb X,Q)$ values on $\mb X$ within treatment group $a$.

All fitted probabilities were truncated to $[0.01,0.99]$ before forming the influence-function estimates. The ACTG 175 analysis used the \texttt{BART} package only to access the publicly available dataset. 

\subsection{Estimated treatment and missingness propensities diagnostics}\label{subapp:ps-diag-data}

\begin{table}[H]
\centering
\caption{Empirical diagnostics for fitted treatment and missingness
probabilities in the ACTG~175 analysis.}
\label{tab:actg-propensity-diagnostics}
\setlength{\tabcolsep}{4pt}
\renewcommand{\arraystretch}{1.05}
\begin{tabular}{lrrrrrrr}
\toprule
Probability
& Min & $q_{01}$ & Median & $q_{99}$ & Max
& $N_{\mathrm{clip}}$ & \% clipped \\
\midrule
$\widehat\pi(1\mid\mb X)$
& 0.476 & 0.620 & 0.755 & 0.839 & 0.881 & 0 & 0.00 \\
$\widehat\pi(A_i\mid\mb X)$
& 0.151 & 0.180 & 0.741 & 0.834 & 0.881 & 0 & 0.00 \\
$\widehat{\widetilde\pi}(1\mid\mb X,Q)$
& 0.223 & 0.486 & 0.758 & 0.932 & 0.990 & 1 & 0.05 \\
$\widehat{\widetilde\pi}(A_i\mid\mb X,Q)$
& 0.041 & 0.127 & 0.729 & 0.931 & 0.990 & 1 & 0.05 \\
$\widehat{\widetilde r}(\mb X,Q)$
& 0.247 & 0.377 & 0.636 & 0.794 & 0.910 & 0 & 0.00 \\
$\widehat{\widetilde r}_0(\mb X,Q)$
& 0.229 & 0.358 & 0.622 & 0.780 & 0.903 & 0 & 0.00 \\
$\widehat{\widetilde r}_1(\mb X,Q)$
& 0.256 & 0.382 & 0.640 & 0.796 & 0.908 & 0 & 0.00 \\
$\widehat{\widetilde r}_{A_i}(\mb X,Q)$
& 0.256 & 0.377 & 0.636 & 0.793 & 0.908 & 0 & 0.00 \\
\bottomrule
\end{tabular}
\begin{tablenotes}\scriptsize
\item The diagnostics are based on the $2,139$ participants. Quantiles are computed from the raw cross-fitted predictions before numerical truncation. The fitted probabilities used in the influence-function calculations were truncated to $[0.01,0.99]$. $A_i$ denotes the observed treatment assignment for participant $i$; thus, the corresponding rows evaluate the probability assigned to the observed treatment. Only one fitted value of $\widehat{\widetilde\pi}(1\mid\mb X,Q)$ exceeded $0.99$ and was truncated.
\end{tablenotes}
\end{table}

The diagnostics did not indicate substantial empirical instability in the fitted treatment or missingness models. The minimum fitted probability assigned to the observed treatment was $0.0406$, the minimum estimated missingness probability was $0.2286$, and only one fitted treatment propensity exceeded the truncation threshold of $0.99$. Although no formal overlap condition is imposed on
$\widetilde\pi(a\mid\mb X,Q)$, its empirical distribution is reported as a diagnostic for possible finite-sample instability.

\end{document}